\newcommand{\oldtext}[1]{}
\newcommand{\ot}[1]{}

\newcommand{\newtext}[1]{#1}
\newcommand{\nt}[1]{#1}
\documentclass[11pt,preprint]{aastex}
\usepackage{natbib}
\usepackage{color}
\tightenlines

\newcommand     \um     {$\mu$m}        
\newcommand     \mum    {\,\mu{\rm m}}  

\newcommand     \Angstrom       {\,{\rm \AA}}

\newcommand     \beq    {\begin{equation}}
\newcommand     \beqa   {\begin{eqnarray}}
\newcommand     \cm     {\,{\rm cm}}

\newcommand     \eeq    {\end{equation}}
\newcommand     \eeqa   {\end{eqnarray}}
\newcommand     \erg    {\,{\rm ergs}}

\newcommand     \eV     {\,{\rm eV}}
\newcommand     \fcold  {f_{\rm cold}}

\newcommand     \g              {\,{\rm g}}
\newcommand     \gtsim  {\gtrsim}                

\newcommand     \HH     {{\rm H}_2}

\newcommand     \JCMT   {{\it JCMT }}

\newcommand     \Jy     {\,{\rm Jy}}

\newcommand     \K              {\,{\rm K}}
\newcommand     \kms    {\,{\rm km~s}^{-1}}

\newcommand     \Ldust  {L_{\rm dust}}
\newcommand     \Lsol   {L_{\odot}}
\newcommand     \ltsim  {\lesssim}               
\newcommand     \Mpc    {\,{\rm Mpc}}
\newcommand     \Msol   {M_{\odot}}
\newcommand     \Mdust  {M_{\rm dust}}
\newcommand     \MH     {M_{\rm H}}

\newcommand     \NH     {N_{\rm H}}
\newcommand     \OH     {A_{\rm O}}    
\newcommand     \qpah   {q_{\rm PAH}}

\newcommand     \s      {\,{\rm s}}

\newcommand     \SINGS  {{SINGS}}
\newcommand     \Spitzer {{\it Spitzer}}

\newcommand     \SST    {{\it Spitzer }{\it Space }{\it Telescope}}
\newcommand     \Tcold  {T_{\rm cold}}
\newcommand     \Twarm  {T_{\rm diff}}
\newcommand     \Umax   {U_{\rm max}}
\newcommand     \Umin   {U_{\rm min}}
\newcommand     \XCO    {X_{\rm CO}}
\newcommand     \yr     {\,{\rm yr}}
\newcommand{\btdnote}[1]{}

\newcommand     \Ntot   {65}          
\newcommand     \Nmips  {64}          
\newcommand     \Nscuba {17}          
\newcommand     \Nscubamips {16}      
\newcommand     \Nnoscuba {48}        
\newcommand     \Npah   {61}          

\newlength{\figwidth}
\newlength{\figwidthw}
\newlength{\figwidthww}
\newlength{\figwidthd}
\newcommand    \angl    {270}

\newcommand{\sedfigd}[2]{{\begin{center}
                         \includegraphics[angle=\angl,width=\figwidthd]%
                         {#1}
                         \includegraphics[angle=\angl,width=\figwidthd]%
                         {#2}
                         \end{center}
                         \vspace*{-2em}
			 }
                        }

\countdef\decade=200
\advance\decade by \year
\countdef\hours=201
\advance\hours by \time
\divide\hours by 60
\countdef\mins=202
\advance\mins by \hours
\multiply\mins by 60
\multiply\hours by 100
\countdef\miltime=203
\advance\miltime by \hours
\advance\miltime by \time
\advance\miltime by -\mins

\begin{document}
\title{
        \vspace*{-3.0em}
        {\normalsize\rm To appear in {\it The Astrophysical Journal}}\\ 
        \vspace*{1.0em}
       Dust Masses, PAH Abundances, 
       and Starlight Intensities in the SINGS Galaxy Sample
        }

\author{B.~T.~Draine\altaffilmark{1}, 
        D.~A.~Dale\altaffilmark{2}, 
	G.~Bendo\altaffilmark{3},
        K.~D.~Gordon\altaffilmark{4},
	J.~D.~T.~Smith\altaffilmark{4},
	L.~Armus\altaffilmark{5},
	C.~W.~Engelbracht\altaffilmark{4},
	G.~Helou\altaffilmark{6},
	R.~C.~Kennicutt\altaffilmark{4,7},
        A.~Li\altaffilmark{8},
	H.~Roussel\altaffilmark{9},
	F.~Walter\altaffilmark{9},
	D.~Calzetti\altaffilmark{10},
        J.~Moustakas\altaffilmark{4,11},
        E.~J.~Murphy\altaffilmark{12},
	G.~H.~Rieke\altaffilmark{4},
	C.~Bot\altaffilmark{6},
        D.~J.~Hollenbach\altaffilmark{13},
	K.~Sheth\altaffilmark{6},
	and
        H.~I.~Teplitz\altaffilmark{5}
	}

\altaffiltext{1}{Princeton University Observatory, 
                 Peyton Hall, 
                 Princeton, NJ 08544-1001; 
                 draine@astro.princeton.edu}
\altaffiltext{2}{Dept.\ of Physics \& Astronomy,
                 University of Wyoming,
                 Laramie, WY 82071;
		 ddale@uwyo.edu}
\altaffiltext{3}{Astrophysics Group, Imperial College,
                 Blackett Laboratory,
                 Prince Consort Road,
                 London SW7 2AZ, UK;
                 g.bendo@imperial.ac.uk}
\altaffiltext{4}{Steward Observatory, 
                 University of Arizona,
                 Tucson, AZ 85721;
		 kgordon@as.arizona.edu, 
		 cengelbracht@as.arizona.edu,
		 jdsmith@as.arizona.edu
                 }
\altaffiltext{5}{Spitzer Science Center, MS 220-6, Caltech, 
                 Pasadena CA 91125; 
		 lee@ipac.caltech.edu,hit@ipac.caltech.edu}
\altaffiltext{6}{Caltech, MC 314-6, Pasadena, CA 91125;
                 gxh@ipac.caltech.edu, bot@caltech.edu}
\altaffiltext{7}{Institute of Astronomy,
                 University of Cambridge,
                 Madingley Road, Cambridge CB3 0HA, UK;
                 robk@ast.cam.ac.uk}
\altaffiltext{8}{Department of Physics \& Astronomy,
                 University of Missouri,
                 Columbia, MO 65211;
                 LiA@missouri.edu}
\altaffiltext{9}{Max-Planck-Institut f\"ur Astronomie,
                  K\"onigstuhl 17, D-69117,
		  Heidelberg, Germany;
		  roussel@mpia-hd.mpg.de,
		  walter@mpia.de}
\altaffiltext{10}{Space Telescope Science Institute, 
                 3700 San Martin Drive,
                 Baltimore, MD 21218;
                 calzetti@stsci.edu}
\altaffiltext{11}{Dept.\ of Physics,
                  New York University,
		  4 Washington Place,
		  New York, NY 10003;
		  john.moustakas@nyu.edu}
\altaffiltext{12}{Dept.\ of Astronomy, Yale University,
                  P.O.~Box 208101,
		  New Haven, CT 06520-8101;
		  murphy@astro.yale.edu}
\altaffiltext{13}{NASA/Ames Research Center,
                  MS 256-6,
                  Moffett Field, CA 94035;
                  hollenba@ism.arc.nasa.gov}
\begin{abstract}
Physical dust models are presented for \Ntot\ 
galaxies in the \SINGS\ survey that are strongly detected
in the four IRAC bands
and three MIPS bands.
For each galaxy we estimate
(1) the total dust mass,
(2) the fraction of the dust mass contributed by PAHs, and
(3) the intensity of the starlight heating the dust grains.
We find that spiral galaxies have dust properties
resembling the dust in the local region of the Milky Way,
with similar dust-to-gas ratio, and
similar PAH abundance.
The observed SEDs, including galaxies with SCUBA photometry, can
be reproduced by dust models that do not require ``cold'' ($T\ltsim10\K$) dust.

The dust-to-gas ratio is observed to be dependent on metallicity.
In the interstellar media of galaxies with 
$\OH \equiv 12 + \log_{10}({\rm O/H}) >8.1$,
grains contain a substantial fraction of interstellar
Mg, Si and Fe.
Galaxies with $\OH<8.1$ and extended \ion{H}{1} envelopes
in some cases appear to have global dust-to-gas ratios that are
low for their measured oxygen abundance, but the dust-to-gas ratio
{\it in the regions where infrared emission is detected} generally
appears to be consistent with a substantial fraction of interstellar
Mg, Si, and Fe being
contained in dust.
The PAH index $\qpah$ -- 
the fraction of the dust mass in the form of PAHs -- 
correlates with metallicity.  
The nine galaxies in our sample
with $\OH<8.1$ have a median $\qpah=\oldtext{1.1}\newtext{1.0}\%$, whereas
galaxies with $\OH>8.1$ have a median $\qpah=\oldtext{3.5}\newtext{3.55}\%$.
The derived dust masses favor a value 
$\XCO\approx4\times10^{20}\cm^{-2}(\K\kms)^{-1}$
for the CO to H$_2$ conversion factor.
\newtext{Except for some starbursting systems (Mrk~33, Tolo~89, 
NGC~3049), dust in the diffuse ISM dominates the IR power.}
\end{abstract}


\section{\label{sec:intro}
         Introduction}

Interstellar dust affects the appearance of galaxies,
by attenuating short wavelength radiation from stars and ionized gas, and
contributing infrared (IR), far-infrared (FIR), submm, mm, and microwave emission
[for a recent review, see \citet{Draine_2003a}].
Dust also is an important agent in the
fluid dynamics, chemistry, heating, cooling, and even ionization balance
in some interstellar regions,
with a major role in the process of star formation.
Despite the importance of dust, determination of the physical properties of
interstellar dust grains has been a challenging task -- even the overall amount
of dust in other galaxies has often been very uncertain.

The \SINGS\ galaxy sample \citep{Kennicutt+Armus+Bendo_etal_2003}
comprises 75 galaxies with a wide range of morphological
types, all at distances allowing the galaxies to be resolved even at 160\um.
Observations of this sample with all instruments on \SST\ 
\citep{Werner+Roellig+Low_etal_2004} are providing a new
perspective on the dust abundances and 
dust properties in a diverse set of galaxies.

For
67 of the \SINGS\ galaxies we have positive detections 
in the 4 bands of the 
{\it Infrared Array Camera} 
(IRAC; \citet{Fazio+Hora+Allen_etal_2004}) and the three bands of the
{\it Multiband Imaging Photometer for Spitzer} 
(MIPS;  \citet{Rieke+Young+Engelbracht_etal_2004}), spanning 
wavelengths from 3.6\um\ to 160\um, the wavelength range within which
dust radiates most of its power, including the PAH emission
features that are prominent in many galaxies;
65 of these galaxies are used in the present study.
For \oldtext{nine}\newtext{11} 
galaxies we have global fluxes measured with the IRS~16\um\
peak-up filter.
For \Nscuba\ of the \Ntot\ galaxies, global \newtext{850\um} 
flux measurements 
\oldtext{at submillimeter wavelengths}%
are also available from the SCUBA camera on the \JCMT.

The IR emission from dust depends not only on the amount of dust
present, but also on the rate at which it is heated by starlight.
We develop a technique for estimation of dust masses and starlight
intensities in galaxies, and apply it to estimate the total dust mass
$\Mdust$ in each of the \Ntot\ 
galaxies in this study.
For \Npah\ galaxies we are also able to estimate a ``PAH index'' $\qpah$ -- 
defined here to be the mass fraction of the dust that is
contributed by polycyclic aromatic hydrocarbon (PAH) particles 
containing $<10^3$ C atoms.
In addition, we obtain information on the
starlight intensity distribution -- the mean starlight intensity scale
factor $\langle U\rangle$, and 
the fraction of the dust heating that takes place in regions of
high starlight intensities, including photodissociation regions.

The paper is organized as follows.
The sample selection and observational data are summarized in \S\ref{sec:data},
the physical dust models that are used are described in
\S\ref{sec:dust models}, and
the model-fitting procedure is given in \S\ref{sec:model fitting}.
In order to validate the model-fitting procedure, we first limit attention to
the subset of \SINGS\ galaxies for which SCUBA data are available (we will
refer to these as the \SINGS-SCUBA galaxies).
In \S\ref{sec:SCUBA sample} we demonstrate that the dust modeling procedure
works well for most of the \SINGS-SCUBA galaxies, and we present
dust masses, dust-to-gas mass ratios,
PAH index values, and starlight intensity parameters obtained
from these model fits.
For the \SINGS-SCUBA galaxies, our models 
do not require any contribution from the very cold
($<10\K$) dust that has been reported by
\citet{Krugel+Siebenmorgen+Zota+Chini_1998} for the spiral galaxies
NGC~6156 and NGC~6918, or the
5~--~9~K dust that 
\citet{Galliano+Madden+Jones_etal_2003,Galliano+Madden+Jones_etal_2005}
reported for low metallicity dwarf galaxies,
or the 4--6~K dust that 
\citet{Dumke+Krause+Wielebinski_2004} reported for the edge-on-spiral
NGC~4631.
The global SEDs do not show evidence for the submillimeter excess seen in
the faint outer regions of
NGC~4631 \citep{Bendo+Dale+Draine_etal_2006}.

Submillimeter measurements are valuable to help constrain the
dust models, but are unfortunately not available for most galaxies.
In \S\ref{sec:descuba}
we use the \SINGS-SCUBA galaxies to develop a
``restricted'' fitting procedure that can be applied to galaxies for
which submillimeter fluxes are unavailable.
For the \Nscuba\ galaxies with 
global 850\um\ fluxes from SCUBA, we show that the
restricted fitting procedure, without employing SCUBA data, 
recovers dust masses and starlight
intensities 
that are usually within a factor of two 
of the values obtained when the SCUBA data are used.

The restricted fitting technique is 
then applied to the remaining \Nnoscuba\ 
galaxies to estimate dust masses and starlight
intensities.
Dust-to-gas mass ratios for the overall sample are presented in 
\S\ref{sec:dust to gas ratio}, where it is found that some of the
\SINGS\ galaxies have
very low dust-to-gas ratios, and that the 
dust-to-gas ratios are correlated with
both gas-phase metallicity and galaxy type.
The dust-to-gas ratios depend on the assumed value of
$\XCO$, the ratio of H$_2$ column density to CO line intensity;
in \S\ref{sec:XCO} we argue that the
distribution of derived dust-to-gas ratios supports 
$\XCO\approx4\times 10^{20}\cm^{-2}(\K\kms)^{-1}$.

Values of the PAH index $\qpah$ are estimated for \Npah\ galaxies in
\S\ref{sec:PAH abundances - scuba plus nonscuba}.
Many galaxies are found to have low values of $\qpah$, and
$\qpah$ is found to be strongly correlated with gas-phase
metallicity. 
The starlight parameters for the \Ntot\ galaxy sample are
presented and discussed in 
\S\ref{sec:starlight parameters - scuba plus nonscuba}.
The results are discussed in \S\ref{sec:discussion},
and summarized in \S\ref{sec:summary}.

\section{\label{sec:data}
         Data}
\subsection{IRAC and MIPS}

\SST\ was used to observe all 75 galaxies in the \SINGS\ sample, following
the observing strategy described in \citet{Kennicutt+Armus+Bendo_etal_2003}.
The observations in the 4 IRAC bands (3.6, 4.5, 5.7, 7.9\um) and 
3 MIPS bands (24, 71, 160\um) have been reduced following the
procedures described in \citet{Dale+Bendo+Engelbracht_etal_2005}.
The extended source aperture corrections found 
in \citet{Reach+Megeath+Cohen_etal_2005} were applied to the IRAC data.
Global IRAC and MIPS photometry for these galaxies is taken from
\citet{Dale+GildePaz+Gordon+etal_2007}, which included (usually modest)
corrections for aperture size and nonlinearity of the 71\um\ detectors.

Ten of the SINGS galaxies were excluded from the present study:
\begin{itemize}
\item NGC~584 (E4) may be contaminated by a background source, and
also has low surface brightness in the MIPS bands
compared to the foreground cirrus.
\item NGC~3034~=~M~82 (I0, starburst): Saturation of the bright core of this
galaxy precludes
global photometry.
\item NGC~1404 (E1): An off-center IR source contributes a significant fraction of the
MIPS flux, but it is not
known whether or not it is a foreground/background object.
\item NGC~4552 (E0) has low surface brightness compared to the foreground
cirrus emission,
and the MIPS 160\um\ photometry is very uncertain.
\item The dwarf galaxies M81dwA and Ho~IX are not detected at 
$\lambda\ge5.8\mum$.
\item The dwarf galaxies DDO~154 and DDO~165 are only marginally ($<3\sigma$)
detected at 71 and 160\um, and M81dwB is a $<2\sigma$ detection at
160\um.
\item The dwarf starburst galaxy 
NGC~1377 shows strong silicate absorption at 9.8\um\ and 18\um\ 
\citep{Roussel+Helou+Smith_etal_2006}.
The objective of the present study is to model the IR spectra using
a simple model that assumes the galaxy to be optically-thin at
$\lambda>3.5\mum$;
because NGC~1377 is clearly not optically thin, it
has been removed from the sample.
\end{itemize}

The sample studied here therefore consists of 65 galaxies. 
MIPS 71\um\ photometry for NGC~7552 is not usable because of detector
saturation, but NGC~7552 is retained in the sample because it is well-observed
at other wavelengths and with other instruments.

\begin{figure}[h]
  \begin{center}
    \includegraphics[angle=0,width=\figwidth]%
      {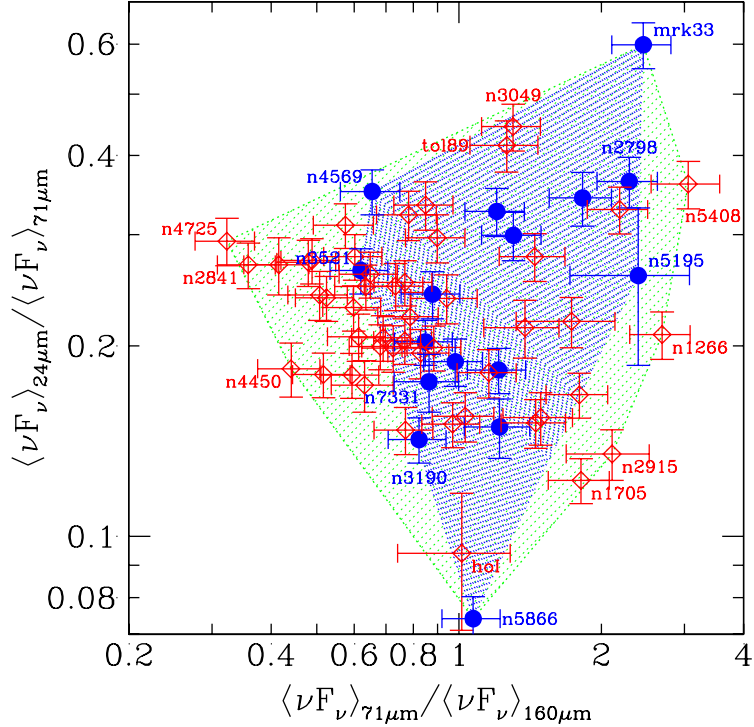}
    \caption{\label{fig:fluxratios}\footnotesize
      MIPS color-color plot for the \Ntot\ galaxies.
      The blue shaded area shows the region sampled by the
      \Nscuba\ galaxies for which global SCUBA fluxes are known 
      (plotted as blue solid symbols) -- these only partially cover the 
      color-color space populated by the \SINGS\ sample (green shaded
      area). Galaxies without SCUBA fluxes are shown in red.
    Galaxies with relatively extreme flux ratios are labelled.
    }
   \end{center}
\end{figure}

Most of the dust mass in galaxies is relatively cool, radiating primarily
at wavelengths in the MIPS 71$\mum$ and 160$\mum$ bands.
Dust grains that are warmer, either because they are in regions
with intense starlight, or because they are very 
small grains immediately following
absorption of a stellar photon, may radiate strongly in 
the MIPS 24$\mum$ band and shortward.
The variety of dust conditions found in the \SINGS\ sample is
evident from the distribution of the \Nmips\ galaxies in MIPS color-color
space, shown in Figure \ref{fig:fluxratios}:
$\langle\nu F_\nu\rangle_{71\mu{\rm m}}/
\langle\nu F_\nu\rangle_{160\mu{\rm m}}$ varies by over
a factor of 10 within our sample, and 
$\langle\nu F_\nu\rangle_{24\mu{\rm m}}/
\langle\nu F_\nu\rangle_{71\mu{\rm m}}$ varies by a
factor of 8.
These variations in MIPS colors presumably reflect galaxy-to-galaxy
differences in the intensity of the starlight heating the dust grains,
although galaxy-to-galaxy differences in dust composition may also
play a role.
It is evident from the scatter in Figure \ref{fig:fluxratios} that the
MIPS 24/71 color is not simply a function of the 71/160 color: the
MIPS colors cannot be described by a one-parameter sequence of, e.g.,
``dust temperature''.  Realistic dust models will therefore need to vary
at least two parameters to account for the range in observed MIPS colors.

\oldtext{The}
\newtext{The SINGS sample does not include any powerful AGN, and in all
         SINGS galaxies the global flux in the}
IRAC 3.6\um\ band is starlight-dominated.  Neglecting extinction, we can
estimate the ``nonstellar'' flux in longer-wavelength bands by subtracting
the estimated stellar contribution, assuming a 5000 K blackbody.
For IRAC band 4, the IRS 16\um\ Peakup band (see \S\ref{sec:IRS16} below)
and the MIPS 24\um\ band, the nonstellar fluxes are:
\beqa
\label{eq:7.9ns}
\langle F_\nu^{\rm ns}\rangle_{7.9} &=& 
\langle F_\nu\rangle_{7.9} - 0.260 \langle F_\nu\rangle_{3.6}
~~~,
\\
\label{eq:16ns}
\langle F_\nu^{\rm ns}\rangle_{16} &=& 
\langle F_\nu\rangle_{16} - 0.0693 \langle F_\nu\rangle_{3.6}
~~~,
\\
\label{eq:24ns}
\langle F_\nu^{\rm ns}\rangle_{24} &=& 
\langle F_\nu\rangle_{24} - 0.0326 \langle F_\nu\rangle_{3.6}
~~~.
\eeqa
The coefficients 0.260 and 0.0326 in eq.\ (\ref{eq:7.9ns}) and (\ref{eq:24ns})
are close to the values 0.232 and 0.032 estimated by 
\citet{Helou+Roussel+Appleton+Frayer+Stolovy_etal_2004}.
Some stars (e.g., cool red giants, and AGB stars) have circumstellar
dust that is in many cases warm enough to radiate strongly at wavelengths from
8-30\um.  If such stars are sufficiently numerous, the result will be
a ``stellar'' contribution to the 8, 16, and 24\um\ fluxes that will be
larger than assumed in eq.\ (\ref{eq:7.9ns}--\ref{eq:24ns}).
In the absence of extinction at 3.6\um, 
equations (\ref{eq:7.9ns}--\ref{eq:24ns}) will therefore {\it overestimate}
the flux contributed by interstellar dust.

\subsection{\label{sec:IRS16}
            IRS 16\um}

In addition to IRAC and MIPS photometry, nine of the \SINGS\ galaxies
have been imaged using the IRS ``blue peak-up'' detector array,
with nominal wavelength $\lambda=16\mum$.
\newtext{%
    The IRS blue peak-up observations were taken as part of the IRS
    instrument calibration program and consist of a grid of peakup
    images covering the full extent of each galaxy.  The reduced
    images, processed by IRS pipeline version S14, were mosaicked
    using the MOPEX program (ver.\ 030106) to remove cosmic rays after
    subtraction of a custom dark image.  This custom dark image was
    created using images taken beyond the visible extent of the
    galaxy.  In some cases, the best dark subtraction was found to
    require allowing the dark image to vary with time.  ``Sky'' (zodiacal light
    and Galactic cirrus) contributions were also removed by this custom dark
    subtraction step.  The total 16~\um\ flux was measured using the
    same methods used for the IRAC and MIPS fluxes
    \citep{Dale+Bendo+Engelbracht_etal_2005}.  The measured global
    fluxes were then scaled down by $\sim$15\%, to account for a
    recent SSC calibration update ($\sim$10\%), and the appropriate
    aperture correction factor ($\sim$5\%).
    }%
Fluxes for these \oldtext{9}\newtext{11} galaxies at 16\um\ and 24\um\
are given in Table \ref{tab:IRSPU16}.
The flux density ratio 
$\langle F_\nu\rangle_{16}/\langle F_\nu\rangle_{24}$
varies considerably among these galaxies, from
a low of
\oldtext{$0.442\pm0.032$}%
\newtext{$0.376\pm0.033$} for NGC~7552
to a high of 
$\oldtext{1.15\pm0.08}\newtext{1.21\pm0.12}$ for NGC~3190.

\begin{deluxetable}{l c c c}
\tabletypesize{\scriptsize}
\tablewidth{0pt}
\tablecolumns{4}
\tablecaption{\label{tab:IRSPU16}
              Global Fluxes (Jy) in IRS 16\um\ Peak-Up Band
		}
\tablehead{
  \colhead{galaxy} &
  \colhead{$\langle F_\nu\rangle_{16}$} &
  \colhead{$\langle F_\nu\rangle_{24}$} &
  \colhead{$\langle F_\nu\rangle_{16}/\langle F_\nu\rangle_{24}$}\\
  &
  \colhead{Jy} &
  \colhead{Jy} &
  }
\startdata
\nt{NGC~0337} & \nt{$0.375\pm0.029$} & \nt{$0.678\pm0.028$} & \nt{$0.553\pm0.049$} \cr
\nt{NGC~3190} & \nt{$0.322\pm0.030$} & \nt{$0.267\pm0.011$} & \nt{$1.21 \pm0.12$} \cr
\nt{NGC~3351} & \nt{$1.39\pm0.23$  } & \nt{$2.583\pm0.12$ } & \nt{$0.538\pm0.091$} \cr
\nt{NGC~3521} & \nt{$4.86\pm0.59$  } & \nt{$5.51 \pm0.22$ } & \nt{$0.88 \pm0.11$} \cr
\nt{NGC~3627} & \nt{$4.92\pm0.58$  } & \nt{$7.42 \pm0.30$ } & \nt{$0.663\pm0.082$} \cr
\nt{NGC~3938} & \nt{$0.849\pm0.068$} & \nt{$1.087\pm0.045$} & \nt{$0.781\pm0.070$} \cr
\nt{NGC~4321} & \nt{$2.64\pm0.43$  } & \nt{$3.34 \pm0.13$ } & \nt{$0.79 \pm0.13$} \cr
\nt{NGC~4536} & \nt{$1.74\pm0.14$  } & \nt{$3.46 \pm0.14$ } & \nt{$0.503\pm0.044$} \cr
\nt{NGC~5055} & \nt{$5.09\pm0.43$  } & \nt{$5.73 \pm0.23$ } & \nt{$0.888\pm0.083$} \cr
\nt{NGC~6946} & \nt{$12.8\pm1.02$  } & \nt{$20.37\pm0.81$ } & \nt{$0.628\pm0.056$} \cr
\nt{NGC~7552} & \nt{$4.01\pm0.31$  } & \nt{$10.66\pm0.44$ } & \nt{$0.376\pm.033$} \cr
\enddata
\end{deluxetable}

\begin{figure}[h]
  \begin{center}
    \includegraphics[angle=0,width=\figwidthd]%
      {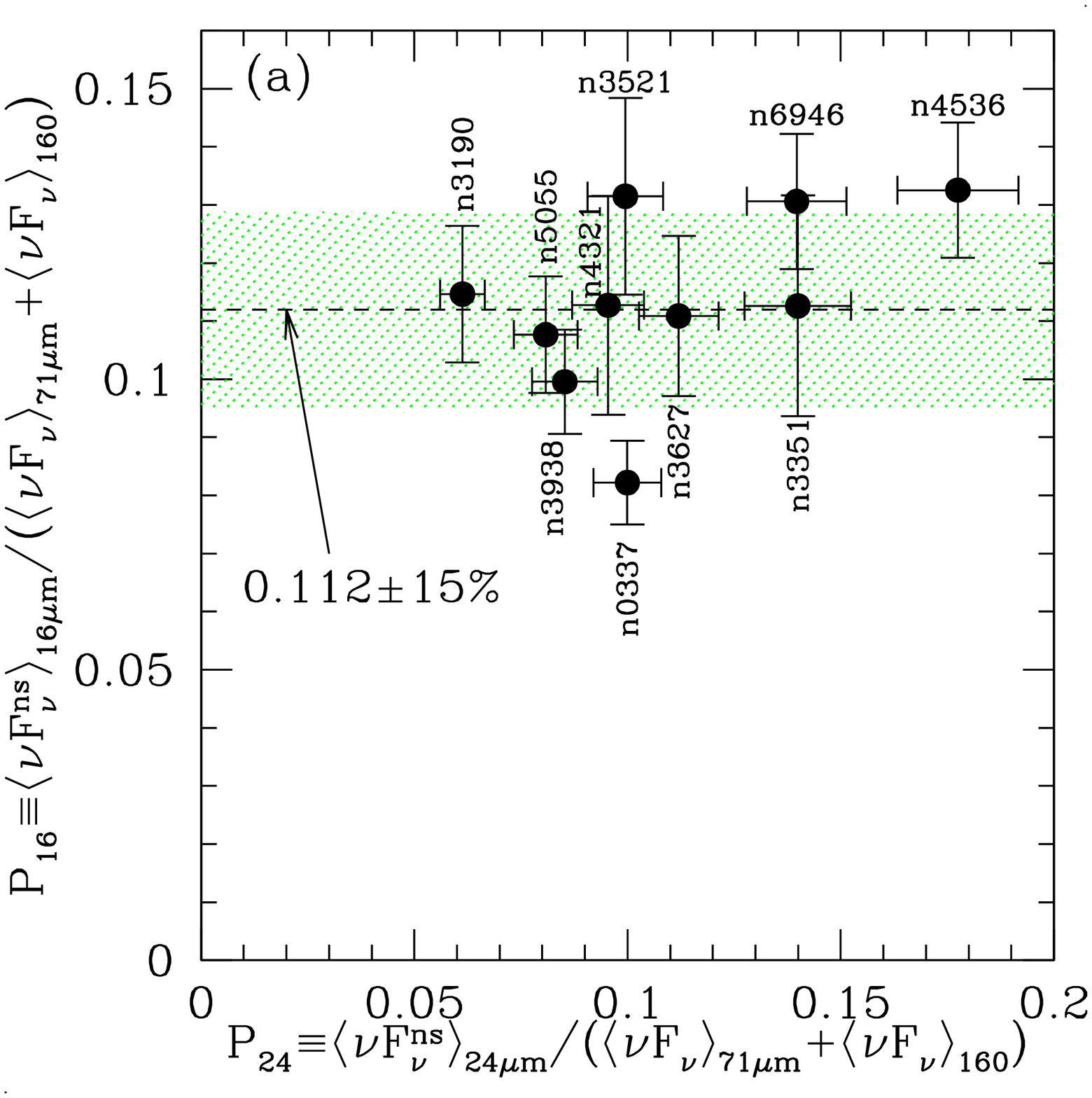}
    \includegraphics[angle=0,width=\figwidthd]%
      {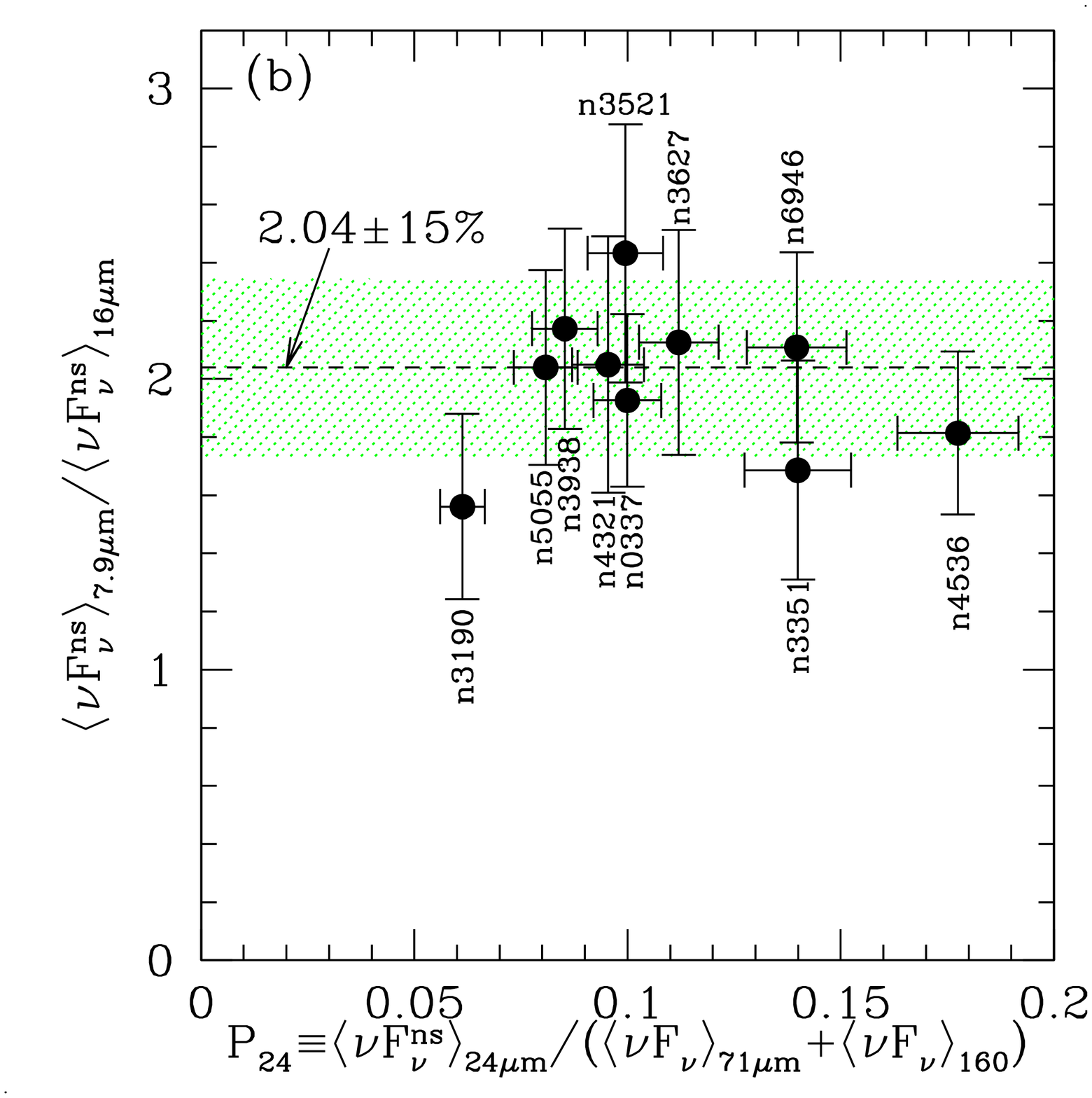}
    \caption{\label{fig:P16vsP24}\footnotesize
      (a) 
      $P_{16}\equiv\langle \nu F_\nu^{\rm ns}\rangle_{16}/
      (\langle \nu F_\nu\rangle_{71}+\langle \nu F_\nu\rangle_{160})$
      versus
      $P_{24}\equiv\langle \nu F_\nu^{\rm ns}\rangle_{24}/
       (\langle \nu F_\nu\rangle_{71}+\langle \nu F_\nu\rangle_{160})$
      for the \oldtext{8}\newtext{10} 
      galaxies imaged with the IRS 16\um\ peak-up filter
      for which we also have global MIPS fluxes (NGC~7552 is excluded).
      (b) $\langle\nu F_\nu^{\rm ns}\rangle_{7.9}/
      \langle\nu F_\nu^{\rm ns}\rangle_{16}$ vs
      $P_{24}$,
      Both the ratio of nonstellar 16\um\ emission to total IR power
      and 
      the ratio of nonstellar 7.9\um\ emission to nonstellar 16\um\ emission 
      remain constant to within
      $\sim15\%$, even as $P_{24}$ varies by a factor of 3.
      }
   \end{center}
\end{figure}
Figure \ref{fig:P16vsP24}a shows
$P_{16}\equiv \langle\nu F_\nu^{\rm ns}\rangle_{16}/
(\langle\nu F_\nu\rangle_{71}+\langle\nu F_\nu\rangle_{160})$
vs.\ $P_{24}\equiv \langle\nu F_\nu^{\rm ns}\rangle_{24}/
(\langle\nu F_\nu\rangle_{71}+\langle\nu F_\nu\rangle_{160})$
for the \oldtext{eight}\newtext{ten} 
galaxies for which we have 16, 24, 71, and 160\um\
photometry,
where $F_\nu^{\rm ns}$ is the nonstellar contribution
from eq.\ (\ref{eq:7.9ns}-\ref{eq:24ns}).
The flux ratio $P_{16}$ remains essentially constant
at $P_{16}\approx\oldtext{0.095\pm0.012}\newtext{0.112\pm0.017}$ while
$P_{24}$ varies by almost a factor of 3, from 0.06 to 0.18.
This suggests that the 16\um\ flux is not emitted by the same grains
as the 24\um\ flux.  
In the dust model of \citet[][hereafter DL07]{Draine+Li_2007}, 
a minimum level of 24\um\ emission
is produced by single-photon heating of PAHs or ultrasmall grains
(with $\langle\nu L_\nu\rangle_{24}/L_{\rm TIR} \approx 0.03-0.06$,
depending on $\qpah$ -- see Fig.\ 15 of DL07), but
$P_{24}$ increases as the starlight intensity increases.
Figure \ref{fig:P16vsP24}a suggests that the 16\um\ emission may be 
dominated by single-photon heating of PAHs, remaining relatively constant
even as $P_{24}$ rises due to increased starlight heating.
This is supported by Figure \ref{fig:P16vsP24}b, which shows 
that the ratio of 7.9\um\ emission to 16\um\ emission
is nearly constant for this sample, 
varying by only $\sim\pm15\%$ over the eight galaxies,
consistent with the idea that both are primarily the result of single-photon
heating of PAHs.
This is consistent with the finding by 
\citet{ForsterSchreiber+Roussel+Sauvage+Charmandaris_2004}
that, when averaged over disks of spiral galaxies, 
the 15\um\ continuum intensity is proportional to the
PAH intensity in the 5--8\um\ region.

\subsection{SCUBA}

The dust properties in a galaxy will be best determined if submillimeter
photometry is available to constrain the quantity of cool dust present.
SCUBA observations at 450\um\ and/or 850\um\
are available for
26 of the \Ntot\ galaxies,\footnote{%
  The 25 galaxies in Table 3 of \citet{Dale+Bendo+Engelbracht_etal_2005},
  plus NGC~1482}
but not all of these are usable in the present study:
\begin{enumerate}
\item NGC~4254, NGC~4579, NGC~5194, and NGC~6946 were scan-mapped, and the
  data processing removes an unknown amount of extended emission.  Thus 
  global 850\um\ (or 450\um) fluxes are not available for these galaxies.  
  We restrict ourselves to data obtained in ``jiggle map'' 
  \btdnote{George B: should a reference be given for ``jiggle map''?}
  mode,
  except for the scan-map observations of NGC~5195, for which background
  subtraction is thought to be more reliable because of the smaller
  angular extent of NGC~5195.
\item For some galaxies the SCUBA map did not fully
  cover the region where 160\um\ emission is detected.
  \citet{Dale+Bendo+Engelbracht_etal_2005} multiplied the observed 
  SCUBA fluxes by
  a correction factor based on the fraction of the 160\um\ emission coming
  from the region mapped by SCUBA.  Because the 850\um/160\um\ flux ratio
  may be spatially variable, we choose here to not use 
  the 850\um\ flux estimates
  for galaxies 
  where the corrrection factor exceeds 1.6 (NGC~1097, NGC~4321,
  NGC~4736, and NGC~5033).
\item For the Sombrero galaxy, NGC~4594, the 850\um\ flux is dominated by
  the AGN \citep{Bendo+Buckalew+Dale_etal_2006}, 
  and the extranuclear 850\um\ flux is not
  reliably measured, so we cannot use the SCUBA data
  to constrain the dust model for this galaxy.
\end{enumerate}
We thus have \Nscuba\ galaxies where 
the dust models are constrained by 
\newtext{850\um} SCUBA data 
\newtext{(with $450\micron$ fluxes for 3 of the \Nscuba)}
in addition to
IRAC and MIPS data.
The global fluxes $F_{{\rm obs},b}$ and 1-$\sigma$ uncertainties for
each band $b$ are taken from 
\citet{Dale+GildePaz+Gordon+etal_2007}.

The \Nscubamips\footnote{%
    NGC~7552 is absent from Fig.\ 1 because of detector saturation at 71\um.} 
galaxies for which we have global SCUBA and MIPS photometry are indicated in
Figure \ref{fig:fluxratios}.  
While the galaxies with SCUBA photometry span a considerable portion of
the populated region in
MIPS color-color space, a number of the \SINGS\ galaxies have
MIPS colors that are outside of the range seen among the 
galaxies for which SCUBA
data are available.
For example, NGC~4725 or NGC~5408, with 
$\langle\nu F_\nu\rangle_{71\mu{\rm m}}/
\langle\nu F_\nu\rangle_{160\mu{\rm m}}=0.33$ and
3.1, respectively, fall outside the range 0.6--2.4 observed for this
ratio among the \SINGS-SCUBA galaxies.
The dust models discussed below are able to reproduce the IRAC and
MIPS fluxes for all of the \SINGS\ galaxies.
Galaxies such as NGC~4725 or NGC~2841, with the lowest values
of $\langle\nu F_\nu\rangle_{71\mu{\rm m}}/
\langle\nu F_\nu\rangle_{160\mu{\rm m}}$,
require cooler dust (i.e., lower starlight intensities) than any
of the \SINGS-SCUBA galaxies.
NCG~5408, with a very high value of 
$\langle\nu F_\nu\rangle_{71\mu{\rm m}}/
\langle\nu F_\nu\rangle_{160\mu{\rm m}}$,
requires hotter dust (higher starlight intensities).
Submillimeter observations would be of great value to test the
predicted submm fluxes from these extreme cases.

\subsection{IRAS and ISO}

IRAS photometry is obtained from the latest SCANPI data for most
SINGS sources, and from HiReS maps for the largest SINGS targets
(optical diameters $\gtsim 10\arcmin$),
except in some bands (e.g., no IRAS 12\um\ flux for NGC~0024).
A 20\% calibration uncertainty estimate was applied.
In the case of NGC~5195 (=M51b), the proximity of NCG~5194 (=M51a) made flux
determination more difficult, and 50\% uncertainty estimates were adopted.
\btdnote{Danny -- is this an accurate summary?}

Ths IRAS~25\um\ flux densities are not used, as they are 
superseded by the MIPS~24\um\ photometry
at essentially the same wavelength, but we employ the
IRAS~12\um, 60\um, and 100\um\ photometry in our model-fitting.

When available, we show ISOCAM~6.75\um\ and 15\um\ photometry in the
figures showing the spectral energy distributions (SEDs), 
but they are not used in the model-fitting.

\section{\label{sec:dust models}
         Physical Dust Models}

A direct estimate for the dust-to-hydrogen mass ratio $\Mdust/\MH$ 
in the local
Milky Way can be obtained from the difference between total
interstellar abundances -- taken to be equal to current estimates for
solar abundances --
and observed gas phase abundances in diffuse interstellar clouds, where
the intensively studied diffuse cloud on the line-of-sight to
$\zeta$Oph is taken to be representative.
Using current estimates of solar abundances, this mass inventory
results in $\Mdust/\MH\approx \oldtext{0.0070}\newtext{0.0073}$ 
(see Table \ref{tab:MW depleted mass}).

We will apply (and thereby also test) the dust models developed by 
\citet[hereafter WD01]{Weingartner+Draine_2001a}
and 
\citet[hereafter LD01]{Li+Draine_2001b}, 
and updated by DL07.
These dust models consist of specified mixtures of carbonaceous grains and 
amorphous silicate grains, with the smallest carbonaceous grains having
the physical properties of polycyclic aromatic hydrocarbon (PAH) particles;
the size distributions are chosen to reproduce the observed
wavelength-dependent extinction in the local Milky Way, in the
Large Magellanic Cloud, and the Small Magellanic Cloud bar region
\citep{Weingartner+Draine_2001a}.
\citet{Li+Draine_2001b,Li+Draine_2002c} 
showed that these grain models appeared to be consistent with 
the observed
IR emission from diffuse clouds in the Milky Way and the Small
Magellanic Cloud.

\begin{deluxetable}{l c c c}
\tabletypesize{\footnotesize}
\tablewidth{0pt}
\tablecolumns{4}
\tablecaption{\label{tab:MW depleted mass}
              Dust Mass per H from Milky Way Abundances}
\tablehead{
  \colhead{$X$} &
  \colhead{$(N_{X}/N_{\rm H})_\odot$(ppm)$^a$} &
  \colhead{$(N_{X}/N_{\rm H})_{\rm gas}/(N_{X}/N_{\rm H})_\odot$~$^a$ }&
  \colhead{$M_{X,{\rm dust}}/M_{\rm H}$}
          }
\startdata
C   & 247 & 0.57    & 0.0013 \cr
\newtext{N}   & \newtext{85}  & \newtext{0.72}    & \newtext{0.0003} \cr
O   & 490 & 0.73    & 0.0021 \cr
Mg  & 38  & 0.08    & 0.0008 \cr
Al$$  &  3$^b$  & $\ltsim$0.1$^c$    & 0.0001 \cr
Si  & 32  & 0.05    & 0.0009 \cr
Ca  &  2$^b$  & 0.0002$^d$  & 0.0001 \cr
Fe  & 29  & 0.007   & 0.0016 \cr
Ni  &  2  & 0.004   & 0.0001 \cr
\hline
total &   &         & 
                      \newtext{0.0073}\cr
\hline
\multicolumn{4}{l}{$^a$ \citet{Jenkins_2004} except as noted.}\cr
\multicolumn{4}{l}{$^b$ $(N_{X}/N_{\rm H})_\odot$ from \citet{Grevesse+Sauval_1998}}\cr
\multicolumn{4}{l}{$^c$ assumed}\cr
\multicolumn{4}{l}{$^d$ \citet{Savage+Sembach_1996}}\cr
\enddata
\end{deluxetable}
\begin{deluxetable}{l l c c c}
\tabletypesize{\footnotesize}
\tablewidth{0pt}  
\tablecolumns{4}
\tablecaption{\label{tab:dust models}
              Physical Dust Models}
\tablehead{
  \colhead{$j_M$} & 
  \colhead{Model} & 
  \colhead{$\qpah$} & 
  \colhead{$\Mdust/\MH^{a,b}$} &
  \colhead{$A_V/\NH$} \\
  \colhead{}&
  \colhead{}&
  \colhead{\%} & &
  \colhead{${\rm mag}\cm^2$/H}
  }
\startdata
%
%
1 & MW3.1\_00 & 0.47  & 0.0100  & $5.3\times10^{-22}$ \\
2 & MW3.1\_10 & 1.12  & 0.0100  & $5.3\times10^{-22}$\\
3 & MW3.1\_20 & 1.77  & 0.0101  & $5.3\times10^{-22}$\\
4 & MW3.1\_30 & 2.50  & 0.0102  & $5.3\times10^{-22}$\\
5 & MW3.1\_40 & 3.19  & 0.0102  & $5.3\times10^{-22}$\\
6 & MW3.1\_50 & 3.90  & 0.0103  & $5.3\times10^{-22}$\\
7 & MW3.1\_60 & 4.58  & 0.0104  & $5.3\times10^{-22}$\\
8 & LMC2\_00  & 0.75  & 0.00343 & $1.2\times10^{-22}$\\
9 & LMC2\_05  & 1.51  & 0.00344 & $1.2\times10^{-22}$\\
10 & LMC2\_10 & 2.40  & 0.00359 & $1.2\times10^{-22}$\\
11 & SMCbar   & 0.010  & 0.00206 & $6.2\times10^{-23}$\\
\hline
\multicolumn{4}{l}{$^a$ $\MH\equiv M({\rm H\,I + H_2})$}\\
\multicolumn{4}{l}{$^b$ $\Mdust/M_{\rm gas}=[\Mdust/\MH]/1.36$}\\
\enddata
\end{deluxetable}

The DL07 dust models used here adopt
PAH optical properties 
that are slightly different from those adopted by LD01.
As discussed by DL07, the changes are in part to better conform to recent
theoretical estimates for band strengths, and in part to improve
agreement with band profiles determined observationally by 
\citet{Smith+Draine+Dale+etal_2007}.


The PAH emission features at 3.3\um, 6.2\um, 7.7\um, 8.6\um, 11.3\um, 12.0\um,
12.7\um, 16.4\um, and 17\um\ are prominent in many galaxy spectra, and
the dust models considered here include various abundances of PAHs;
in each model the PAHs
have a wide range of sizes.
Because PAH emission is mainly the result of single-photon heating, the PAH
emission spectrum depends on the size (heat capacity) of the PAH.
The prominent 6.2\um, 7.7\um, and 8.6\um\ emission features are primarily
due to PAHs with $\ltsim 10^3$ C atoms [see Fig.\ 7 of \citet{Draine+Li_2007}].
Following DL07, 
we characterize the PAH abundance by a ``PAH index'' $\qpah$, defined to be
the percentage of the total grain mass contributed by PAHs containing
less than $10^3$ C atoms:
\beq
\qpah \equiv
\frac{M({\rm PAHs~with~}N_{\rm C} < 10^3)}{\Mdust}
~~~,
\eeq
where $\Mdust$ includes the mass of all silicate and carbonaceous dust,
including the PAHs.

WD01 put forward seven different size distributions for dust in the
diffuse ISM of the Milky Way.  The models are all consistent with the
average interstellar extinction law, but have different PAH abundances,
ranging from $\qpah=0.47\%$ to 4.6\%.
Some properties for these seven models are given in 
Table \ref{tab:dust models}, where the dust masses are calculated
assuming densities of $3.5\g\cm^{-3}$ for silicates,
$2.24\g\cm^{-3}$ for carbonaceous grains (WD01).
The WD01 Milky Way dust models have $\Mdust/\MH\approx0.010$, about 40\%
larger than the value \oldtext{0.0070}\newtext{0.0073} estimated from observed depletions
(see Table \ref{tab:MW depleted mass}), assuming
solar abundances of C, Mg, Si, and Fe in the local interstellar medium.
It is possible that current estimates of solar abundances of C, Mg, Si,
and Fe may be lower than the actual elemental abundances in the
interstellar medium today; 
a $\sim$25\% increase in the solar abundances would bring
$\Mdust/\MH$ estimated from depletions 
into accord with the WD01 model.  Alternately, it is
possible that the WD01 model overestimates the mass of interstellar dust
by a factor 1.4 or so.

The PAHs constitute the small size end of the carbonaceous grain population;
the varying PAH abundances are the result of varying the grain size
distribution, while holding the overall extinction constant.
The model with the largest PAH abundance appears to reproduce the
observed IR-submm emission from dust in the local interstellar medium
(LD01, DL07).
Because it is likely that the PAH abundance varies from galaxy-to-galaxy,
here we take linear combinations of these models to produce a sequence of
43 MW models
with $\qpah=0.4\%$ to 4.6\% in steps of 0.1\%.

Similarly, we take linear combinations of the
3 WD01 models (as modified by DL07) for dust in the LMC2 region
(with $\qpah=0.75$, 1.49, and 2.37\%) to produce a sequence of 18 LMC models
with $\qpah=0.7\%$ to 2.4\%.

The dust models include amorphous silicate grains.
We assume optically-thin dust when modeling the global SEDs, and therefore
the models presented here do not include silicate absorption at 9.8\um.
Such absorption is in fact seen in observed spectra of the nuclei of a few
of the galaxies in this sample -- e.g., NGC~3198 
\citep{Smith+Draine+Dale+etal_2007} --
but we neglect dust absorption in modeling the global emission.

The dust models employed here do not include H$_2$O and other ices.
H$_2$O ice is known to be present in dark clouds in the Milky Way, 
in regions where
$A_V\gtsim 3.3\,{\rm mag}$ \citep{Whittet+Bode+Longmore_etal_1988}.
In these regions, ices will modify the optical properties of the dust.
However, it appears that such dark regions contain at most a modest fraction
of the total dust in a galaxy; furthermore, the weak starlight in such
regions implies that the dust grains located there radiate only a very small
fraction of the total IR power.
As will be seen below, an ice-free dust model appears to be satisfactory
for modeling the IR emission from normal galaxies.

\section{\label{sec:model fitting}
         Constraining the Dust Model and Starlight Intensities}

In addition to choosing a physical dust mixture (relative numbers of
carbonaceous grains and silicate grains of different sizes, including the
fraction $\qpah$ of the dust mass that is in PAHs), 
we must specify the intensity
of the radiation that is heating the dust grains.  
Almost all of the heating of
dust grains in a normal galaxy is from absorption of $h\nu<13.6\eV$ photons.
In a normal star-forming galaxy 
a significant (but sub-dominant) fraction of the stellar luminosity
may be at $h\nu>13.6\eV$, but 
most of these photons end up 
photoionizing hydrogen or helium rather than being absorbed by dust.
The $h\nu<13.6\eV$ recombination radiation -- 
especially Lyman $\alpha$ -- may be largely
absorbed by dust.
However, the total power in ionizing photons 
is a minor fraction of the total stellar luminosity in
a galaxy with a more-or-less steady rate of star formation.
Only in galaxies that have undergone an extreme burst of star formation within
the last $\sim10^7$ yrs will the dust heating be dominated by
Lyman $\alpha$ and Lyman continuum radiation.
Only in rare cases, e.g., NGC~1377 \citep{Roussel+Helou+Smith_etal_2006}, 
where the massive
stars formed in a starburst are surrounded by dense, dusty \ion{H}{2} regions,
will the dust grains directly absorb a major fraction 
of the $h\nu>13.6\eV$ radiation.

The IR emission from
dust is relatively insensitive to the detailed
spectrum of the $h\nu<13.6\eV$ photons,
and the DL07 models 
simply adopt the spectrum of the local interstellar radiation field
as a reasonable representation for
the average spectrum of the interstellar radiation in a normal spiral galaxy.
The specific energy density of starlight is therefore taken to be
\beq \label{eq:ISRF}
u_\nu = U\times u_\nu^{(\rm MMP83)}
~~~,
\eeq
where $u_\nu^{(\rm MMP83)}$ is the specific energy density estimated
by \citet{Mathis+Mezger+Panagia_1983} (hereafter MMP83) for the local
\oldtext{interstellar medium,}%
\newtext{Galactic interstellar radiation field,}
and $U$ is a dimensionless scale factor.
The intensity of interstellar UV radiation is
often characterized by $G_0$, the ratio of the 6--13.6~eV energy density
relative to the value $5.288\times10^{-14}\erg\cm^{-3}$ for the
radiation field estimated by \citet{Habing_1968}.
The MMP83 ISRF has 
$u(0-13.6\eV~{\rm starlight})=8.65\times10^{-13}\erg\cm^{-3}$,
and $u(6-13.6\eV)=6.01\times\times10^{-14}\erg\cm^{-3}$.
Thus $U=0.88G_0$.

Dust at different locations in the galaxy will be exposed to different 
radiation intensities.  A physically correct model would specify the locations
of the stellar sources and the absorbing dust, and would solve the radiative
transfer problem [e.g., \citet{Tuffs+Popescu+Volk_etal_2004}], 
but this would require
many uncertain assumptions about the relative distributions of stars and
dust, as well as massive numerical computations.

Following DL07, 
we suppose that a large fraction of the dust in a galaxy is
located in the diffuse interstellar medium, exposed to starlight
with an intensity that does not vary much within the galaxy.
This ``diffuse ISM''
dust is idealized as being heated by a single radiation field,
with intensity factor $U=\Umin$.
In addition, DL07 suppose that
a small fraction of the dust is located in regions where
the radiation field is more intense, with a wide range of intensities
ranging from $U=\Umin$ to $U=\Umax\gg\Umin$.
This would include dust in photodissociation regions, so we refer
to this as the ``PDR'' component, and we use the power-law
distribution function used by DL07.
\newtext{%
Although this ``PDR'' component contains 
only a small fraction of the total dust
mass, in some galaxies it contributes a substantial fraction of the total power
radiated by the dust.
The models do not include a component corresponding to cold dust in dark
clouds; as discussed in \S\ref{sec:dark clouds}, such cold dust
appears to account for a minor fraction of the dust mass, and a very small
fraction of the total dust luminosity.
         }
Thus, following DL07,
we take the dust mass
$d\Mdust$ exposed to radiation intensities in $[U,U+dU]$ to be a
combination of a $\delta$ function and a power-law,
\beq \label{eq:Udist}
\frac{d\Mdust}{dU} = 
(1-\gamma)\Mdust \delta(U-\Umin)
+
\gamma \Mdust\frac{(\alpha-1)}{\Umin^{1-\alpha}-\Umax^{1-\alpha}} U^{-\alpha} 
~~~~{\rm for}~ \Umin \leq U \leq \Umax ~,~\alpha\neq 1
\eeq
where $\delta$ is the Dirac $\delta$ function.
This distribution function has
five parameters: the total dust mass $\Mdust$, the
fraction $\gamma$ of the dust mass that is associated with 
the power-law part of the starlight intensity distribution, and 3 parameters
($\Umin$, $\Umax$, $\alpha$) characterizing the distribution of starlight
intensities in the high intensity regions.
The $\delta$--function term represents the fraction $(1-\gamma)$ of the
dust that is heated by a general diffuse interstellar radiation field
with approximately uniform intensity $\sim \Umin$.
The power-law part of the distribution function, introduced by
\citet{Dale+Helou+Contursi_etal_2001} and used by
\citet{Li+Draine_2002c},
represents the fraction $\gamma$ of the dust mass that is close to
OB associations and is exposed to radiation
with intensity $U>\Umin$; this includes
the dust in ``photodissociation regions'' (PDRs),
where radiation intensities can be orders of magnitude higher than the
``average'' dust grain is exposed to.
Equation (\ref{eq:Udist}) is a very simplistic description of the
distribution of dust grain environments; it neglects the possibility
that the dust properties (composition, size distribution) might be
variable and correlated with $U$,
and also neglects regional variations in the spectrum of the starlight
irradiating the dust.
Despite its simplicity, we will see here that eq. (\ref{eq:Udist}) appears
to capture the essential elements of the distribution of starlight intensities
found in galaxies, yielding models with SEDs that approximate what is
observed.

Very small particles cool almost completely between photon absorptions,
and therefore the shape of the emission spectrum from these grains is 
almost independent of $U$.  
The larger grains, that account for more than half
of the total absorption of starlight in the Milky Way, do not cool
significantly between photon absorptions, and their temperature 
depends on the intensity $U$.  
The observed FIR emission spectrum is sensitive to the temperature
of these larger grains, and therefore allows us to 
constrain the distribution function for $U$.
While \SST\ provides unprecedented sensitivity to
FIR emission, the spectral coverage is somewhat sparse,
with a factor of 3 gap in wavelength between IRAC 7.9\um\ 
and the MIPS 24\um\ bands, 
and a factor of 2.9 gap between the MIPS 24\um\ and 71\um\ bands.
The IRS ``Peak-Up'' mode observations can be used for imaging at
16\um\ and 26\um, but these observations are not generally available.
The large gaps in wavelength coverage leave considerable
uncertainty in the spectral energy distribution at wavelengths between the 
bands.\footnote{
  In principle, the IRS instrument can provide 6--36\um\ photometry, and
  the MIPS SED mode covers 55--95\um,
  but observing time considerations preclude using these observing modes for
  large areas.}
Furthermore, the longest wavelength band, at 160\um, falls on the
short wavelength side of the emission peak for dust temperatures
$T_d \ltsim (1/6)hc/(160\mum) = 15\K$.  As a result, \Spitzer\ photometry
does not strongly constrain the abundance of dust with temperatures
$T_d \ltsim 15\K$.
To determine the amount of cool dust, global submillimeter observations are
required.  
Fortunately, such data exist for \Nscuba\ of the \Ntot\ galaxies
for which we have complete \Spitzer\ photometry
(or near-complete in the case of NGC~7552), and this subsample will
be analyzed first, in \S\ref{sec:SCUBA sample}.

We select a model for the IR emission from a galaxy as follows.
For each size and composition of dust, 
we precompute and tabulate the temperature
distribution function for the dust grain illuminated by the radiation field
of eq.\ (\ref{eq:ISRF}) for selected values of $U$ 
(with $0.1\leq U \leq 10^7$),
using heat capacities, absorption cross sections, and numerical methods
described by 
\citet{Draine+Li_2001},
LD01,
and
DL07.
Neutral and ionized PAHs are assumed to have different absorption cross
sections. 
The ionized fraction $\phi_i(N_{\rm C})$ (a function of the number
$N_{\rm C}$ of carbon atoms in the PAH) will depend primarily on the ratio
$U\sqrt{T_e}/n_e$, where $n_e$ and $T_e$ are the electron density
and temperature \citep{Bakes+Tielens_1994,Weingartner+Draine_2001c}.
Here we take the ionized fraction $\phi_i(N_{\rm C})$ estimated by
LD01
for the
local diffuse ISM\footnote{
  Following 
  LD01, 
  we assume 43\% of the dust to be
  in the ``cold neutral medium'' 
  (CNM; $n_e/U\approx0.045\cm^{-3}$, $T\approx100\K$),
  43\% to be in the ``warm neutral medium'' 
  (WNM; $n_e/U\approx 0.04\cm^{-3}$, $T\approx6000\K)$,
  and
  14\% to be in the ``warm ionized medium''
  (WIM; $n_e/U\approx 0.1\cm^{-3}$, $T\approx8000\K)$.
  The resulting $\phi_i(N_{\rm C})$ is shown in Fig.\ 7 of
  LD01
  and Fig.\ 8 of
  DL07.
  }
(we are therefore implicitly 
assuming that other galaxies have
the same mix of $U\sqrt{T_e}/n_e$ as the Milky Way).
Our model-fitting procedure assumes that all galaxies have
the same shape for the PAH size distribution, 
the same PAH ionized fraction $\phi_i(N_{\rm C})$, and the
same spectral shape of the illuminating starlight.
As a result, our model fits have fixed ratios of
PAH emission band strengths.
Observations do reveal variations in PAH band ratios from one galaxy
to the next -- for example,
$L(7.7\mum)/L(11.3\mum)$ varies from $\sim$4--5 to $\ltsim$0.7
as we progress from \ion{H}{2}-dominated nuclei to AGN-dominated nuclei
\citep{Smith+Draine+Dale+etal_2007}%
\newtext{, and low values of $L(7.7\mum)/L(11.3\mum)$ are
also seen in some AGN-free galactic and extragalactic
regions \citep[e.g.,][]{Reach+Boulanger+Contursi+Lequeux_2000,
                      Hony+VanKerckhoven+Peeters+Tielens+etal_2001,
                      Pagani+Lequeux+Cesarsky_etal_1999} --}
but most spiral galaxies have global PAH feature ratios that are near-average
and characteristic
of star-forming regions.

With these temperature distribution functions for the dust, 
and the dust absorption cross section
$C_{\rm abs}(\lambda)$, we can calculate
the time-averaged IR emission for a given grain type and size for each of
the discrete values of $U$ considered; we then sum over the grain types
and sizes for a dust model $j_M$ (see Table \ref{tab:dust models}) to 
find the power radiated per unit frequency
per unit mass of dust mixture $j_M$ exposed to starlight intensity $U$:
\beqa
p_\nu^{(0)}(j_M,U)
&=&
\frac{1}{M(j_M)}
\sum_k \int da \left(\frac{dn_{k}}{da}\right)_{j_M}
\int dT \frac{dP(k,a)}{dT} 4\pi B_\nu(T) C_{\rm abs}(a,\lambda) 
\\
M(j_M) &\equiv& \sum_k \int da \left(\frac{dn_k}{da}\right)_{j_M}
\frac{4\pi a^3}{3}\rho_k
~~~,
\eeqa
where the sum runs over the different grain types $k$ (amorphous silicate,
graphite, PAH$^0$, PAH$^+$) in the DL07 dust model,
with $\rho_k$ the density of material $k$, and
$n_k(a)$ being the number of dust grains of type $k$ per H nucleon,
with radii $\leq a$.

We also calculate the specific power per unit dust mass
\beq
p_\nu(j_M,\Umin,\Umax,\alpha)\equiv
\frac{(\alpha-1)}{\Umin^{1-\alpha}-\Umax^{1-\alpha}}
\int_{\Umin}^{\Umax} p_\nu^{(0)}(j_M,U) ~U^{-\alpha}dU 
\eeq
for a power-law distribution of starlight intensities,
$d\Mdust/dU \propto U^{-\alpha}$.
For each dust model $j_M$ 
we consider a range of values
of $\Umin$ and $\Umax$.

At wavelengths $\lambda \ltsim 5\mum$, the emission from a galaxy
is dominated by starlight.  For $\lambda \geq 3.6\mum$ we neglect
reddening, and approximate the stellar contribution by a dilute
blackbody with color temperature $T_\star=5000\K$,
which \citet{Smith+Draine+Dale+etal_2007} found to provide a suitable
approximation to the stellar continuum at $\lambda > 5\micron$.
The model emission spectrum is then
\beq 
F_{\nu,{\rm model}} = \Omega_\star B_\nu(T_\star) + 
        \frac{\Mdust}{4\pi D^2}
        \left[
	(1-\gamma)p_\nu^{(0)}(j_M,\Umin)
        +
        \gamma p_\nu(j_M,\Umin,\Umax,\alpha) 
	\right]
~~~,
\eeq
where $\Omega_\star$ is the solid angle subtended by the stars.
For each $(j_M,\Umin,\Umax,\alpha)$, 
we obtain $\Omega_\star$, $\Mdust$, and $\gamma$ by minimizing
\beq \label{eq:define chi^2}
\chi^2 \equiv \sum_{b}\frac
{\left[F_{{\rm obs},b} - \langle F_{\nu,{\rm model}}\rangle_b \right]^2}
{\sigma_{{\rm obs},b}^2+\sigma_{{\rm model},b}^2}
~~~,
\eeq
where the sum is over observed bands $b$, 
$\langle F_{\nu,{\rm model}}\rangle_b$ is the model spectrum convolved
with the response function for band $b$, and
$\sigma_{{\rm obs},b}$ is the observational uncertainty in the observed flux 
density $F_{{\rm obs},b}$ for
band $b$.
The {\it ad-hoc} insertion of the 
$\sigma_{{\rm model},b}^2$ 
term in the denominator in eq.\ (\ref{eq:define chi^2})
is to allow for the approximate nature of our models and
is motivated by the very high photometric accuracy achieved for 
the MIPS 24\um\ photometry, with
$\sigma_{{\rm obs},24\mum}/F_{{\rm obs},24\mum}= 0.040$ achieved
for many of the \SINGS\ galaxies.
We do not want the fitting procedure to give extreme weight to
one or two bands where $F_{{\rm obs},b}$ may have been measured with very high
precision ({\it i.e.,} small $\sigma_{{\rm obs},b}/F_{{\rm obs},b}$)
because we do not expect our models to fit perfectly in the
limit of perfect observations.  
Our subjective sense is that the present models should be regarded
as satisfactory if they are able to reproduce the actual fluxes
(i.e., perfect data, with $\sigma_{{\rm obs},b}=0$)
to within, say, $\pm10\%$.
We therefore arbitrarily set 
$\sigma_{{\rm model},b}=0.1 \langle F_{\nu,{\rm model}}\rangle_b$
for each band $b$.
It is important to recognize that 
eq.\ (\ref{eq:define chi^2})
differs from the usual
$\chi^2$ used for testing models, and the
numerical values of $\chi^2$ found here
are not rigorous measures of the goodness-of-fit.
Nevertheless, these $\chi^2$ values do appear to provide a practical
guide to distinguish between good fits and poor fits.

The model components must be nonnegative:
we require $\Omega_*\geq 0$, $\Mdust\geq 0$,
and $0\leq \gamma\leq 1$.
We select the dust model $j_M$, and intensity model
$(\gamma,\Umin,\Umax,\alpha)$ giving the best fit
(smallest $\chi^2$) subject to these constraints.

For the MW dust models, there are effectively 7 adjustable parameters:
$\Omega_\star$, $\Mdust$, $\qpah$ (via $j_M$), 
$\gamma$, $\Umin$, $\Umax$, and $\alpha$.
However, we will fix one of these parameters from the outset, by
setting $\alpha=2$.  This choice is based on experimentation that
indicated that the quality of the fit is 
not very sensitive to the precise value
of $\alpha$, with $\alpha\approx 2$ providing satisfactory fits for
diverse galaxies.
Therefore we start with 6 parameters to be adjusted.
For the galaxies with SCUBA data, we will examine the effects of varying
$\Umax$, but we will see in \S\ref{sec:Umax}
that setting $\Umax=10^6$ works well;
therefore, the number of adjustable parameters is, effectively, 5:
$\Omega_\star$, $\Mdust$, $\qpah$, $\Umin$, and $\gamma$.
If the number of observed bands is $N_b$,
the number of degrees of freedom 
$=(N_b-5)$ where 
$N_b$ ranges from $N_b=9$ (e.g., NGC~0024, with 4 IRAC bands, 3 MIPS bands,
and IRAS 60, 100\um)
to
$N_b=12$ (e.g., NGC~3190 with 4 IRAC bands, 3 MIPS bands, IRS~16\um,
IRAS 12,60,100\um, and SCUBA 850\um).
To assess the goodness of fit, we calculate $\chi^2$ per degree of freedom:
\beq \label{eq:chi_r^2}
\chi_r^2 \equiv \frac{\chi^2}{N_b-5}
~~~.
\eeq

\section{\label{sec:SCUBA sample}
          SCUBA-Constrained Dust Models}

\subsection{Spectral Energy Distributions}

The distribution of grain temperatures in a dust cloud or a galaxy is
never determined precisely.
At wavelength $\lambda$
the thermal emission per unit dust mass 
$\propto [e^{hc/\lambda kT_d}-1]^{-1}$.
For 
$\lambda \ltsim 
360\mum(20\K/T_d)$, small errors in
the adopted
$T_d$ lead to large errors in the dust mass inferred from the
observed $F_\nu$.
For example, varying the adopted $T_d$ from 22 to 18K produces
a 150\% increase in the mass of dust required to produce
a given 160\um\ flux, but only a 35\% increase in the mass
required to produce a given 850\um\ flux.
Observations at $\lambda \gtsim 450\mum$
provide dust masses with the least sensitivity
to dust temperature (for dust temperatures $T_d\gtsim 15\K$).
Therefore, the subset of \Nscuba\ \SINGS-SCUBA galaxies
will be those for which we can best determine the dust mass.

For each galaxy Table \ref{tab:scuba galaxies} 
lists the best-fit model for 
$10^3\leq \Umax \leq 10^7$, for fixed $\alpha=2$.
If the best-fit model has either
LMC or SMC dust models (see NGC~5195)
then we also list
the parameters for the best-fit model using MW dust.
Finally, if the best-fit MW dust model has $\Umax\neq10^6$
we also list
the parameters for the best-fit MW dust model with $\Umax=10^6$.

\clearpage
\begin{deluxetable}{c c c c c c c c c c c c c c c}
\tabletypesize{\footnotesize}
\rotate
\tablewidth{0pt}  
\tablecolumns{15}
\tablecaption{\label{tab:scuba galaxies}
         \Nscuba\ \SINGS\ Galaxies with IRAC, MIPS, and SCUBA global fluxes}
\tablehead{
   \colhead{galaxy} & 
   \colhead{morph} &
   \colhead{$D$\tablenotemark{a}} &
   \colhead{$\log[M({\rm HI})]$\tablenotemark{a}} & 
   \colhead{$\log[M({\rm H}_2)]$\tablenotemark{a}} &
   \colhead{$\log(\Mdust)$\tablenotemark{b}} & 
   \colhead{$\log(\Ldust)$\tablenotemark{b}} & 
   \colhead{$\qpah$\tablenotemark{c}} & 
   \colhead{$\langle U\rangle$\tablenotemark{d}} &
   \colhead{dust} &
   \colhead{$\Umin$\tablenotemark{e}} &
   \colhead{$\Umax$\tablenotemark{e}} & 
   \colhead{$\gamma$\tablenotemark{f}} &
   \colhead{$f(U>10^2)$\tablenotemark{g}} &
   \colhead{$\chi_r^2$\tablenotemark{h}}\\
        &
   \colhead{type} &
   \colhead{Mpc} &    
   \colhead{$M_\odot$} &
   \colhead{$M_\odot$} & 
   \colhead{$M_\odot$} &
   \colhead{$L_\odot$} & 
   \colhead{\%} &
        &  
   \colhead{type} &  
        &
	& 
   \colhead{\%} &
   \colhead{\%} &
   }
\startdata
   ngc0337 &   Sd & 24.70 &      --- &      --- &  7.65 & 10.31 
                                                & \ot{2.8}\nt{2.7} 
                                                & \ot{3.40}\nt{3.41} & MW &  2.5 & $10^4$ 
                                                & \ot{4.95}\nt{4.96} 
                                                & \ot{16.7}\nt{16.8} 
                                                & \ot{0.16}\nt{0.21} \cr
           &      &       &          &          &  7.63 
                                                & \ot{10.30}\nt{10.29} 
                                                & \ot{2.7}\nt{2.5} 
                                                & \ot{3.44}\nt{3.43} & MW &  3.0 & $10^6$ 
                                                & \ot{1.25}\nt{1.23} 
                                                & \ot{10.0}\nt{9.9} 
                                                & \ot{0.26}\nt{0.43} \cr
   ngc1482 &   S0 & 22.00 &  $<$8.88 &     9.77 &  7.47 & 10.68 & 2.9 & 11.82 & MW &  8.0 & $10^7$ &     3.66 & 28.5 &  0.25 \cr
           &      &       &          &          &  7.48 & 10.67 & 3.5 & 11.57 & MW &  8.0 & $10^6$ &     4.15 & 26.5 &  0.55 \cr
   ngc2798 &   Sa & 24.70 &     9.29 &     9.77 &  7.29 & 10.58 & 2.4 & 14.26 & MW &  8.0 & $10^5$ &     9.28 & 36.0 &  0.45 \cr
           &      &       &          &          &  7.25 & 10.57 & 2.0 & 15.51 & MW & 10.0 & $10^6$ &     5.24 & 31.1 &  0.46 \cr
   ngc2976 &   Sc &  3.56 &     8.12\tablenotemark{i} 
                                     &     8.08\tablenotemark{j}
                                                &  6.34 &  8.93 & 3.4 &  2.82 & MW &  2.5 & $10^7$ &     0.90 &  9.2 &  0.30 \cr
           &      &       &          &          &  6.35 &  8.93 & 3.6 &  2.79 & MW &  2.5 & $10^6$ &     0.99 &  8.2 &  0.33 \cr
   ngc3190 &   Sa & 17.40 &     8.65 &      --- & \ot{7.18}\nt{7.19} 
                                                & \ot{9.72}\nt{9.73} 
                                                & \ot{3.5}\nt{3.8} 
                                                & \ot{2.57}\nt{2.55} & MW &  2.5 & $10^7$ 
                                                & \ot{0.19}\nt{0.15} 
                                                & \ot{2.1}\nt{1.7} 
                                                & \ot{0.21}\nt{0.38} \cr
           &      &       &          &          & \ot{7.18}\nt{7.19} 
                                                & \ot{9.72}\nt{9.73} 
                                                & \ot{3.6}\nt{4.6} 
                                                & \ot{2.56}\nt{2.54} & MW &  2.5 & $10^6$ 
                                                & \ot{0.19}\nt{0.14} 
                                                & \ot{1.8}\nt{1.3} 
                                                & \ot{0.21}\nt{0.39} \cr
    Mark33 &   Im & 21.70 &     8.77 &      --- &  6.54 &  9.83 & 1.9 & 14.33 & MW &  4.0 & $10^5$ &     28.3 & 54.6 &  0.37 \cr
           &      &       &          &          &  6.47 &  9.81 & 1.3 & 15.96 & MW &  7.0 & $10^6$ &     11.8 & 47.6 &  0.58 \cr
   ngc3521 &  Sbc &  9.00 &     9.75 &     9.79 & \ot{7.82}\nt{7.83} 
                                                & \ot{10.30}\nt{10.32} & 4.5 
                                                & \ot{2.24}\nt{2.26} & MW &  2.0 & $10^7$ 
                                                & \ot{0.82}\nt{0.90} 
                                                & \ot{8.4}\nt{9.2} 
                                                & \ot{1.31}\nt{1.00} \cr
           &      &       &          &          & \ot{7.83}\nt{7.84} 
                                                & \ot{10.31}\nt{10.32} & 4.5 
                                                & \ot{2.23}\nt{2.24} & MW &  2.0 & $10^6$ 
                                                & \ot{0.94}\nt{1.00} 
                                                & \ot{7.8}\nt{8.2} 
                                                & \ot{1.37}\nt{1.15} \cr
   ngc3627 &   Sb &  8.90 &     8.88 &     9.76 & \ot{7.68}\nt{7.69} 
                                                & \ot{10.37}\nt{10.38} 
                                                & \ot{4.6}\nt{4.3} 
                                                & \ot{3.67}\nt{3.62} & MW &  3.0 & $10^5$ 
                                                & \ot{2.36}\nt{2.20} 
                                                & \ot{13.4}\nt{12.6} 
                                                & \ot{1.12}\nt{0.74} \cr
           &      &       &          &          & \ot{7.57}\nt{7.58} 
                                                & \ot{10.36}\nt{10.37} 
                                                & \ot{4.6}\nt{4.3} 
                                                & \ot{4.62}\nt{4.58} & MW &  4.0 & $10^6$ 
                                                & \ot{1.35}\nt{1.28} 
                                                & \ot{10.8}\nt{10.3} 
                                                & \ot{1.27}\nt{0.77}\cr
   ngc4536 &  Sbc & 25.00 &     9.71 &    10.01 &  7.80 
                                                & \ot{10.79}\nt{10.80} 
                                                & \ot{3.5}\nt{3.8} 
                                                & \ot{7.30}\nt{7.38} & MW &  5.0 & $10^5$ 
                                                & \ot{5.16}\nt{5.35} 
                                                & \ot{24.4}\nt{25.0} 
                                                & \ot{1.09}\nt{0.68}\cr
           &      &       &          &          & \ot{7.81}\nt{7.82} 
                                                & \ot{10.78}\nt{10.79} 
                                                & \ot{3.2}\nt{3.5} 
                                                & \ot{6.74}\nt{6.85} & MW &  5.0 & $10^6$ 
                                                & \ot{3.10}\nt{3.31} 
                                                & \ot{21.2}\nt{22.2} 
                                                & \ot{1.37}\nt{0.79} \cr
   ngc4569 &  Sab & 20.00 &     8.80 &     9.97\tablenotemark{j}
                                                &  7.75 & 10.30 & 4.0 &  2.60 & MW &  2.0 & $10^7$ &     2.09 & 18.5 &  0.75 \cr
           &      &       &          &          &  7.76 & 10.30 & 4.4 &  2.55 & MW &  2.0 & $10^6$ &     2.27 & 16.4 &  0.79 \cr
   ngc4631 &   Sd &  9.00 &    10.09 &     9.49 &  8.11 & 10.55 & 3.7 &  2.02 & MW &  1.0 & $10^3$ &     17.2 & 19.7 &  0.33 \cr
           &      &       &          &          &  7.95 & 10.53 & 3.6 &  2.76 & MW &  2.5 & $10^6$ &     0.89 &  7.4 &  0.48 \cr
   ngc4826 &  Sab &  5.60 &     8.49\tablenotemark{i}
                                     &     9.03\tablenotemark{j}
                                                &  6.89 &  9.66 & 3.0 &  4.35 & MW &  4.0 & $10^7$ &     0.63 &  6.7 &  0.16 \cr
           &      &       &          &          &  6.89 &  9.66 & 3.1 &  4.32 & MW &  4.0 & $10^6$ &     0.70 &  5.9 &  0.17 \cr
   ngc5195 &   Im &  8.20 &      --- &     8.55 &  6.61 &  9.47 & 2.5 &  5.35 & MW &  1.2 & $10^4$ &     43.1 & 44.5 &  0.33 \cr
           &      &       &          &          &  6.45 &  9.43 & 2.4 &  7.16 & MW &  5.0 & $10^6$ &     3.85 & 24.8 &  0.70 \cr
\tablebreak
   ngc5713 &  Sbc & 26.60 &     9.93 &    10.02 &  7.95 & 10.72 & 2.2 &  6.63 & LMC & 4.0 & $10^4$ &     9.62 & 26.8 &  0.41 \cr
           &      &       &          &          &  7.94 & 10.72 & 3.1 &  4.46 & MW &  3.0 & $10^5$ &     5.17 & 24.0 &  0.49 \cr
           &      &       &          &          &  7.96 & 10.71 & 2.8 &  4.19 & MW &  3.0 & $10^6$ &     3.39 & 22.4 &  0.55 \cr
   ngc5866 &   S0 & 12.50 &  $<$8.28 &     8.93 &  6.65 &  9.48 & 2.0 &  5.00 & MW &  5.0 & $10^3$ &        0 &  0.0 &  0.97 \cr
           &      &       &          &          &  6.65 &  9.48 & 2.0 &  5.00 & MW &  5.0 & $10^6$ &        0 &  0.0 &  0.97 \cr
   ngc7331 &   Sb & 14.72\tablenotemark{k} &     9.96\tablenotemark{i}
                                     &    10.14\tablenotemark{j}
                                                &  8.05 & 10.69 & 4.2 &  3.22 & MW &  3.0 & $10^7$ &     0.53 &  5.6 &  0.54 \cr
           &      &       &          &          &  8.06 & 10.69 & 4.3 &  3.20 & MW &  3.0 & $10^6$ &     0.57 &  4.9 &  0.55 \cr
   ngc7552 &  Sab & 22.30 &     9.68 &      --- &  7.84 & 11.07 & 1.5 
                                                & \ot{12.41}\nt{12.43} & MW &  8.0 & $10^7$ 
                                                & \ot{4.22}\nt{4.24} 
                                                & \ot{31.4}\nt{31.5} 
                                                & \ot{0.51}\nt{0.50} \cr
           &      &       &          &          &  7.84 & 11.07 & 2.2 
                                                & \ot{12.53}\nt{12.55} & MW &  8.0 & $10^6$ 
                                                & \ot{5.28}\nt{5.30} 
                                                & \ot{31.0}\nt{31.1} 
                                                & \ot{0.64}\nt{0.63} \cr
\enddata
\tablenotetext{a}{$D$, 21~cm flux, and CO~1-0 flux from \citet{Kennicutt+Armus+Bendo_etal_2003} unless otherwise noted.}
\tablenotetext{b}{Estimated from dust model.}
\tablenotetext{c}{Fraction of dust mass contributed by PAHs with $N_C<10^3$ C atoms.}
\tablenotetext{d}{Dust-weighted mean starlight intensity.}
\tablenotetext{e}{$\Umin$, $\Umax$ = lower, upper cutoff for starlight intensity scale factor $U$.}
\tablenotetext{f}{Fraction of dust mass in regions with $U>\Umin$.}
\tablenotetext{g}{fraction of dust luminosity from regions with $U>10^2$}
\tablenotetext{h}{$\chi^2/(N_b-5)$, where $N_b=$ number of bands used for fitting.}
\tablenotetext{i}{21~cm flux from \citet{Walter_2005}.}
\tablenotetext{j}{CO~1-0 flux from \citet{Sheth+Vogel+Regan_etal_2005}.}
\tablenotetext{k}{\citet{Freedman+Madore+Gibson+Ferrarese+Kelson_etal_2001}.}
\end{deluxetable}


\begin{figure}[h]
\begin{center}
\sedfigd{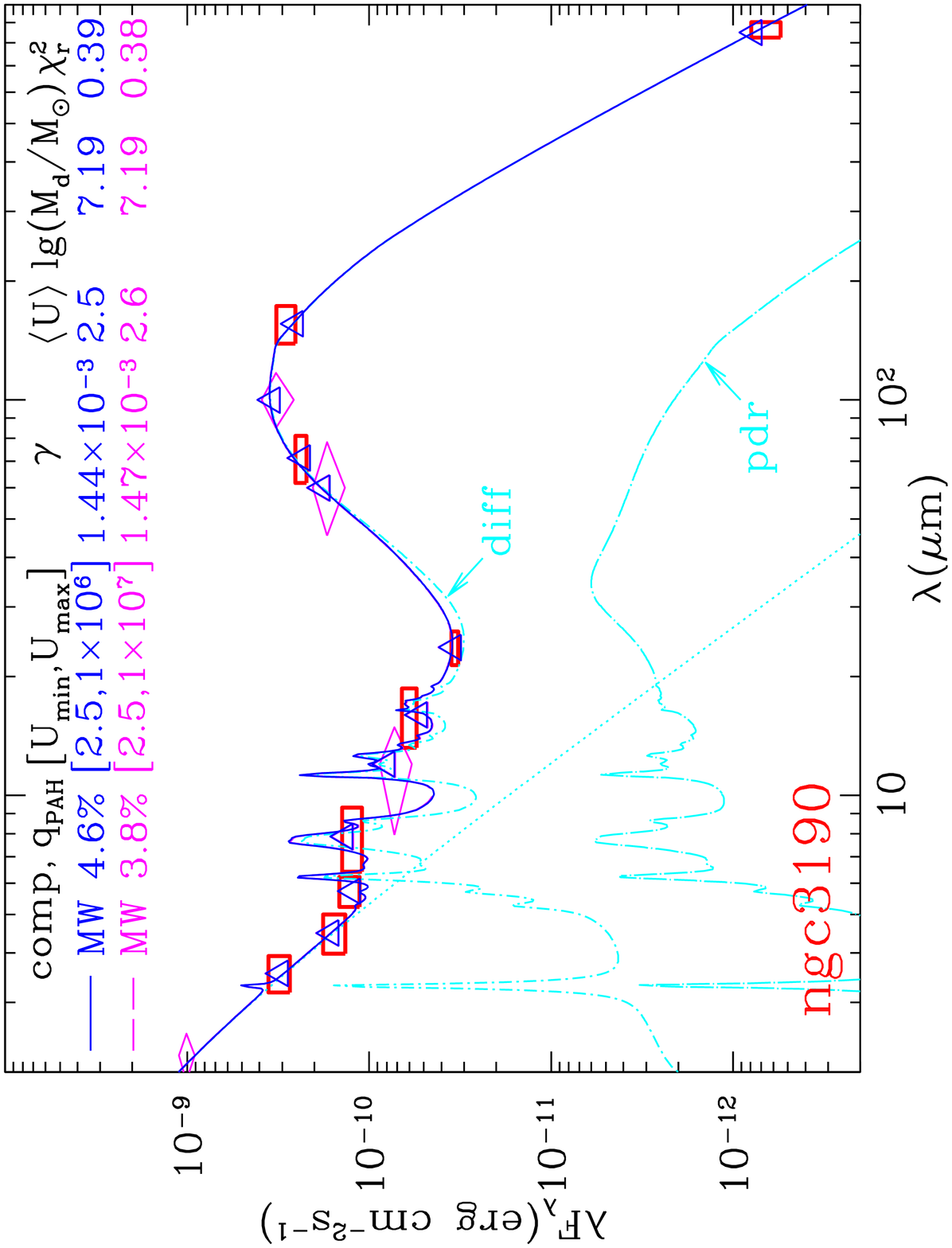}{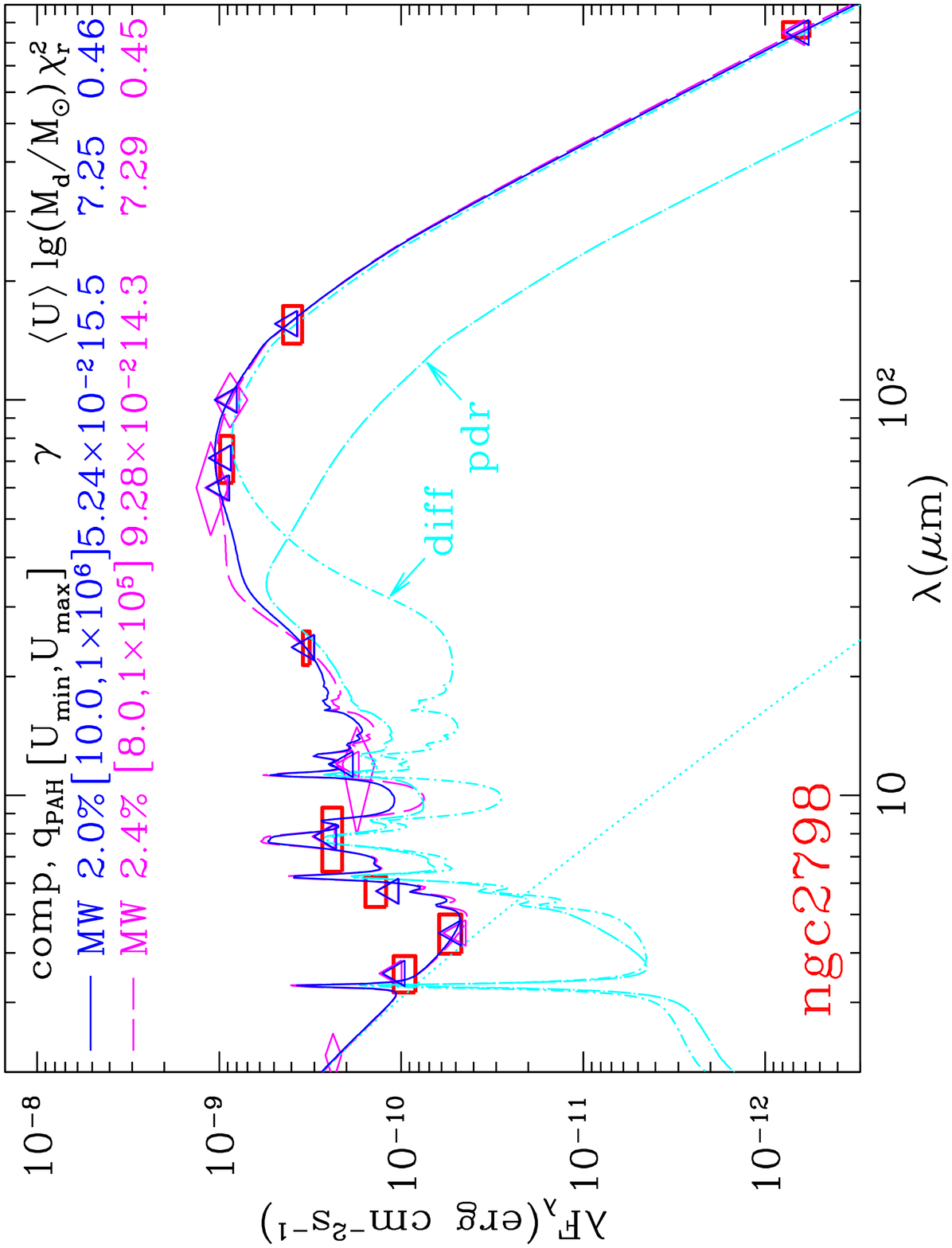}
\vspace*{1.0cm}
\caption{\label{fig:n3190}
         \label{fig:n2798}
	 \footnotesize
	 (a) SED for the SAap galaxy NGC~3190.  
	     Rectangles show observed fluxes in the IRAC, MIPS, 
	     SCUBA bands, and the IRS~16\um\ band;
	     diamonds show 2MASS~2.2\um, and IRAS 12, 60, and 100\um, 
	     fluxes.
	     Vertical extent of rectangles and diamonds corresponds to
	     $\pm1\sigma$ range, and width corresponds to nominal
	     width of band.
   	     Solid line shows best-fit model $\Umax=10^6$ model, 
	     with IRAC, MIPS, IRS~16\um, IRAS,
	     and SCUBA data used to constrain the fit; $N_b=12$ for NGC~3190.
	     Triangles show model convolved with the IRAC, MIPS, IRS~16\um, 
	     IRAS, and SCUBA bands.
	     Dot-dashed curves show separate contributions of starlight 
	     and emission from dust heated by $U=\Umin$ (labelled ``diff''),
	     and dust heated by $\Umin<U<\Umax$ (labelled ``pdr'').
	     Long-dashed curve shows best-fit model when $\Umax$ is
	     unconstrained -- for this case the model spectrum is nearly
	     indistinguishable for the simple reason that there is relatively
	     little dust 
             ($\gamma\approx \ot{0.002}\nt{0.0015}$), 
             and therefore relatively
	     little power, in the ``pdr'' component.
	  (b) As for (a), but for the SBa galaxy NGC~2798, with
	     $N_b=11$ (
	     global imaging in the IRS~16\um\ band is unavailable).
	     }
\end{center}
\end{figure}
The models are generally very successful in fitting the observed SEDs.
The SED of NGC~3190, together with a model fit, appears in 
Figure \ref{fig:n3190}a. 
This is an
example where the model
(with $\Umin=2.5$, $\Umax=10^7$, $\gamma=\ot{1.9}\nt{1.5}\times10^{-3}$) 
provides an excellent fit to the observations,
with $\chi^2_r=\ot{0.21}\nt{0.38}$.
The inferred dust mass is $\Mdust=10^{\ot{7.18}\nt{7.19}}\Msol$,
and the mean starlight intensity is estimated to be $\langle U\rangle=2.6$.
Only a small fraction $\gamma\approx\ot{0.002}\nt{0.0015}$ of the dust 
is in regions with starlight intensities $U>\Umin$.
Also shown in Figure \ref{fig:n3190}a (solid line) is a fit with the 
upper cutoff fixed at $\Umax=10^6$.  The model SED is nearly identical
(the solid curve and dash-dot curves coincide in the plot) and the
derived dust mass is unchanged.

Figure \ref{fig:n2798}b shows the SED of NGC~2798, a galaxy where the dust is
considerably warmer than in NGC~3190.
The best-fit dust model ($\Umin=8$, $\Umax=10^5$, $\gamma=0.09$)
provides a good fit ($\chi_r^2=0.45$) to the observations,
but the model with $\Umax=10^6$ provides nearly as good a fit
($\chi_r^2=0.46$), a derived dust mass differing by only 0.04 dex,
and similar $\qpah$ (2.0\% vs.\ 2.5\%).

The model fits to NGC~2798 both have a noticeable ``bump''
near $\sim33\mum$, produced
by the warm graphite grains in the DL07 model.
The small but nonzero conductivity for $\vec{E}\parallel c$ in 
graphite\footnote{%
    The $c$-axis is normal to the sheets of hexagonally-organized C atoms.} 
causes the absorption cross section $C_{\rm abs}$ to rise to a broad
local maximum near $\sim33\mum$ 
\citep[DL07]{Draine+Lee_1984}.
When exposed to
starlight with $U\gtsim 10^4$, 
graphite grains produce a broad emission bump peaking in the 30--35$\mum$
region, depending on $U$.
Unfortunately, the MIPS 24\um\ and 71\um\ bands straddle this feature, and
do not constrain it.
Implications of the model 33\um\ feature are discussed below in
\S\ref{sec:dust composition}.

\begin{figure}[h]
 \begin{center}
  \includegraphics[angle=0,width=\figwidth]%
                  {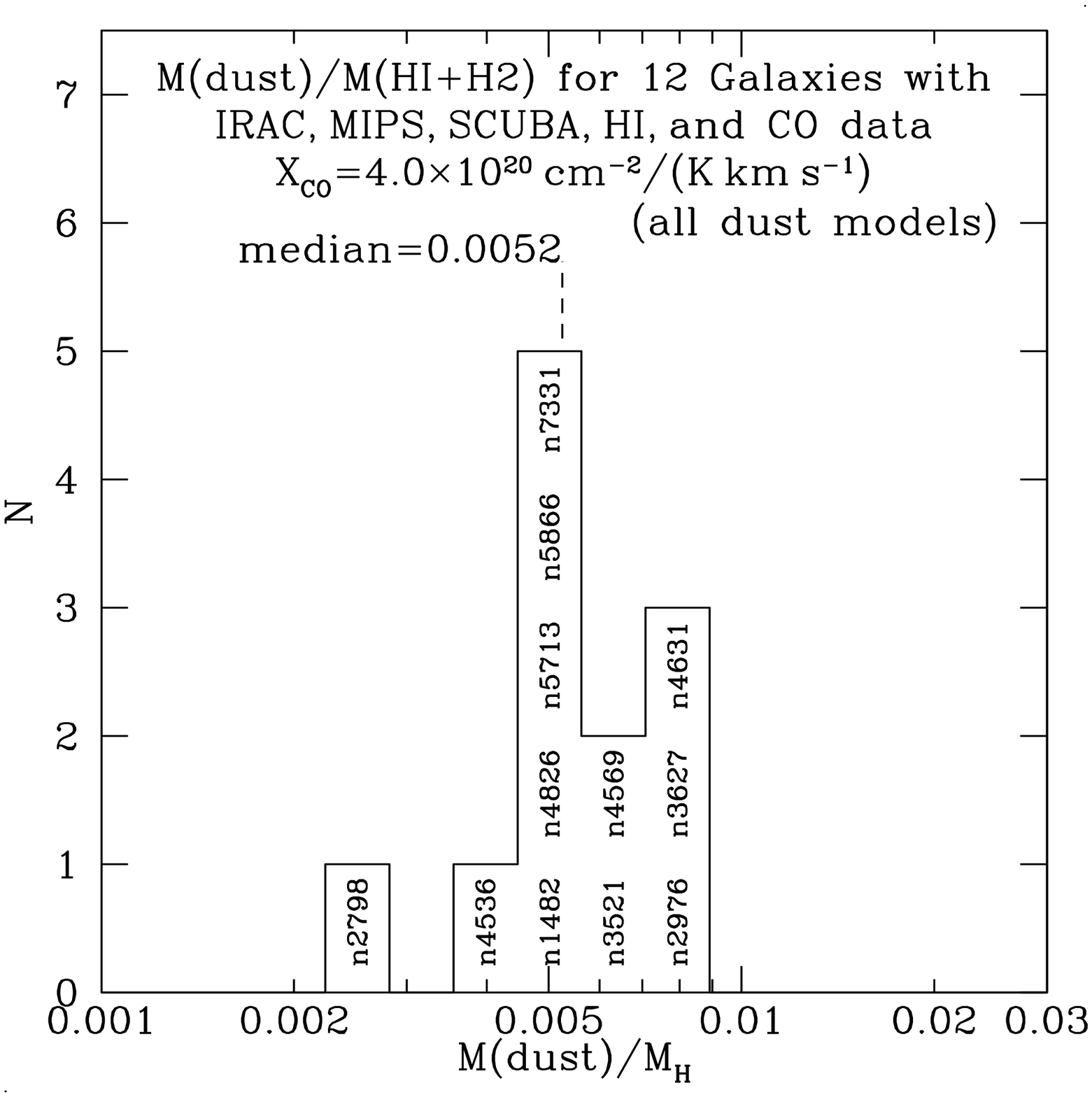}
  \caption{\label{fig:histogram Ubar_scuba_all}
           \label{fig:histogram MH/Md_scuba_all}\footnotesize
	   Histograms of $\Mdust/M({\rm H~I+H}_2)$ for
	   the 12 galaxies in the sample for which both \ion{H}{1}~21~cm and CO~1-0 
	   have been measured, using
	   $\XCO=4\times10^{20}\cm^{-2}(\K\kms)^{-1}$ to estimate
	   $M(\HH)$.
           }
 \end{center}
\end{figure}

\subsection{Dust-to-Gas Ratio}

If the interstellar 
abundances of carbon and all heavier elements were proportional
to the gas-phase oxygen abundance, 
and if the same fraction of the major condensible
elements (C, O, Mg, Si, Fe) were in solid form as in the Milky Way, 
then all galaxies would be
expected to conform to
\beq \label{eq:Md/MH envelope}
\frac{\Mdust}{\MH} \approx 
0.010 \times \frac{{\rm (O/H)}}{{\rm (O/H)}_{\rm MW}} ~~~,
\eeq
where (O/H)$_{\rm MW}$ is the oxygen abundance in the local Milky Way, and
the factor 0.010 is from the MW dust models in 
Table \ref{tab:dust models}.

Table \ref{tab:scuba galaxies} 
gives the estimated dust mass $\Mdust$ for the \Nscuba\ galaxies.
For many of the galaxies there are measurements of 21 cm emission and CO
emission from which the masses of \ion{H}{1} and H$_2$ can be estimated
(see Table \ref{tab:scuba galaxies}).
The usual assumption is made that the H$_2$ mass is proportional to the CO~1-0 luminosity:
\beq \label{eq:XCO}
M(\HH)=1.57\times10^4\Msol
      \left( \frac{\XCO}{4\times10^{20}\cm^{-2}(\K\kms)^{-1}} \right)
      \left( \frac{D}{\Mpc} \right)^2
      \left( \frac{S_{\rm CO}}{\Jy\kms} \right) ~~~.
\eeq
where $\XCO$ is the ratio of H$_2$ column density to antenna temperature
integrated over the CO 1-0 line,
and $S_{\rm CO}\equiv\lambda_{1-0}\int F_\nu(1-0) d\nu$, where
$\int F_\nu(1-0) d\nu$ is the CO 1--0 line flux.
Use of the so-called CO 1-0 luminosity to estimate H$_2$ masses
remains suspect, 
and the 
value to use for the proportionality constant $\XCO$ is uncertain,
with estimates for the Milky Way ranging from 
$3.0\times10^{20}\cm^{-2}(\K\kms)^{-1}$ \citep{Young+Scoville_1991}
to 
$1.56\times10^{20}\cm^{-2}(\K\kms)^{-1}$ 
\citep{Hunter+Bertsch+Catelli+etal_1997}.
Here we use the value $\XCO=4\times10^{20}\cm^{-2} (\K\kms)^{-1}$,
the value recommended by
\citet{Blitz+Fukui+Kawamura_etal_2006} for
giant molecular clouds in Local Group galaxies.
We discuss the value of  $\XCO$
in \S\ref{sec:XCO}, where we argue that dust mass estimates for
the SINGS galaxies favor $\XCO=4\times10^{20}\cm^{-2} (\K\kms)^{-1}$.

Figure \ref{fig:histogram MH/Md_scuba_all} shows the dust/hydrogen
mass ratio for the 12 \SINGS-SCUBA galaxies for which the total (atomic and
molecular) gas masses are known.
The derived dust-to-gas ratios are tightly clustered, with
median $\Mdust/\MH\approx0.005$.
The lowest dust-to-gas ratio is for
NGC~2798, with $\Mdust/\MH\approx0.0023$; the highest is
for NGC~2976, with $\Mdust/\MH\approx 0.0089$.

The dust to gas mass ratios in Figure \ref{fig:histogram MH/Md_scuba_all}
are broadly comparable to the value $\sim 0.007$ 
estimated from gas phase abundances
in the local Milky Way (Table \ref{tab:MW depleted mass}) or 0.010
estimated from models for the observed Milky Way extinction
(Table \ref{tab:dust models}).
The observed range $0.003 \ltsim \Mdust/\MH \ltsim 0.01$ is
consistent with galactic metallicities between $\sim$0.3$\times$ and 
$\sim$1$\times$~solar,
if a similar fraction of the heavy elements is in the form of dust as
in the Milky Way.

Based on Figure \ref{fig:histogram MH/Md_scuba_all}, we conclude that the
model-fitting procedure provides dust mass
estimates for this
sample of galaxies that are 
in reasonable agreement with the dust masses expected if depletions
are similar to depletions in the Milky Way.
Furthermore, the model appears to be able to reproduce the observed
emission spectrum (see, eg., Fig.\ \ref{fig:n3190}).
The metallicity dependence of the dust-to-gas ratio will be examined in 
\S\ref{sec:dust to gas ratio}.

\subsection{Role of LMC and SMC Dust Models}

We now ask whether it is necessary to use the LMC
and SMC dust models in the determination of dust masses -- 
do the dust models
developed for the LMC and SMC lead
to improved fits to the observed spectra and, if so, how
different are the inferred dust masses and PAH abundances?
For each galaxy in the \SINGS-SCUBA subset, we searched for
the model that minimized $\chi_r^2$; for 16 of the \Nscuba\ galaxies,
a MW dust model was preferred, for 
\oldtext{NGC~5195 (=M51b)}\newtext{NGC~5713}
an LMC dust model was favored, but in no case did the SMC bar dust model
give the best fit.
For 
\oldtext{NGC~5195}\newtext{NGC~5713},
where an LMC dust model gave the smallest
$\chi_r^2$, we also give in
Table \ref{tab:scuba galaxies} the parameters for the
best-fit MW model;
the derived $\Mdust$ differs from that obtained using the LMC model by
only 0.01 dex,
and $\qpah=\ot{2.6}\nt{3.1}\%$, vs.\ \ot{2.3}\nt{2.2}\% for the LMC model;
\nt{when the MW model fit is constrained to have $\Umax=10^6$, the derived
$\qpah=2.8\%$.}
Therefore, at least for the \Nscuba\ galaxies with SCUBA photometry, 
the model fitting can be limited to MW-type dust
models with minimal impact on the inferred dust masses or
PAH index.

In \S\ref{sec:descuba} below we demonstrate that we can obtain good
estimates for $\Mdust$ and $\qpah$ even for
galaxies lacking submm photometry, by limiting the search to
models with $\Umin\geq0.7$.
Here we
compare estimates for $\qpah$ and $\Mdust$ obtained using LMC and
SMC dust models with those obtained using MW dust models.

\subsubsection{PAH Abundance $\qpah$}
Let $\qpah({\rm MW})$ be the best-fit estimates
for $\qpah$ when fitting with only MW dust models,
and let $\qpah(x{\rm MC})$ be the estimate obtained using  
only LMC+SMC dust models.
As discussed below (see \S\ref{sec:PAH abundances - scuba plus nonscuba})
we exclude 4 galaxies for which $\qpah$ cannot be
reliably estimated.
Because the LMC and SMC dust models are limited to $\qpah\leq2.4\%$,
we further limit consideration to galaxies
for which $\qpah({\rm MW})\leq 2.4\%$.
The MW models have $\qpah({\rm MW})\geq0.4\%$.
Therefore for the 7 galaxies where the best-fit MW model has
$\qpah=0.4\%$, we can conclude only that the actual $\qpah\leq0.4\%$.

\begin{figure}[h]
  \begin{center}
  \includegraphics[angle=0,width=8.0cm]%
                  {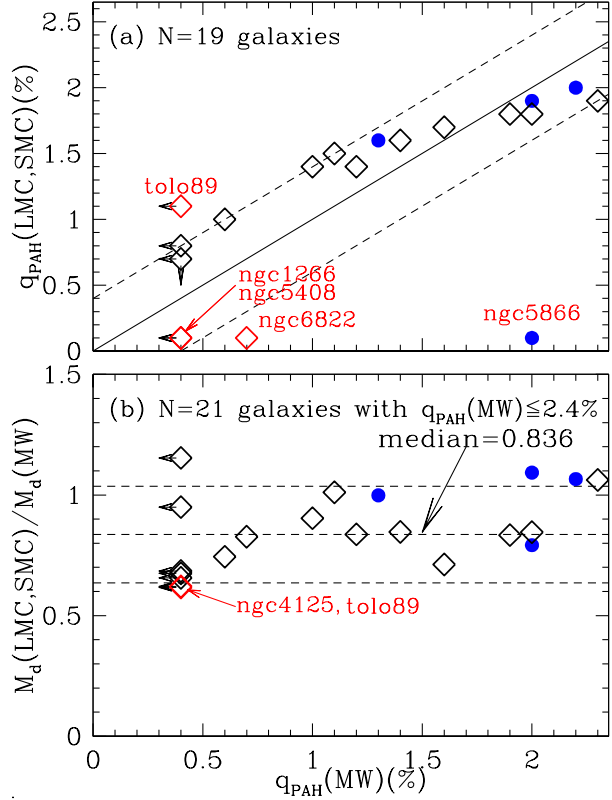}
  \caption{\label{fig:MW_v_LMC params}\footnotesize
           (a) $\qpah$ determined by fitting LMC or SMC models 
               vs.\ $\qpah$ determined by fitting MW models,
	       for galaxies with $\qpah({\rm MW})\leq 2.4\%$.
	       The estimates of $\qpah$ are in good agreement:
	       16/19 galaxies have 
	       $|\qpah({\rm MW})-\qpah(x{\rm MC})|\leq0.4\%$ (region
	       bounded by broken lines).
           (b) Ratio of dust masses for best-fit LMC,SMC model and best-fit
	       MW dust model,
	       for galaxies with $\qpah({\rm MW})\leq2.4\%$.
	       The dust masses are in good agreement: the median ratio is 0.836,
	       and 15/21 galaxies are within $\pm0.20$ of the median.
	     }
  \end{center}
\end{figure}

As seen in Figure \ref{fig:MW_v_LMC params}, the best-fit
MW models and the best-fit LMC+SMC models generally have
similar values of $\qpah$ -- 16/19 galaxies have 
$|\qpah({\rm MW})-\qpah(x{\rm MC})|\leq0.4\%$.
There is one extreme outlier:
the S0 galaxy NGC~5866.
NGC~5866 has relatively little dust -- 
the flux in IRAC band 3 is consistent with starlight alone,
and there is only a modest excess due to dust in IRAC band 4.
As a result, $\qpah$ is not very securely determined for NGC~5866.
When the fitting is limited to LMC and SMC dust models,
the best fit is produced by the SMC model, with $\qpah=0.1\%$.
However, this fit, with $\chi_r^2=1.92$, 
is poor compared to $\chi_r^2=0.97$ for the MW dust model, so the
MW fit would be strongly-favored over the SMC model.
Spectroscopy of the center of NGC~5866
\citep{Smith+Draine+Dale+etal_2007} shows strong PAH emission features,
consistent with the value $\qpah({\rm MW})\approx 2.0\%$.

NGC~6822 is another case of a large discrepancy between
$\qpah=0.7\%$ obtained from MW models versus the value
$\qpah=0.1\%$ obtained when fitting is limited to LMC,SMC dust models.
Once again, the MW dust model produces a significantly better fit
($\chi_r^2=1.13$) than the SMC model ($\chi_r^2=2.04$).

For Tololo 89 (=NGC~5398), on the other hand,
$\qpah\leq0.4\%$ is obtained using the MW dust models, 
but the best-fit LMC,SMC model is an LMC dust model
with $\qpah=1.1\%$.
In this case the fitting is confused by the apparent lack of dust emission in
IRAC band 3, despite a strong dust excess in IRAC band 4.
The MW fit has $\chi_r^2=2.50$, and the
LMC fit has $\chi_r^2=2.47$ -- an indication that there is something wrong
with either the data or the dust model.

Since 16 of the 19 galaxies have 
$\qpah({\rm MW})-\qpah(x{\rm MC})|\leq0.4\%$,
we conclude that the estimates for $\qpah$
are generally not sensitive to the details of the trial dust models, and
restricting the fitting to the MW dust models will
give reasonable results for $\qpah$.

\subsubsection{Dust Mass}

Figure \ref{fig:MW_v_LMC params}b shows, for the
21 galaxies with $\qpah({\rm MW})<2.4\%$, the ratio of dust masses
obtained using the two different dust models.
The median ratio is 0.84, with 15/21 falling within the interval
$\Mdust({\rm LMC,SMC})/\Mdust({\rm MW})=0.85\pm0.20$.
It is clear that the inferred dust mass is relatively insensitive to
the details of the model dust mixture used in the modeling.

Given that we obtain similar values for the dust mass $\Mdust$
and the PAH abundance $\qpah$ using either the MW or the LMC+SMC
dust models, we will henceforth limit the modeling to the MW dust models.
This has the advantage
of providing a set of models with
$\qpah$ ranging from 0.4\% to 4.6\% (see Table \ref{tab:dust models})
allowing systematic comparisons of the level of PAH emission among different
galaxies.  The resulting estimates for dust
masses, starlight intensity distributions should be relatively
robust, and even the PAH abundances appear to also reliable to
within $\Delta\qpah=0.004$. 
From this point on we will limit the model-fitting to the
MW dust models.

\begin{figure}[h]
  \begin{center}
  \includegraphics[angle=0,width=\figwidthd]%
                  {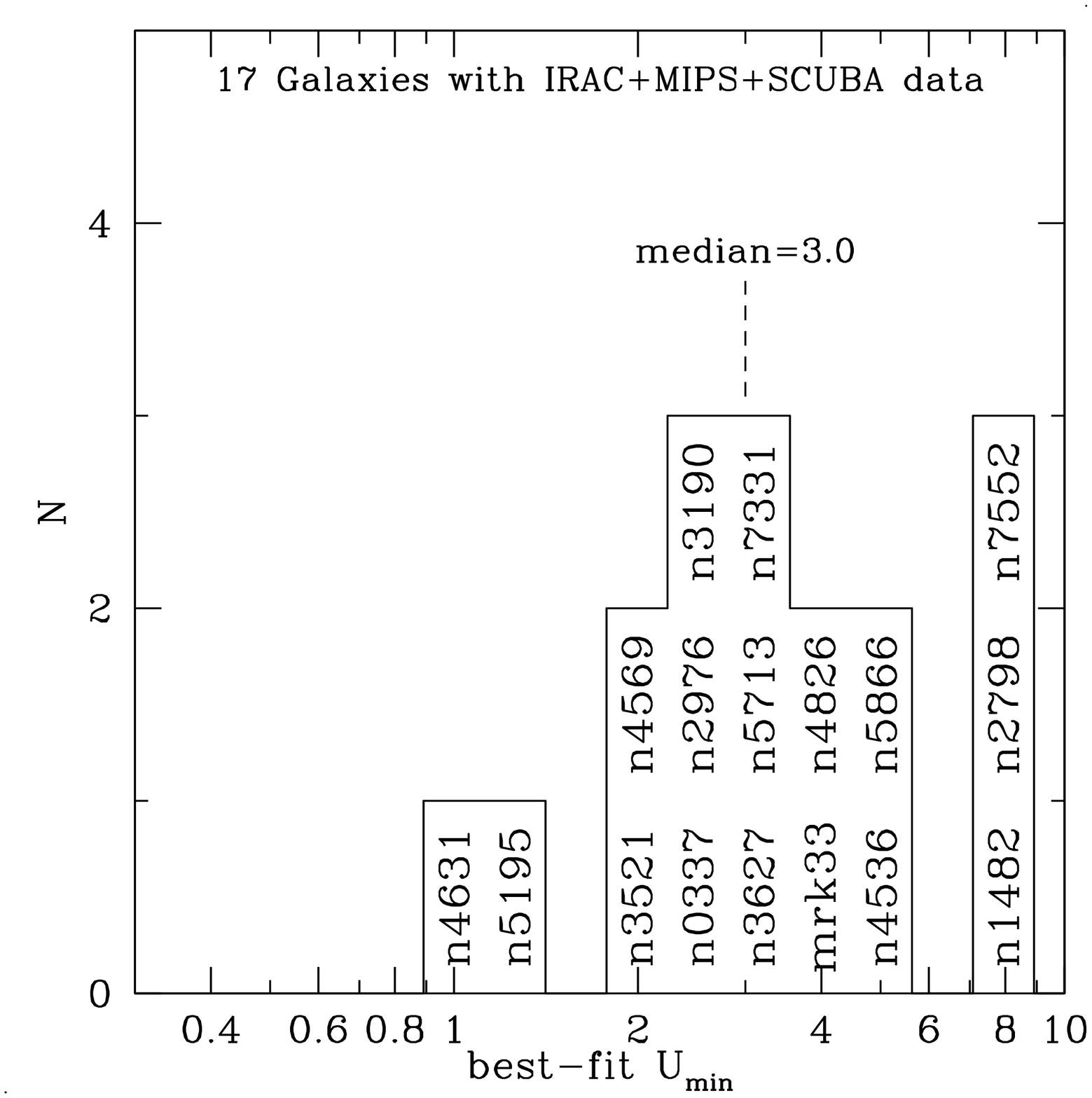}
  \includegraphics[angle=0,width=\figwidthd]%
                  {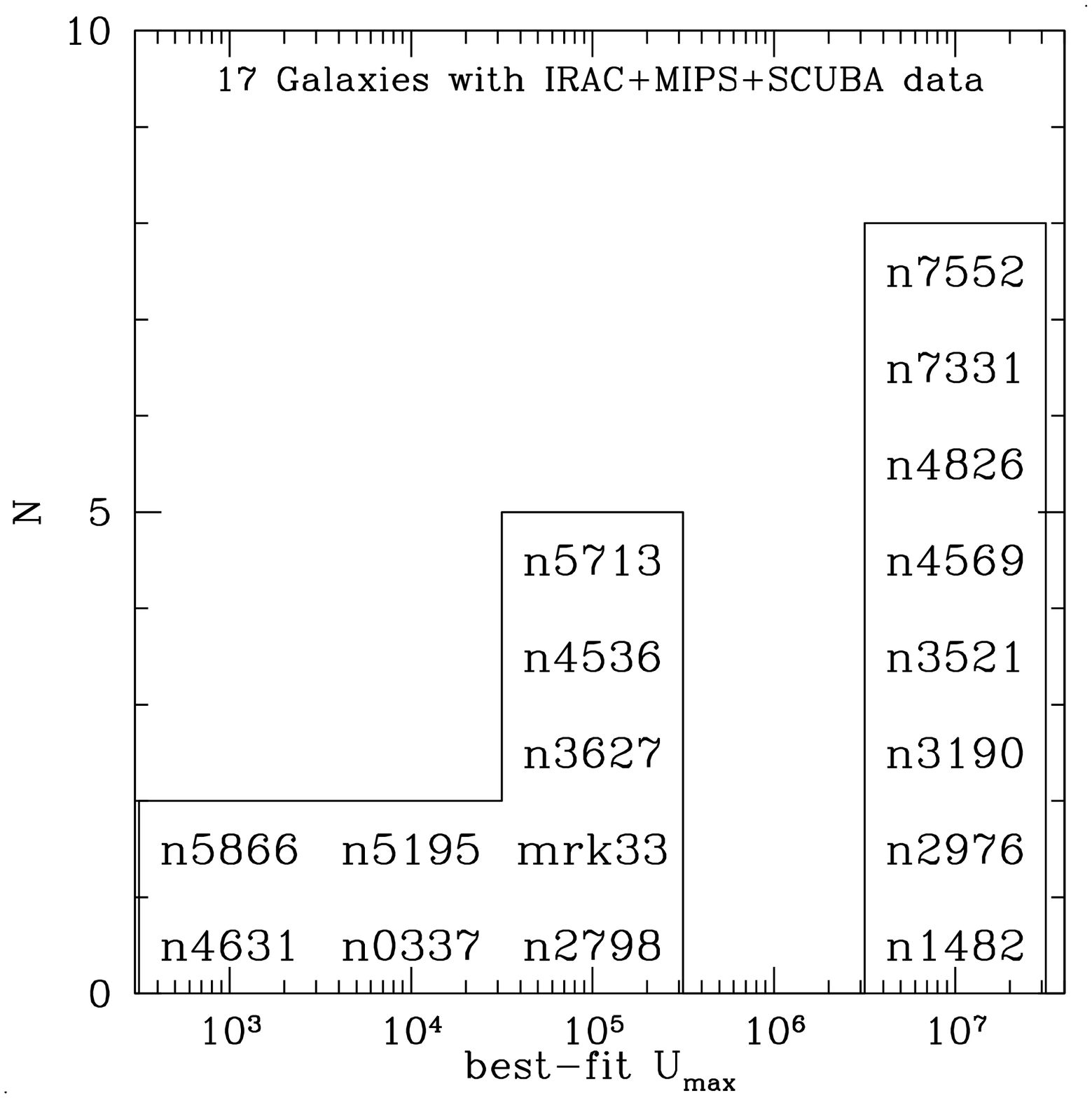}
  \caption{\label{fig:scuba umin umax}\footnotesize
           (a) Distribution of 
           minimum starlight intensity scale factor $U_{\rm min}$.
           (b) Distribution of 
	   maximum starlight intensity scale factor $U_{\rm max}$.
	   The fits were limited to MW dust models.
	   }
 \end{center}
\end{figure}

\subsection{Radiation Field Parameters}
\subsubsection{Lower Cutoff $\Umin$}
\begin{figure}[h]
  \begin{center}
  \includegraphics[angle=0,width=\figwidthd]%
                  {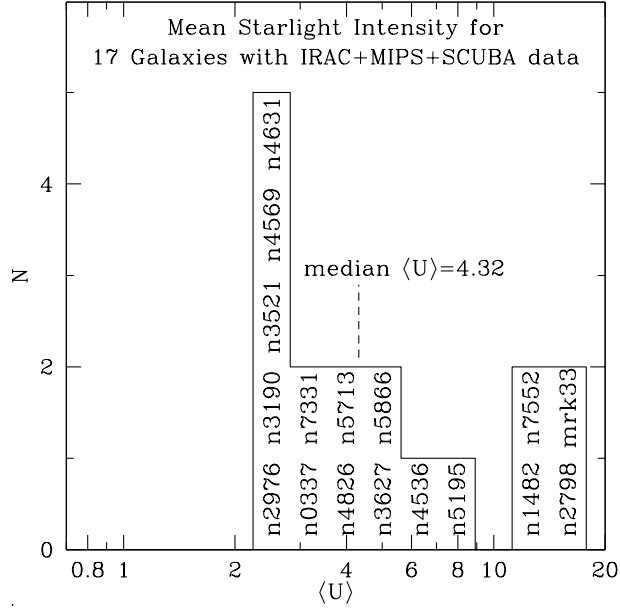}
  \caption{\label{fig:ubar_scuba_all}\footnotesize
           Distribution of $\langle U\rangle$ for \Nscuba\ galaxies
	   with SCUBA data (fits limited to MW dust with $\Umax=10^6$,
	   with
	   adjustable $\gamma$ and $\Umin$.)
           }
  \end{center}
\end{figure}

Figure \ref{fig:scuba umin umax}a shows the distribution
of the best-fit lower cutoff $\Umin$ for the SINGS-SCUBA galaxies.  
In no case did the best-fit
model have $U_{\min}<1$, even though the model-fitting procedure
considered values as small as 0.1.  We therefore conclude that -- at
least in the \SINGS-SCUBA sample -- only a very small fraction of the
overall FIR emission is produced by cool grains within dark
clouds (grains inside dark clouds, shielded from external starlight,
will have $U\ll 1$ and will be cool, unless embedded star formation is
present).  The 160\um\ and 850\um\ emission appears to be produced
primarily by dust in diffuse regions with $U\gtsim 1$.  
\oldtext{%
Indeed, we
will see below that our derived dust mass is generally close
to the value expected based on the observed gas mass,
further showing} 
\newtext{%
    Cool grains in dark clouds could, in principle, contain a substantial
    fraction of the dust mass while contributing only a minor fraction
    of the total IR luminosity.
    However, we will see below (\S\ref{sec:dark clouds})
    that the derived mass of dust heated by starlight with
    $U\gtsim0.7$ is generally close
    to the value expected based on the observed total (atomic and molecular)
    gas mass, indicating}
that the model-fitting procedure is not
overlooking substantial amounts of cooler dust in dark regions.

\subsubsection{\label{sec:Umax}
               Upper Cutoff $\Umax$}

Figure \ref{fig:scuba umin umax}b also shows that 
the model-fitting procedure does not strongly favor any particular value
of $\Umax$.
We have experimented with fixing $\Umax$, and we find that fixing
$\Umax=10^5$ results in only an increase in $\chi^2$ per galaxy of 0.47,
while the number of degrees of freedom per galaxy is increased by 1.
Fixing $\Umax=10^6$ works almost as well -- the mean increase in
$\chi^2$ per galaxy is $0.70$ -- but fixing $\Umax$ at $10^4$ 
or $10^7$ results in larger increases in average $\chi^2$ per galaxy
(in both cases $>1$) relative to the fits where $\Umax$ is allowed to vary.

Because of the insensitivity of the results to the precise value of
$\Umax$, we will
simply set $\Umax=10^6$ and reduce the number
of adjustable parameters from 6 to 5: $\Omega_\star$, $\Mdust$,
$\qpah$, $\Umin$, and $\gamma$.
Having $\Umax$ as large as $10^6$ is not surprising.
The ultraviolet intensity in the Orion Bar photodissociation region, for example,
is $\sim3\times10^4$ times stronger than the local starlight background
\citep{Marconi+Testi+Natta+Walmsley_1998,Allers+Jaffe+Lacy_etal_2005},
and the dust within compact and ultracompact \ion{H}{2} regions, and in the
surrounding PDRs,
is expected to be heated by radiation fields as large as $\sim10^6$.
We therefore expect the IR spectrum of a star-forming galaxy
to have a noticeable contribution from dust heated by $U\approx10^6$.
From this point onward, we will discuss only MW dust models restricted
to $\Umax=10^6$.

\subsubsection{Mean Intensity $\langle U\rangle$ 
               and Fraction 
               of Power from High-Intensity Regions}
Let $P_0$ be the power absorbed per unit dust mass in a radiation
field $U=1$.
For the intensity distribution (\ref{eq:Udist}) with $\alpha=2$,
the dust luminosity $\Ldust$ is obtained by integrating
\beq
d\Ldust = U~ P_0~ d\Mdust = 
P_0 \Mdust
\left[
(1-\gamma)U\delta(U-\Umin)
+
\frac{\gamma}{\Umin^{-1}-\Umax^{-1}}U^{-1}
\right] dU ~~~.~~~
\eeq
Thus the total dust luminosity is
\beq
\Ldust = \langle U\rangle P_0 \Mdust ~~~,
\eeq
where
the dust-weighted mean starlight intensity scale factor is
\beqa
\langle U\rangle &=& 
\left[
(1-\gamma)\Umin
+
\frac{\gamma \ln(\Umax/\Umin)}{\Umin^{-1}-\Umax^{-1}}
\right]
~~~.
\eeqa
Figure \ref{fig:ubar_scuba_all} shows the distribution of
$\langle U\rangle$ for the \Nscuba\ \SINGS-SCUBA galaxies,
fitted with MW dust models with $\Umax=10^6$.
These galaxies are clearly not all alike: $\langle U\rangle$ varies
by a factor of 6 over the sample,
from $\langle U\rangle=2.6$ (for NGC~4569) to 
$\langle U\rangle =16$ (for Mrk~33),
with median $\langle U\rangle = 4.3$.

\begin{figure}[h]
 \begin{center}
   \includegraphics[angle=0,width=\figwidthd]%
                   {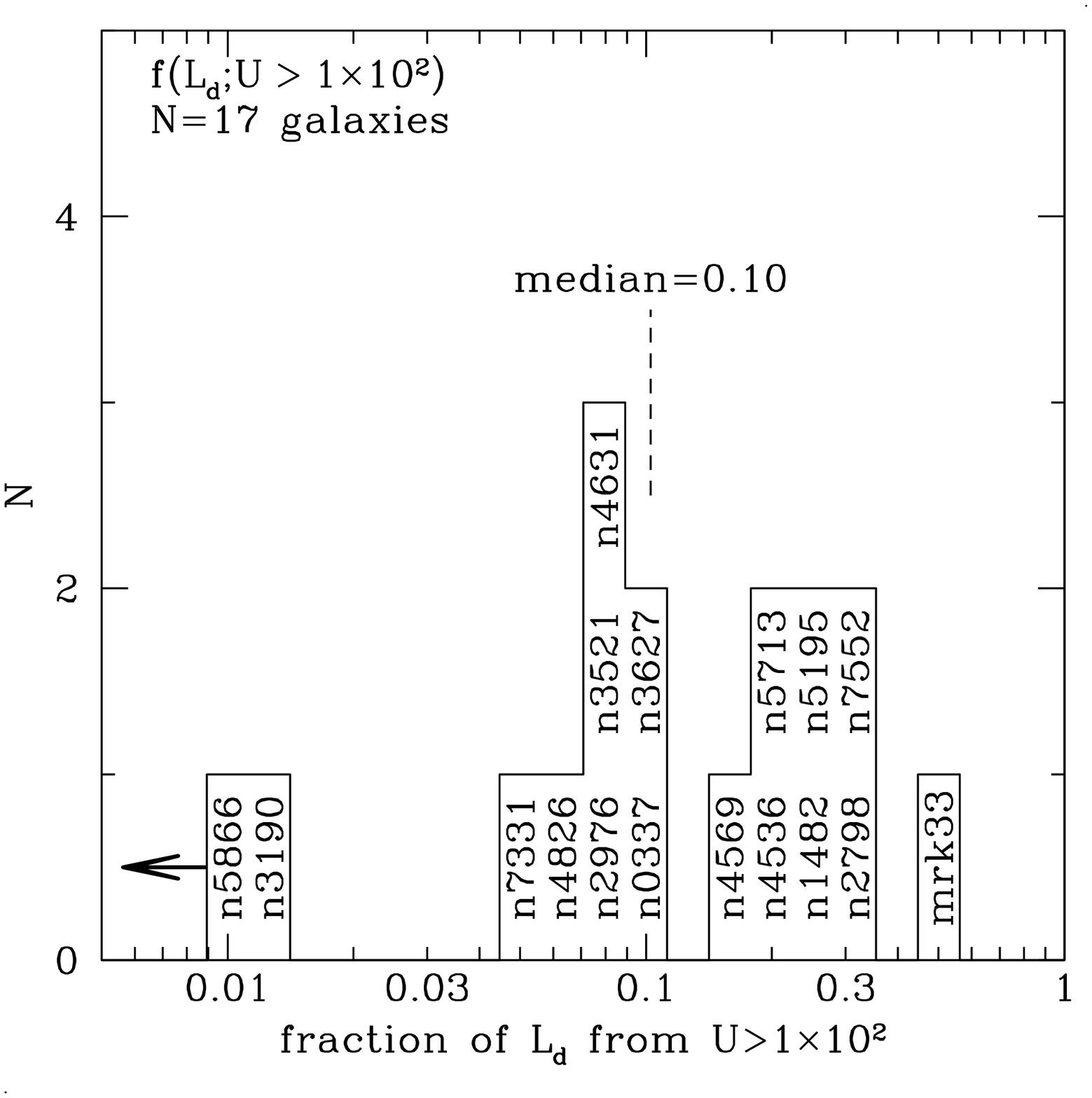}
   \includegraphics[angle=0,width=\figwidthd]%
                   {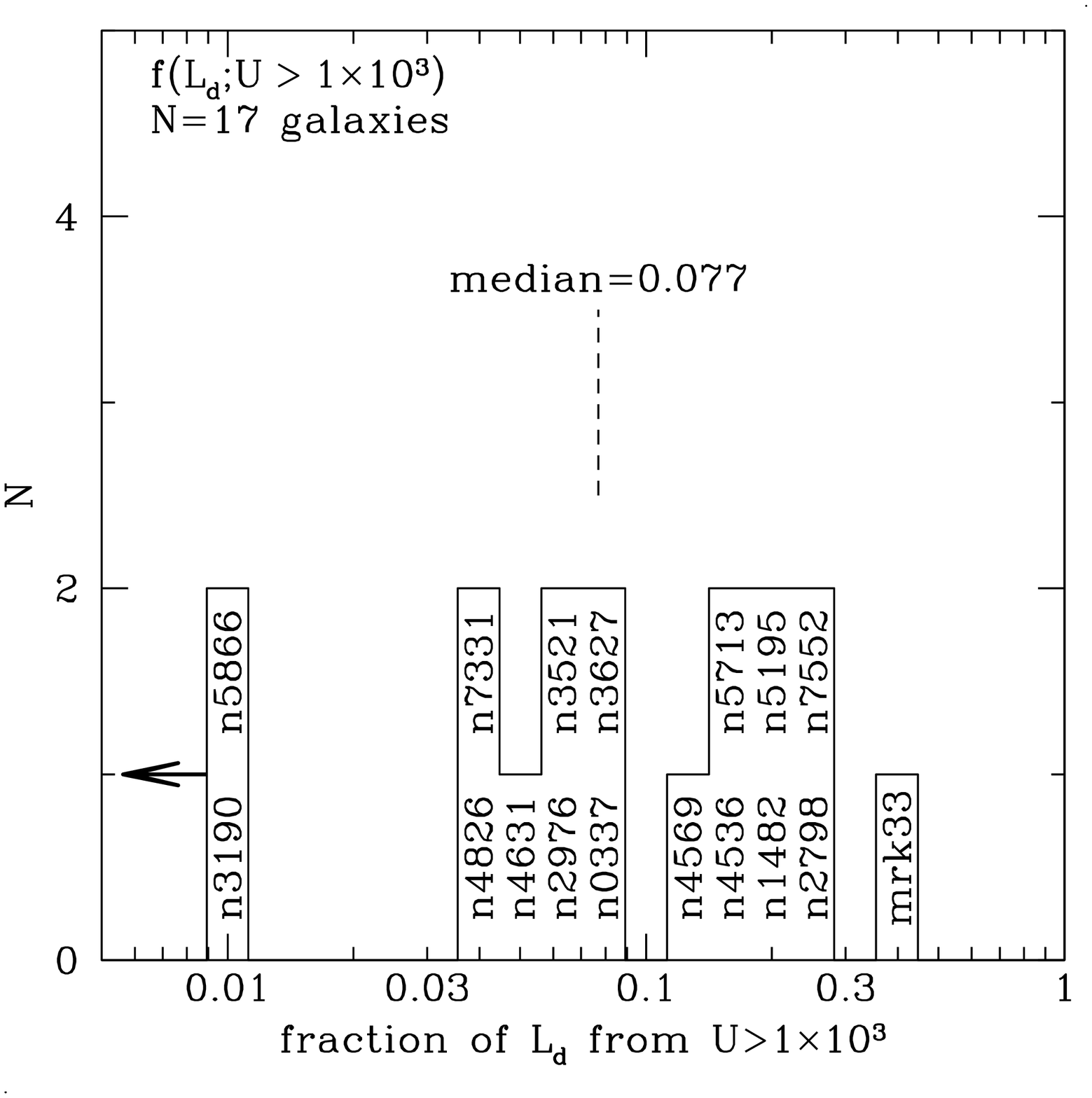}
   \caption{\label{fig:pdr power - scuba}\footnotesize
            Distribution of the
	    fraction of the dust luminosity radiated by dust exposed
	    to starlight intensities $U>10^2$ and $U>10^3$.
	    MW dust models are used for the fits,
	    with fixed $\Umax=10^6$, and
	    adjustable $\gamma$, $\Umin$.
	    Galaxies with $f\leq 0.01$ have been grouped in a bin
	    at 0.01.
	    }
 \end{center}
\end{figure}

The fraction of the dust luminosity radiated from regions with $U>U_c$ (for
$\Umin<U_c<\Umax$) is
\beq
f(\Ldust;U>U_c) = \frac{\gamma \ln(\Umax/U_c)}
           {
           (1-\gamma)(1-\Umin/\Umax)
           + 
           \gamma \ln(\Umax/\Umin) 
           }
~~~.
\eeq
Figure \ref{fig:pdr power - scuba} shows the distribution of
$f(\Ldust;U>U_c)$ for $U_c=10^2$ and $10^3$ for the \Nscuba\ 
galaxies with SCUBA data.
The median galaxy in 
\oldtext{this}%
\newtext{the SINGS-SCUBA} 
sample has $\sim\oldtext{11}\newtext{10}\%$ of the dust power originating
in regions with $U>10^2$.
In extreme cases, this fraction can be considerably larger,
e.g., Mrk~33, with $f(\Ldust;U>10^2)\approx48\%$,
and $>30\%$ for NGC~2798 and NGC~7552.
At the other extreme, the best-fit model for NGC~5866 
(see \S\ref{sec:NGC5866} below)
has $\gamma=0$ and $f(\Ldust;U>10^2)=0$.
Values of $U\gtsim 10^2$ are expected to arise primarily
in star-forming regions where dust is found near luminous stars.
The values of $f(\Ldust;U>U_c)$ estimated from this fitting
procedure are only approximate, but values of $f(\Ldust;U>10^2)\gtsim 0.03$
should be indicative of significant star formation rates.

\begin{figure}[h]
 \begin{center}
  \includegraphics[angle=0,width=\figwidth]%
                  {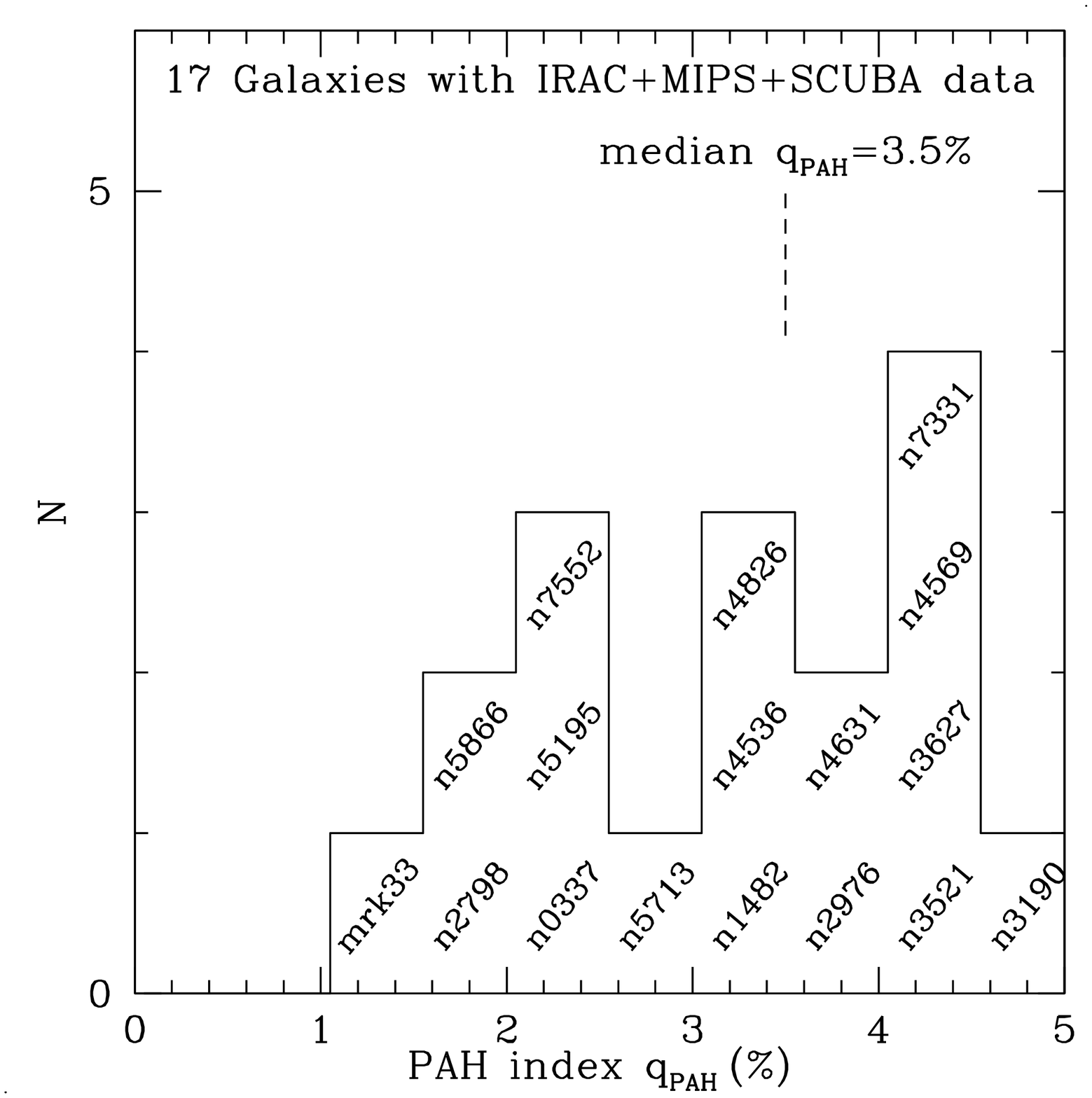}
  \caption{\label{fig:scuba pah}\footnotesize
           PAH index for \Nscuba\ galaxies with SCUBA data,
	   estimated by fitting MW dust models, with $\Umax$ allowed
	   to vary.
	  }
  \end{center}
\end{figure}

\subsection{\label{sec:PAH abundances-SCUBA}
            PAH Abundances}

Observations with {\it ISO} showed that normal star-forming galaxies
routinely have strong 6--9\um\ PAH emission 
\citep{Dale+Helou+Contursi_etal_2001}.
This is confirmed for the present sample of \Nscuba\ \SINGS-SCUBA galaxies.
Figure \ref{fig:scuba pah} shows a histogram of $\qpah$ for these
galaxies.
All of the galaxies show PAH emission, with median 
$\qpah=\oldtext{3.2}\newtext{3.5}\%$.
None of these galaxies have $\qpah<1.3\%$.
The distribution of PAH abundances over a total of \Npah\ \SINGS\ galaxies
will be discussed in \S\ref{sec:PAH abundances - scuba plus nonscuba} below.

\subsection{Discussion of Selected Galaxies}

\subsubsection{\label{sec:model failures?}\label{sec:NGC5866}
               Discrepant Cases: NGC~3521 and NGC~\oldtext{4536}\newtext{5866}}

Most of the \Nscuba\ galaxies in the \SINGS-SCUBA sample have
SEDs that are satisfactorily reproduced by the adopted dust model
(see Fig.\ \ref{fig:n3190}).
If we consider only MW dust models, with fixed $\Umax=10^6$,
the \Nscuba\ galaxies have median $\chi_r^2 = 0.55$; 
\oldtext{14}\newtext{16}/17 of the galaxies
have $\chi_r^2<1$.
\oldtext{However, a few of the galaxies are not well-fit.}
The SINGS-SCUBA galaxies with the worst fits (largest $\chi_r^2$) 
\oldtext{for MW models with $\Umax=10^6$}
are NGC~3521 ($\chi_r^2=1.37$),
and NGC~\oldtext{4536}\newtext{5866} ($\chi_r^2=\oldtext{1.37}\newtext{0.97}$).
Here we consider
why the model has difficulty fitting these two galaxies.
\begin{figure}[h]
\begin{center}
\sedfigd{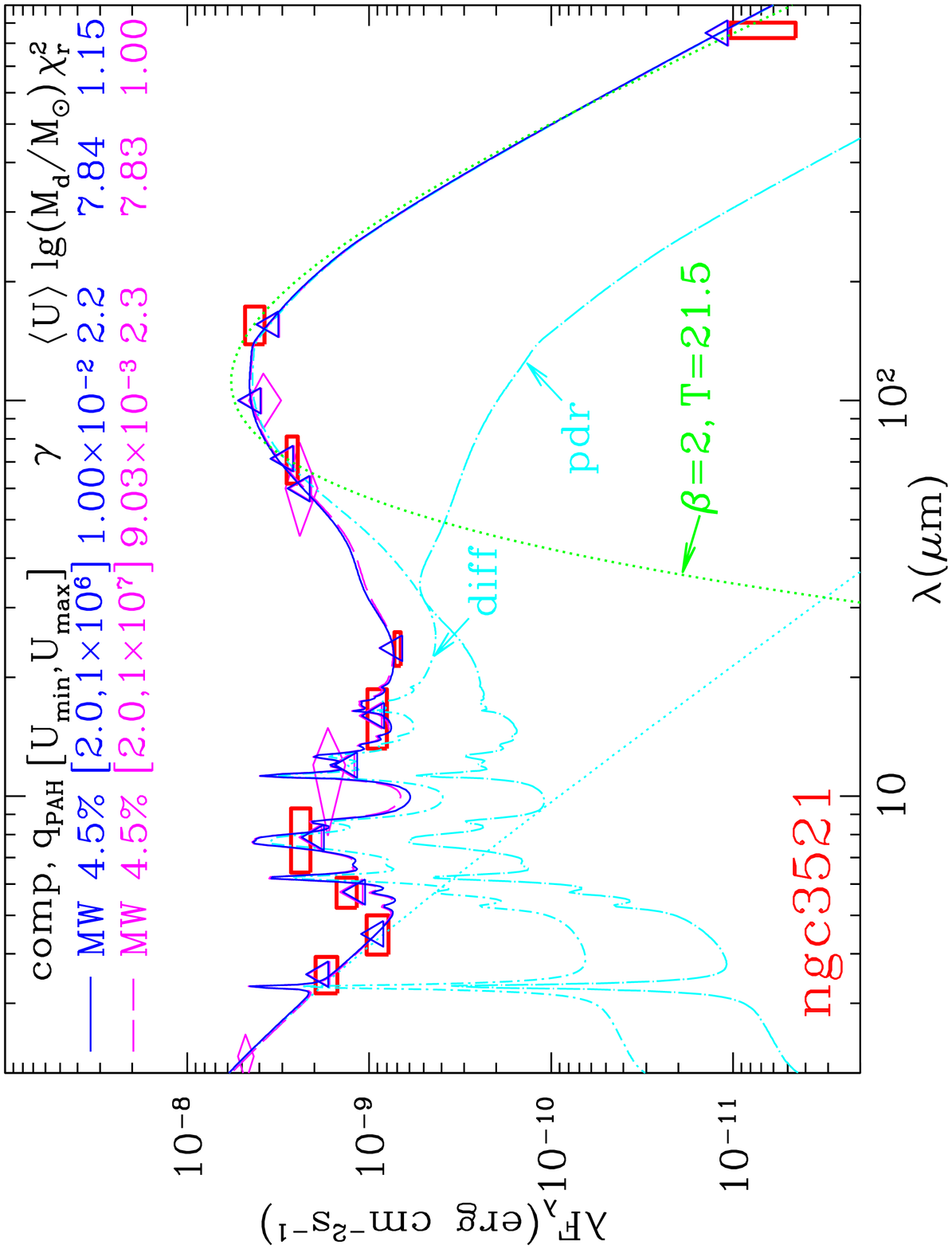}{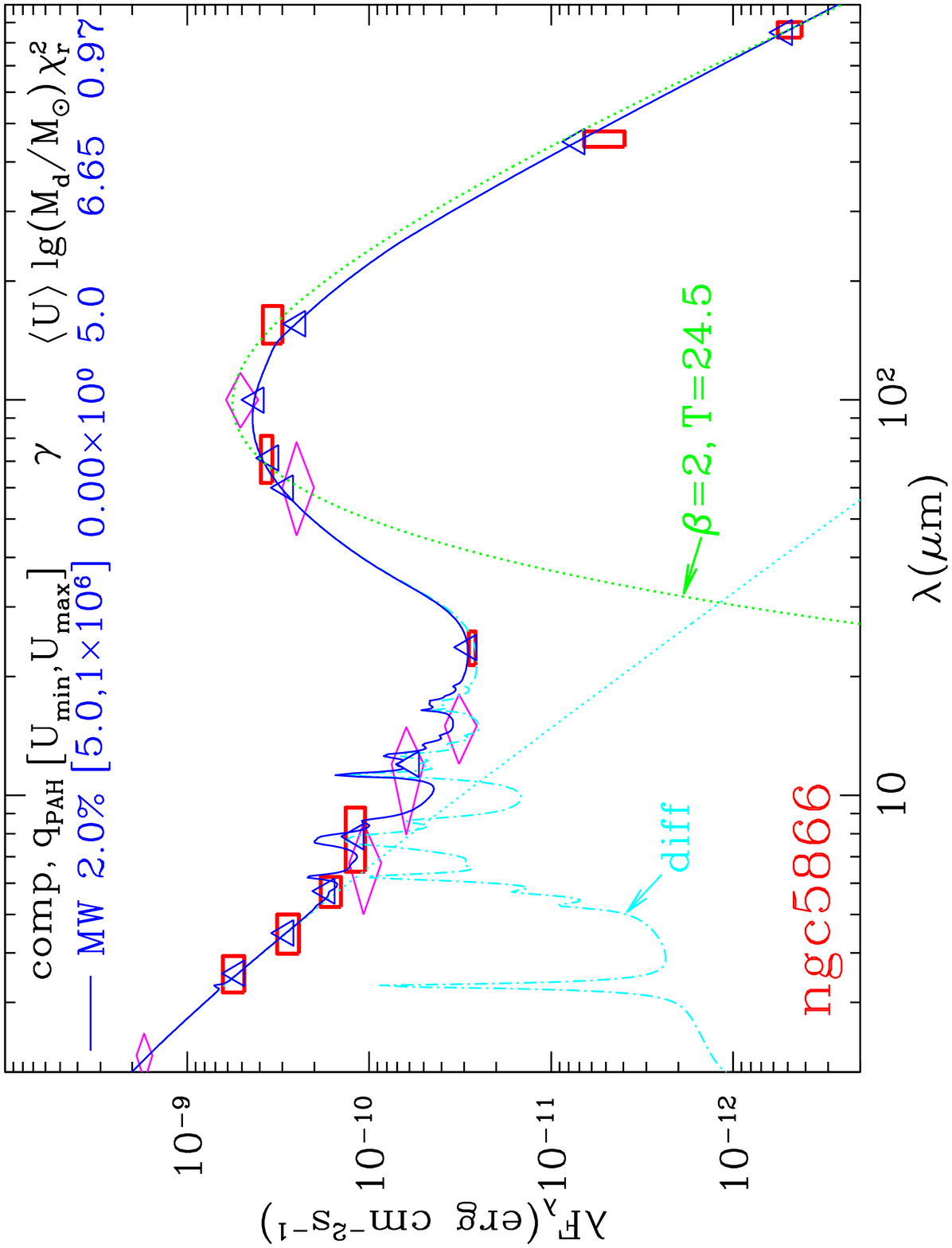}
\vspace*{1.0em}
\caption{\label{fig:n3521}
         \label{fig:n5866}
	 \footnotesize
	 (a) SED of the SABbc galaxy NGC~3521, with
	 $\chi_r^2=1.37$ for the best-fit MW dust model
	 with $\Umax=10^6$.
	 The dotted curve
	 shows emission from hypothetical
	 $T=21.5\K$ dust with $\kappa_\nu\propto\nu^2$,
	 showing that dust with $\kappa_\nu\propto\nu^2$ 
	 cannot reproduce the observed 160\um\
	 emission without exceeding either the 71\um\ flux (if $T>21.5\K$)
	 or the 850\um\ flux (if $T<21.5\K$).
	 Curves labelled ``diff'' and ``pdr'' show contributions of
	 dust heated by $U=\Umin$ and $U>\Umin$, respectively, for the
	 MW model with $\Umax=10^6$.
         (b) 
         \oldtext{SED for the Sbc galaxy NGC~4536.
	 The best-fit MW dust model with $\Umax=10^6$ (solid line) has
	 $\chi_r^2=1.37$.
	 The dotted curve shows emission from hypothetical $T=24.5\K$ dust
	 with $\kappa_\nu\propto\nu^2$, showing that
	 dust with $\kappa_\nu\propto\nu^2$ cannot reproduce the
	 observed 160\um\
	 emission without exceeding either the 71\um\ flux (if $T>24.5\K$)
	 or the 850\um\ flux (if $T<25\K$).
	 For both NGC~3521 and NGC~4536, 
	 the IRAS 100\um\ flux appears to be low, contributing to
	 the relatively large $\chi^2$.}
	 \newtext{SED for the S0 galaxy NGC~5866.
	 The best-fit MW dust model -- with a single starlight
	 intensity $U=5$ --  has $\chi_r^2=0.97$.
	 The model has difficulty reproducing the sharpness of
	 the peak in $\lambda F_\lambda$ near 100\um\, and
	 overestimates the flux at 450\um.}
	 }
\end{center}
\end{figure}

\oldtext{Comparison of the observations (rectangles and diamonds) of the
SABbc galaxy NGC~3521 in
Figure \ref{fig:n3521}a
with the band-convolved model (triangles) shows that much of
the contribution to
$\chi^2$ comes from MIPS 160\um\ (model is low) and 
SCUBA 850\um\ (model is high).

To see the difficulty in fitting the observations, for both
NGC~3521 and 4536 we have plotted
the emission from dust with opacity $\kappa_\nu \propto \nu^2$
with the dust mass adjusted to reproduce the emision at 160\um.
For NGC~3521, such dust with $T=21.5\K$ would account for all of the
flux observed at 71\um, but would exceed the 850\um\ measurement by about
1~$\sigma$.
If the dust temperature is lowered to $T<21.5\K$, 
it would
produce even more emission at 850\um; if raised to $T>21.5\K$
so as not to violate the
850\um\ constraint, it would overproduce the 71\um\ emission.
This is for a single temperature, whereas the galaxy itself undoubtedly
contains dust with a distribution of temperatures, which
must broaden the emission spectrum, leading to greater
discrepancies with the 71\um\ and 850\um\ data.
In short, the reported fluxes at 71\um, 160\um, and 850\um\ do not appear
to be consistent with dust emission, {\it unless the dust opacity is even
steeper than $\nu^2$.}
We note that a modest ($\sim$20\%) 
increase in the 71\um\ flux, or a comparable decrease in the 160\um\ flux,
or a modest increase in the SCUBA flux
would result in an SED that could be comfortably reproduced.
We also note that the IRAS 100\um\ photometry appears to be somewhat
low relative to the 70 and 160\um\ photometry.

In the case of NGC~3521, the best-fit model underpredicts the 6.2 and 7.9\um\
emission, while overpredicting the 16\um\ emission.  
This suggests that the PAH emission spectrum in NGC~3521 may differ
slightly from the average of the band ratios seen in other objects.
NGC~3521 has
$\langle\nu F_\nu^{\rm ns}\rangle_{7.9}/[\langle\nu
F_\nu\rangle_{71}+\langle\nu F_\nu\rangle_{160}]=0.32$, the highest
value in the SINGS-SCUBA sample. 
Further study of NGC~3521, including submm observations,
is called for.

The best-fit model for the SED of the Sbc galaxy NGC~4536
also falls short at 160\um\, while exceeding the SCUBA 850\um\
photometry.  As was the case with NGC~3521, the three longest-wavelength
points (71, 160, and 850\um) appear to be incompatible with
even single-temperature 
dust with $\kappa_\nu\propto\nu^2$.
Unless one or more of the photometric measurements is incorrect,
the 850\um\ photometry appears to require that the dust opacity
be even steeper than $\nu^2$.
}
\newtext{In the case of NGC~3521, the best-fit model underpredicts
         the 7.9\um\ emission, but the more significant
	 discrepancies are at long wavelengths, where the
	 model is high at 71\um, low at 160\um\, and high at 850\um.
	 To illustrate the difficulty with reproducing these
	 three fluxes, Figure \ref{fig:n3521} also shows
	 a single-temperature component with a $\nu^2$ opacity
	 with the temperature and power chosen to fit the 71 and 160\um\
	 data.
	 Note that even this single-temperature component is too
	 ``broad'' -- it exceeds  
	 the 850\um\ datum.
	 If a range of dust temperatures is present, it will
	 broaden the spectrum, and worsen the fit.
	 Either there is a problem with
	 some of the photometric data, or the dust in this
	 galaxy has an opacity that is steeper than $\nu^2$.

	 The second-worst fit is for NGC~5866, 
         an S0 galaxy with a substantial edge-on disk, 
         and the lowest value of
         $\langle\nu F_\nu\rangle_{24}/\langle\nu L_\nu\rangle_{71}$ among the
         \SINGS-SCUBA galaxies
         (see Fig.\ \ref{fig:fluxratios}). The best-fit model has $\gamma=0$ --
all of the dust is exposed
to starlight with intensity $U=\Umin=5$.
The small value of $\gamma$ is presumably because the illumination
of the dusty disk is dominated by the distributed old stellar population,
with little contribution from star formation in the disk.
This is consistent with nondetection of H$\alpha$, with
SFR$<0.1\Msol\yr^{-1}$ \citep{Kennicutt+Armus+Bendo_etal_2003}
(although dust extinction may also be important).
The nuclear spectrum of NGC5866 \citep{Smith+Draine+Dale+etal_2007}
includes ionic emission lines (e.g., [\ion{Ne}{2}], [\ion{Si}{2}], [\ion{S}{3}]) but they
are relatively weak, consistent with only low levels of star formation;
the low ratio [\ion{S}{3}]/[\ion{Si}{2}] flux ratio also suggests a deficiency of
massive stars.

\cite{Roberts+Hogg+Bregman_etal_1991} estimated the dust mass
to be $1.4\times10^6\Msol$ (for our adopted distance of 12.5 Mpc).
Here 
we estimate the dust mass to be $4.5\times10^6\Msol$ -- larger by a factor 
three.
Our estimate for $\Mdust$ results in 
$\Mdust/\MH\approx0.005$, consistent
with $\OH\equiv\log_{10}({\rm O/H})+12=\oldtext{8.38\pm0.18}\newtext{8.43\pm0.18}$ 
\citep{Moustakas+etal_2007},
$\sim$\oldtext{50}\newtext{55}\% of solar ($(\OH)_\odot=8.69$ 
from Table \ref{tab:MW depleted mass}).

	 Once again,
	 the difficulty is in fitting the long-wavelength data.
	 In this case, a single-temperature $T=24.5\K$
	 component with a $\nu^2$ opacity can fit the
	 71, 160, and 850\um\ photometry, but exceeds the
	 450\um\ datum.
	 Once again, a distribution of temperatures
	 would exacerbate this problem, indicating either
	 problems with the photometry or requiring an
	 opacity substantially steeper than $\nu^2$
	 between 160 and 450\um.
	 }
 
\subsubsection{\label{sec:Mrk33}
                Markarian 33}
 
\begin{figure}[h]
\begin{center}
\includegraphics[angle=0,width=14cm]{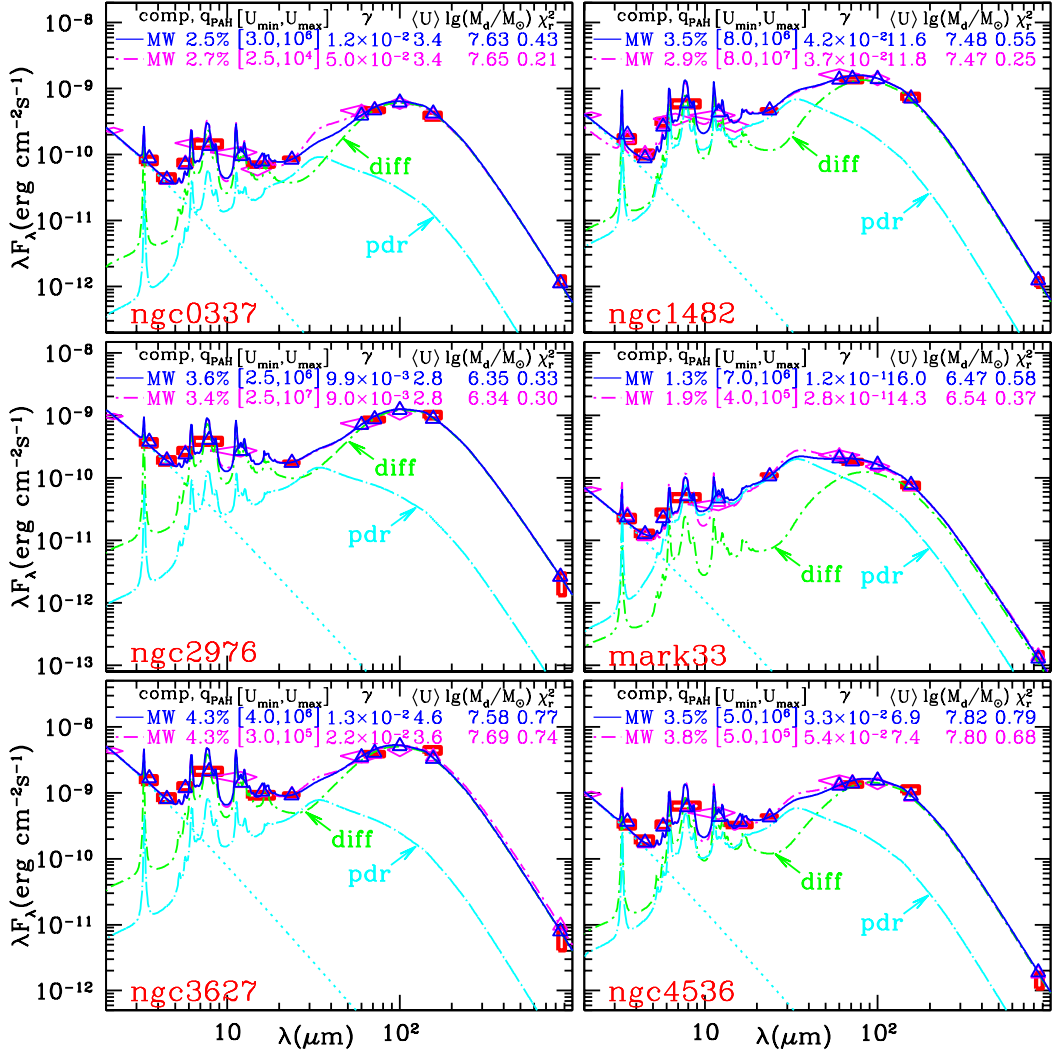}
\caption{\footnotesize
         \label{fig:scubagals}
	 \label{fig:n0337}
	 \label{fig:n1482}
	 \label{fig:n2976}
	 \label{fig:mark33}
	 \label{fig:n3627}
	 \label{fig:n4536}
         \label{fig:n4569}
         \label{fig:n4631}
         \label{fig:n4826}
	 \label{fig:n5195}
	 \label{fig:n5713}
	 \label{fig:n7331}
	 \label{fig:n7552}
	 Observed fluxes and model emission spectra
         for galaxies with IRAC, MIPS, and IRAS data,
	 and at least one submm flux measured by SCUBA.
	 Some galaxies (e.g., NGC~3627) 
	 have also been measured in
	 the IRS~16\um\ peakup filter.
	 Rectangles and diamonds show observed fluxes; vertical extent
	 is $\pm1\sigma$.
	 Curves show models fitted to IRAC, MIPS, SCUBA, and 
	 IRAS 12, 60, and 100\um\ data,; triangles are models convolved
	 with bandpass used for observations.
	 Solid curves are models with MW dust and
	 $\Umax=10^6$; broken curves are used for other models
	 that for some galaxies give smaller $\chi_r^2$ (see text).
	 Dashed-dotted curves show separate contributions of
	 stars, dust with $U=\Umin$ (labeled ``diff'')
	 and dust with $U>\Umin$ (labeled ``pdr'').
         See Fig.\ \ref{fig:n2798} for NGC~2798 and 3190, and
         Fig.\ \ref{fig:n3521} for NGC~3521 and NGC~\oldtext{4536}\newtext{5866}.}
\end{center}
\end{figure}

The dwarf starburst galaxy Mrk~33 (see Fig.\ \ref{fig:mark33})
has the highest value of
$\langle\nu F_\nu\rangle_{24}/\langle \nu F_\nu\rangle_{71}$ among the
\SINGS-SCUBA galaxies (see Fig.\ \ref{fig:fluxratios}).
H$\alpha$ observations of Mrk~33 indicate
a SFR$\approx 1.5\Msol\yr^{-1}$, giving it the third-highest
SFR/$L_{\rm opt}$ ratio in the \SINGS\ sample 
\citep{Kennicutt+Armus+Bendo_etal_2003}.
The best-fit $\Umax=10^6$ model
has a very high mean starlight intensity $\langle U\rangle\approx16$,
and $f(\Ldust;U>10^2)\approx48\%$, the highest value in the sample.

\begin{figure}[ht]
\vspace*{-1em}
\begin{center}
\includegraphics[angle=0,width=14cm]{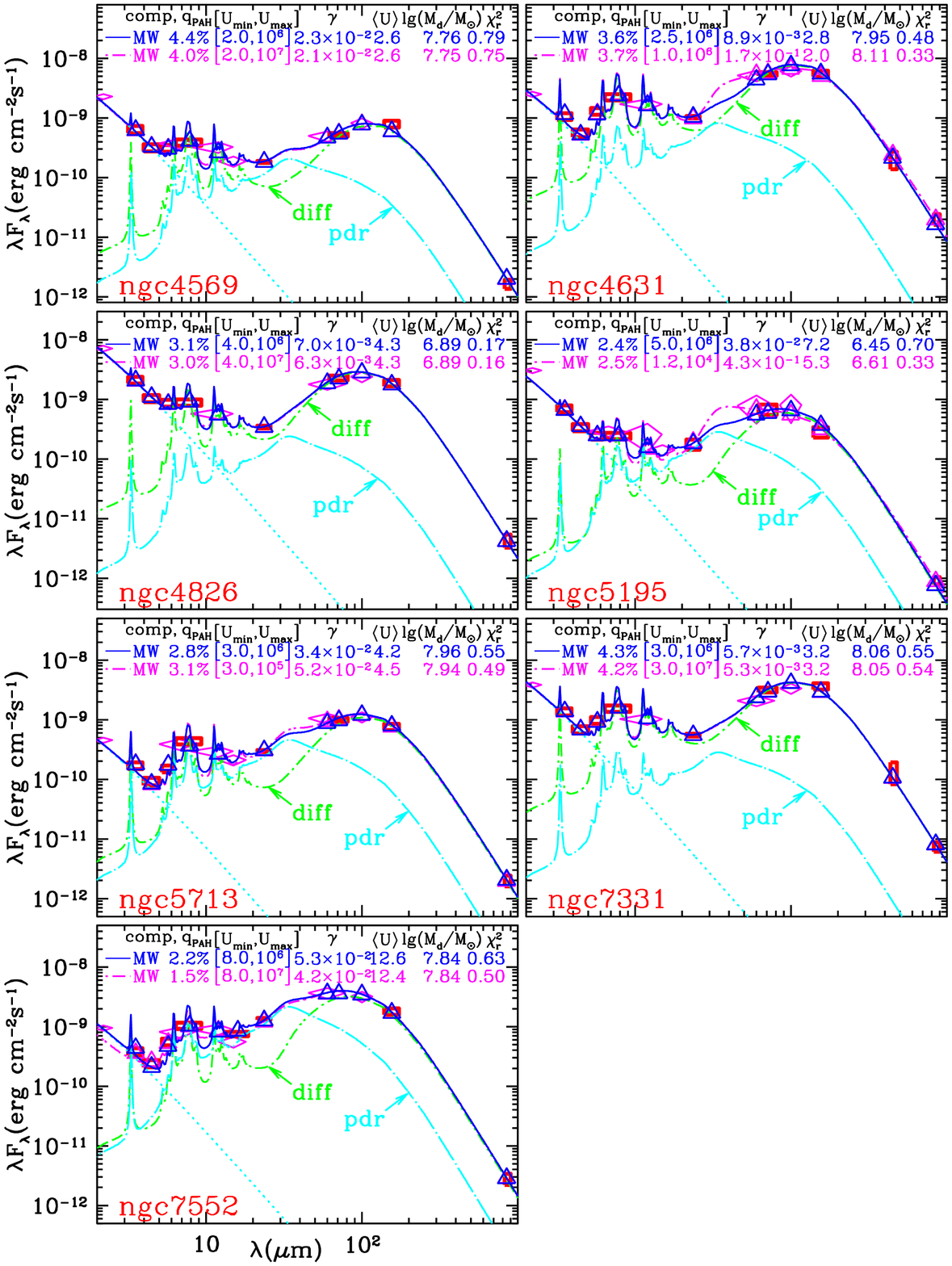}

{\footnotesize Fig. \ref{fig:scubagals}, continued.}
\end{center}
\end{figure}

%
%


\oldtext{
\subsubsection{\label{sec:NGC5866}
               NGC~5866}
               
NGC~5866, an S0 galaxy with a substantial edge-on disk, 
has the lowest value of
$\langle\nu F_\nu\rangle_{24}/\langle\nu L_\nu\rangle_{71}$ among the
\SINGS-SCUBA galaxies
(see Fig.\ \ref{fig:fluxratios}).
The best-fit model has $\gamma=0$ --
all of the dust is heated by starlight with a single intensity $U=5$.
The small value of $\gamma$ is presumably because the illumination
of the dusty disk is dominated by the distributed old stellar population,
with little contribution from star formation in the disk.
This is consistent with nondetection of H$\alpha$, with
SFR$<0.1\Msol\yr^{-1}$ \citep{Kennicutt+Armus+Bendo_etal_2003},
although dust extinction may also be important.
The nuclear spectrum of NGC5866 \citep{Smith+Draine+Dale+etal_2007}
includes ionic emission lines (e.g., [\ion{Ne}{2}], [\ion{Si}{2}], [\ion{S}{3}]) but they
are relatively weak, consistent with only low levels of star formation;
the low ratio [\ion{S}{3}]/[\ion{Si}{2}] flux ratio also suggests a deficiency of
massive stars.

\cite{Roberts+Hogg+Bregman_etal_1991} estimated the dust mass
to be $1.4\times10^6\Msol$ (for our adopted distance of 12.5 Mpc).
Here 
we estimate the dust mass to be $4.5\times10^6\Msol$ -- larger by a factor 
three.
Our estimate for $\Mdust$ results in 
$\Mdust/\MH\approx0.005$, consistent
with $\OH\equiv\log_{10}({\rm O/H})+12=\oldtext{8.38\pm0.18}\newtext{8.43\pm0.18}$ 
\citep{Moustakas+etal_2007},
$\sim$\oldtext{50}\newtext{55}\% of solar ($(\OH)_\odot=8.69$ 
from Table \ref{tab:MW depleted mass}).
See \S\ref{sec:dust to gas ratio} for details.
}

\section{\label{sec:descuba}
         Estimating Dust Masses Without Submm Data}

Thus far we have been fitting dust models constrained by IRAC, MIPS, IRAS,
and at least one submm flux measured by SCUBA.
Unfortunately, submillimeter observations are unavailable at this
time for most of the \SINGS\ galaxies.  How well can dust masses be
estimated in the absence of submm constraints?

\begin{figure}[h]
\begin{center}
  \includegraphics[angle=0,width=\figwidthd]%
                  {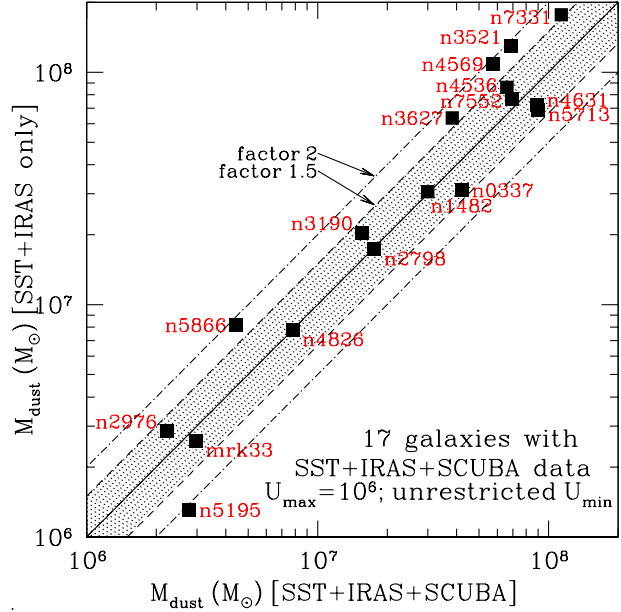}
  \caption{\label{fig:scuba_MW_v_descuba_MW}
           \label{fig:scuba_v_descuba}
	   \footnotesize
           Dust mass $\Mdust$ 
           determined for \Nscuba\ galaxies
	   using IRAS and SST data only,
	   vs. the masses derived using IRAS, SST, and SCUBA 
	   data combined.
	   Data are fit by MW dust models with $\Umax=10^6$ but
	   no restriction on $\Umin$.
	   Without SCUBA data, cool dust is not strongly constrained.
	   Nevertheless, 11/\Nscuba\ of the galaxies fall within a factor 
	   1.5 of the value obtained when SCUBA data are
	   employed, and all 17 galaxies are within a factor of 2.2 .
	   }
 \end{center}
\end{figure}

To address this question, we first repeat the dust modeling procedure
used in \S\ref{sec:SCUBA sample} for the
\Nscuba\ \SINGS-SCUBA galaxies, 
{\it but with the SCUBA data omitted.}
The 
dust masses so obtained are
shown in Figure \ref{fig:scuba_MW_v_descuba_MW},
where they are plotted versus the dust masses estimated when 
SCUBA data are also used to constrain the models.

If the dust masses obtained with the SCUBA fluxes included are regarded
as our ``gold standard'', it is clear that dropping the SCUBA data from
the fitting procedure leads to substantial uncertainties in the model 
fitting.
Five of the \Nscuba\ galaxies have estimated dust masses that exceed the
``gold standard'' estimate by more than a factor of 1.5 
(NGC~3521, NGC~3627, NGC~4536, NGC~4569, and NGC~7331).
In five cases 
the dust mass estimate is lowered when the
SCUBA data are not used, in one case by a factor of more than 1.5
(NGC~5195).

\begin{figure}[h]
\begin{center}
   \includegraphics[angle=0,width=\figwidthd]%
                   {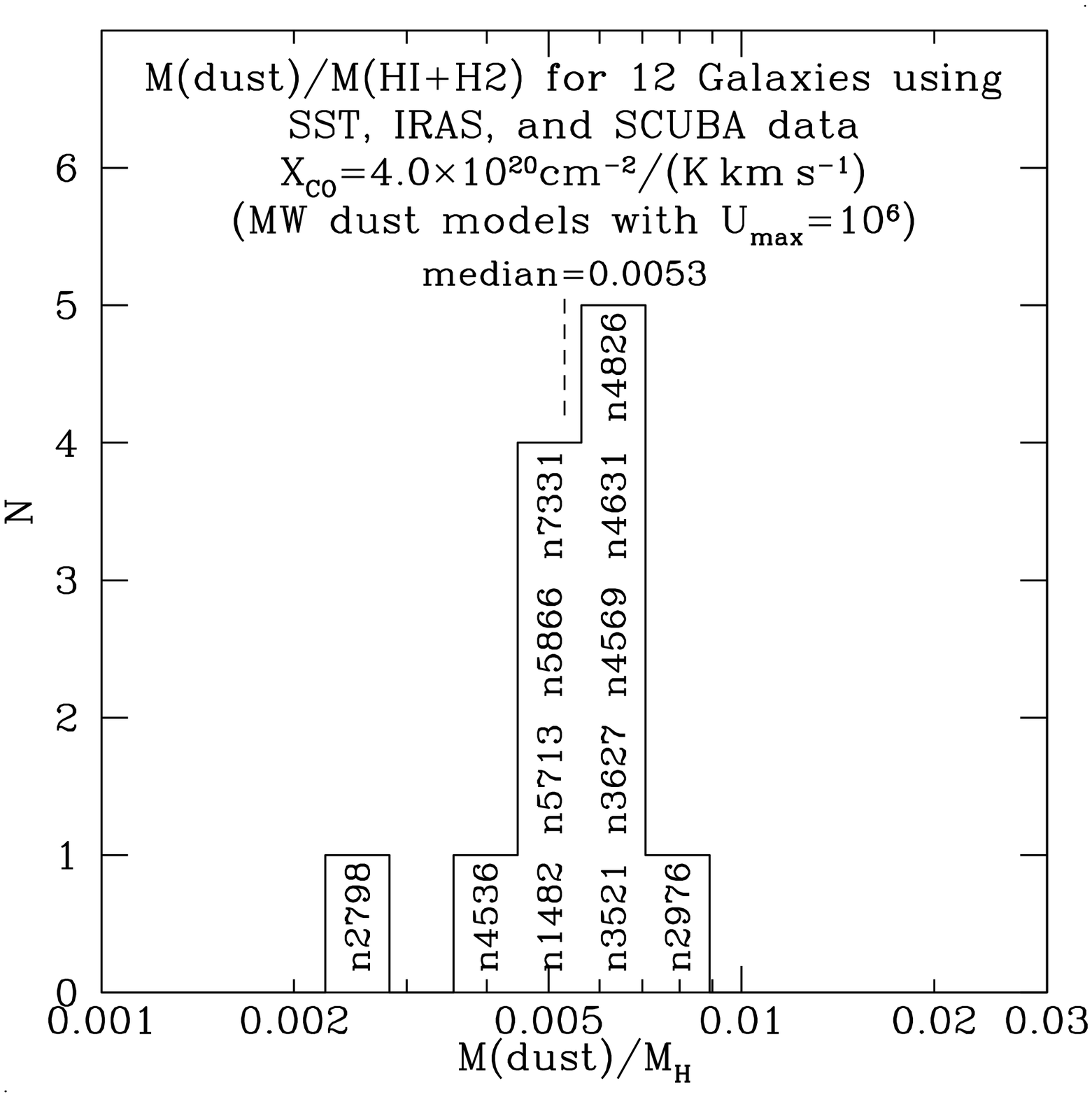}
   \includegraphics[angle=0,width=\figwidthd]%
                   {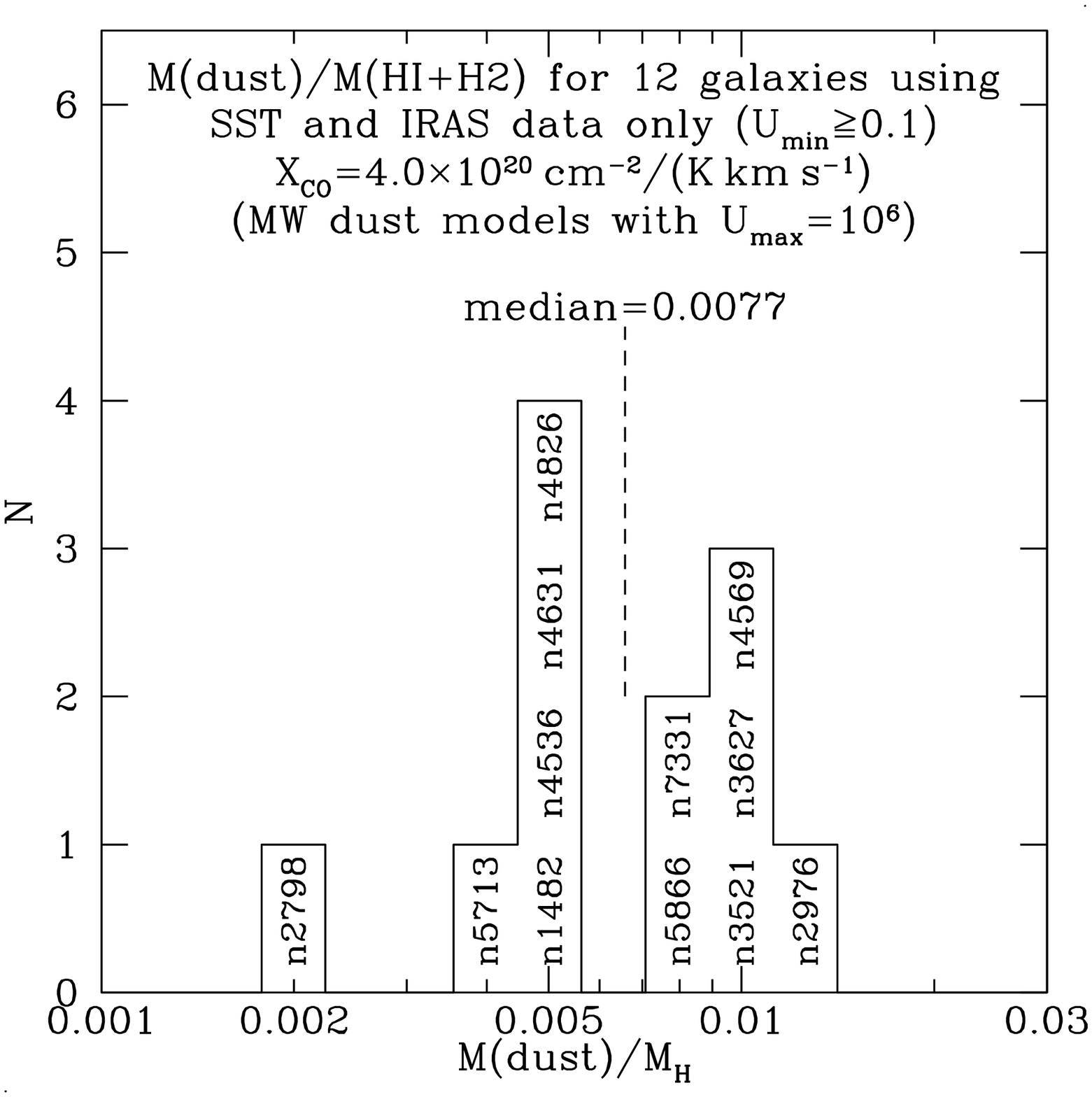}
   \caption{\label{fig:Mgas/Mdust histograms}\footnotesize
            (a) Histogram of $\Mdust/\MH$ obtained using
                IRAC+MIPS+SCUBA data for 12 galaxies for which \ion{H}{1}~21cm and
                CO 1-0 fluxes have been measured.
            (b) Same as (a), but estimated without using SCUBA data, with
		$\Umax=10^6$ but
                no restrictions on $U_{\rm min}$.
	    }
 \end{center}
\end{figure}

Figure \ref{fig:Mgas/Mdust histograms} shows estimated dust-to-gas mass
ratios for the \Nscuba\ galaxies with SCUBA data for which the gas mass is
known.  
Figure \ref{fig:Mgas/Mdust histograms}a shows the dust/gas
ratios estimated when the SCUBA fluxes are used in the fitting: the
median dust/hydrogen mass ratio is 0.0053 for this sample, comparable
to the value $\sim$0.007 expected for solar-metallicity material with
$\sim$2/3 of the C, and most of the Mg, Si, and Fe incorporated into
carbonaceous and silicate material (see Table \ref{tab:MW depleted mass}).
Figure \ref{fig:Mgas/Mdust
histograms}b shows the effect of fitting the SED without using the
SCUBA data -- for most of the galaxies, the estimated dust/hydrogen
ratio rises modestly, with a median value of only 0.0077.

Because these models reproduce the SED fairly well, and the IR power is
mainly shortward of $200\mum$, the total IR power
is relatively unaffected when the SCUBA data are removed.
What is changing
is the estimated mean dust-weighted 
starlight intensity, with
$\langle U\rangle$ often falling
a factor $\sim 1.5$ or more below the
value obtained when the SCUBA data are employed.
In the absence of submm measurements to constrain the mass of
cool dust,
there is a risk that the model-fitting procedure may
invoke a large mass of cool dust, heated by weak starlight.
To prevent this, when submm data are unavailable 
we will restrict the model-fitting to use
radiation fields with $\Umin\geq0.7$.
This of course runs the risk of underestimating the dust mass if there
are galaxies without submm (e.g., SCUBA) observations which
actually contain large amounts of cool dust heated by
starlight with $U\ltsim 0.5$.  However, we note that
all the galaxies with SCUBA data had $\Umin\geq 1$ for the
best-fit model (see Fig.\ \ref{fig:scuba umin umax}a).

Based on the above discussion, we adopt the following ``restricted''
fitting procedure when submm data are unavailable:
\begin{enumerate}
\item Models limited to MW dust models only, with $0.4\leq\qpah\leq4.6\%$;
\item $\Umax=10^6$;
\item $\alpha=2$;
\item $0.7\leq\Umin\leq25$.
\end{enumerate}
This restricted fitting procedure (with $\Umin\geq0.7$)
of course implies that the resulting fits
will necessarily have $\langle U\rangle \geq 0.7$.
This is generally consistent with what we observe for the sample of 
\Nscuba\
galaxies with SCUBA data -- as seen in Figure \ref{fig:ubar_scuba_all},
these galaxies tend to have $\langle U\rangle \gtsim 2$.
However, the \SINGS-SCUBA galaxy sample may be biased
in favor of increased dust mass and associated star formation.
If other galaxies
have very low mean starlight intensities
$\langle U\rangle$, 
the restricted fitting procedure adopted here would {\it underestimate}
the dust mass (because it would overestimate $\langle U\rangle$)
To study the dust in such galaxies, submm observations
are essential.

\section{\label{sec:all 70 galaxies}
         Dust Properties for \Ntot\ \SINGS\ Galaxies}

\subsection{\label{sec:dust to gas ratio}
            Dust to Gas Mass Ratio}

For \Nnoscuba\ galaxies in the SINGS sample
we have complete IRAC and MIPS data, but no SCUBA fluxes.
Dust masses have been estimated for these galaxies using the restricted
fitting procedure described above.
The SEDs for these galaxies, together with the best-fit models,
are given in Appendix \ref{sec:nonscuba SEDs} (Fig.\ \ref{fig:noscuba}).
The results for the dust mass $\Mdust$
are given in Table \ref{tab:nonscuba galaxies}.

For 20 of these galaxies both \ion{H}{1} 21cm and CO 2.6mm fluxes have been
detected (or there is a strong upper limit on the CO flux), 
and we are able to estimate the total gas mass.  For another 24
galaxies \ion{H}{1} 21cm has been measured, but the CO 2.6mm flux is unknown, 
so we have only a lower limit on the gas mass.
For 4 galaxies (the E galaxies NGC~0855 and NGC~4125, 
and the S0 galaxies NGC~1266 and NGC~1316) neither 
\ion{H}{1} nor CO have been detected.

\begin{figure}[h]
\begin{center}
\includegraphics[angle=0,width=\figwidth]%
                {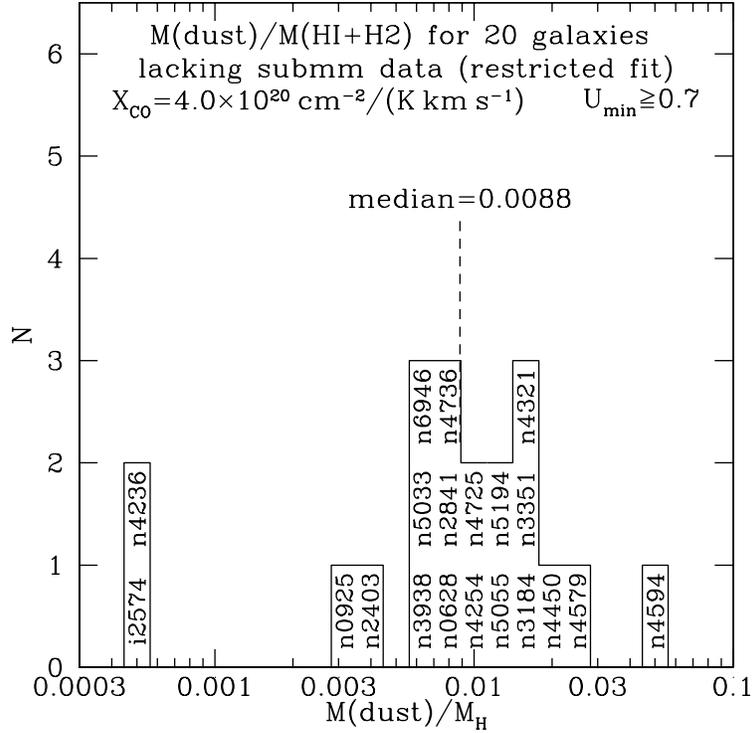}
\caption{\label{fig:noscuba dust to gas histogram}\footnotesize
         Dust to H mass ratio determined from IRAC + MIPS data only,
	 using restricted fits with $\Umin \geq 0.7$,
	 and taking $\XCO=4\times10^{20}\cm^{-2}(\K\kms)^{-1}$.
	 }
\end{center}
\end{figure}

Figure \ref{fig:noscuba dust to gas histogram} shows the distribution of
$\Mdust/\MH$ for the 20 galaxies lacking SCUBA data
for which $\MH$ is known.
The distribution looks similar to the distribution in 
Figure \ref{fig:Mgas/Mdust histograms}b, except for two galaxies -- 
IC~2574 and NGC~4236 --
with $\Mdust/\MH \approx 0.0005$.
Note, however, that many of the sample galaxies do not appear on the
histogram because $\MH$ is not known.

The dust-to-gas mass ratio reflects the enrichment of the gas by
heavy elements (C, O, Mg, Si, Fe) produced in stars,
the production and growth of solid grains from these elements
in supernovae, stellar winds, and in the interstellar medium,
and destruction of grains by supernova blastwaves and other
violent events in the interstellar medium.


If the same fraction of the condensible elements (C, Mg, Si, Fe, ...) is
in grains as in the local Milky Way, then $\Mdust/\MH$ should
conform to eq.(\ref{eq:Md/MH envelope}).
However, it is possible that in some other galaxies, the balance between grain
formation and destruction may be such that most interstellar C, Mg, Si and Fe
is in the gas rather than in grains.
We therefore expect eq.\ (\ref{eq:Md/MH envelope}) to define an
upper envelope to the dust-to-gas mass ratio.

\newcommand     \tnt[2]{\multicolumn{14}{l}{$^{\rm #1}$ #2}\cr}

\clearpage
\begin{deluxetable}{l c c c c c c c c c c c c c}
\tabletypesize{\footnotesize}
\rotate
\tablewidth{0pt}   
\tablecolumns{14}
\tablecaption{\label{tab:nonscuba galaxies}
         \Nnoscuba\ \SINGS\ Galaxies with IRAC and MIPS global fluxes, 
         but no SCUBA data\tablenotemark{a}}
\tablehead{
  \colhead{galaxy} &
  \colhead{morph} &
  \colhead{$D$\tablenotemark{b}} & 
  \colhead{$\log[M({\rm HI})]$\tablenotemark{b}} & 
  \colhead{$\log[M({\rm H}_2)]$\tablenotemark{b,c}} &
  \colhead{$\log(\Mdust)$\tablenotemark{o}} & 
  \colhead{$\log(\Ldust)$\tablenotemark{o}} & 
  \colhead{$\qpah$\tablenotemark{p}} & 
  \colhead{$\langle U\rangle$\tablenotemark{q}} &
  \colhead{dust} &
  \colhead{$\Umin$\tablenotemark{r}} & 
  \colhead{$\gamma$\tablenotemark{s}} &
   \colhead{$f(U>10^2)$\tablenotemark{t}} &
  \colhead{$\chi_r^2$\tablenotemark{u}}\\
      &
  \colhead{type} &
  \colhead{Mpc} &    
  \colhead{$M_\odot$} &
  \colhead{$M_\odot$} & 
  \colhead{$M_\odot$} &
  \colhead{$L_\odot$} & 
  \colhead{\%} &
      & 
   \colhead{type} &
      &
  \colhead{\%} &
  \colhead{\%}
  }
\startdata
%
%
   ngc0024 &   Sc &  8.20 &     9.07 &      --- &  6.54 &  8.78 &  3.1 &  1.26 & MW &  1.2 &     0.40 &  3.5 &  0.43 \cr
   ngc0628 &   Sc & 11.40 &     9.97\tablenotemark{d}
                                     &     9.64\tablenotemark{e}
                                                &  8.02 & 10.29 &  4.3 &  1.36 & MW &  1.2 &     1.04 &  8.4 &  0.37 \cr
   ngc0855 &    E &  9.60 &      --- &      --- &  5.69 &  8.58 &  1.6 &  5.70 & MW &  5.0 &     1.25 & 10.1 &  1.29 \cr
   ngc0925 &   Sd & 10.10 &     9.75\tablenotemark{d}
                                     &     9.18 &  7.35 &  9.71 &  2.9 &  1.67 & MW &  1.5 &     0.94 &  7.7 &  0.76 \cr
   ngc1097 &   Sb & 16.90 &    10.03 &      --- &  8.37 & 10.81 &  3.1 &  2.01 & MW &  1.5 &     2.72 & 18.7 &  0.24 \cr
   ngc1266 &   S0 & 31.30 &      --- &      --- &  7.05 & 10.44 &  0.4 & 18.17 & MW & 15.0 &     2.09 & 15.9 &  0.76 \cr
   ngc1291 & S0/a &  9.70 &     9.19 &      --- &  7.34 &  9.38 &  2.5\tablenotemark{k}
                                                                       &  0.81 & MW &  0.7 &     1.17 &  9.4 &  1.76 \cr
   ngc1316 &   S0 & 26.30 &  $<$8.87 &      --- &  7.63 & 10.03 &  0.4\tablenotemark{k}
                                                                       &  1.82 & MW &  1.5 &     1.71 & 13.0 &  0.77 \cr
   ngc1512 &   Sa & 10.40 &     9.77 &      --- &  7.21 &  9.45 &  3.8 &  1.28 & MW &  1.2 &     0.55 &  4.7 &  0.57 \cr
   ngc1566 &  Sbc & 18.00 &    10.01 &      --- &  8.25 & 10.62 &  4.0 &  1.72 & MW &  1.5 &     1.20 &  9.6 &  0.54 \cr
   ngc1705 & \ot{S0}\nt{Im} &  5.10\tablenotemark{f}
                          &     8.02\tablenotemark{d,f}
                                     &      --- &  4.86 &  7.89 &  0.6 &  7.72 & MW &  7.0 &     0.94 &  7.9 &  0.91 \cr
   ngc2403 &  Scd &  3.50 &     9.48\tablenotemark{d}
                                     &     8.01 &  7.08 &  9.56 &  3.5 &  2.22 & MW &  2.0 &     0.90 &  7.5 &  0.62 \cr
   holmbII &   Im &  3.39\tablenotemark{l}
                          &     8.77\tablenotemark{d} 
                                     &      --- &  5.07 &  7.90 &  0.4 &  4.92 & MW &  4.0 &     2.01 & 15.1 &  4.13 \cr
    ddo053 &   Im &  3.56 &     7.78\tablenotemark{d} 
                                     &      --- &  4.00 &  7.08 &  1.1 &  8.95 & MW &  7.0 &     2.56 & 18.4 &  0.48 \cr
   ngc2841 &   Sb &  9.80 &     9.62\tablenotemark{d}
                                     &     9.45 &  7.74 &  9.73 &  4.3 &  0.72 & MW &  0.7 &     0.21 &  1.8 &  0.66 \cr
   ngc2915 &   Im &  2.70 &     8.25 &      --- &  4.14 &  7.33 &  1.4 & 11.22 & MW & 10.0 &     1.16 &  9.6 &  0.37 \cr
    holmbI &   Im &  3.84\tablenotemark{l}
                          &     8.15\tablenotemark{d} 
                                     &      --- &  4.83 &  7.16 &  1.2 &  1.58 & MW &  1.5 &     0.40 &  3.5 &  0.76 \cr
   ngc3049 &  Sab & 19.60 &     9.10 &      --- &  6.74 &  9.58 &  2.9 &  5.13 & MW &  3.0 &     6.06 & 32.6 &  0.38 \cr
   ngc3031 &  Sab &  3.50 &     9.53\tablenotemark{d} 
                                     &      --- &  7.53 &  9.66 &  4.0 &  1.00 & MW &  1.0 &        0 &  0.0 &  0.33 \cr
   ngc3184 &  Scd &  8.60 &     9.26\tablenotemark{d}
                                     &     9.11 &  7.70 &  9.76 &  4.5 &  0.86 & MW &  0.8 &     0.53 &  4.6 &  0.47 \cr
   ngc3198 &   Sc &  9.80 &     9.71\tablenotemark{d}
                                     &      --- &  7.42 &  9.63 &  3.7 &  1.20 & MW &  1.0 &     1.58 & 12.1 &  0.48 \cr
    ic2574 &   Sm &  4.02\tablenotemark{l}
                          &     9.17\tablenotemark{d} 
                                     &     6.85\tablenotemark{j}
                                                &  5.86 &  8.36 &  0.4 &  2.30 & MW &  2.0 &     1.22 &  9.8 &  0.60 \cr
   ngc3265 &    E & 20.00 &     8.24 &      --- &  6.17 &  9.47 &  2.8 & 14.81 & MW & 10.0 &     4.58 & 28.5 &  0.14 \cr
   ngc3351 &   Sb &  9.30 &     9.01\tablenotemark{d}
                                     &     8.98 & \oldtext{7.45}\newtext{7.46} 
                                                & \oldtext{9.86}\newtext{9.89} 
                                                & \oldtext{2.4}\newtext{3.2}
                                                & \oldtext{1.91}\newtext{1.95} & MW &  1.5 
                                                & \oldtext{2.20}\newtext{2.42} 
                                                & \oldtext{15.9}\newtext{17.1} 
                                                & \oldtext{2.64}\newtext{0.56} \cr
   ngc3621 &   Sd &  6.20 &     9.79\tablenotemark{d}
                                     &      --- &  7.38 &  9.85 &  4.5 &  2.19 & MW &  2.0 &     0.78 &  6.5 &  0.68 \cr
   ngc3773 &   S0 & 12.90 &     7.99 &      --- &  5.90 &  8.86 &  2.3 &  6.82 & MW &  5.0 &     3.25 & 22.0 &  0.40 \cr
   ngc3938 &   Sc & 12.20 &     9.57 &     9.61 &  7.69 &  9.94 &  4.6 &  1.30 & MW &  1.2 &     0.67 &  5.7 
                                                & \oldtext{0.45}\newtext{0.44} \cr
   ngc4125 &    E & 21.40 &      --- &      --- &  6.38 &  9.07 &  0.4\tablenotemark{k}
                                                                       &  3.59 & MW &  3.0 &     1.69 & 13.0 &  0.38 \cr
\tablebreak
   ngc4236 &  Sdm &  4.45\tablenotemark{l}
                          &     9.44 &  $<$8.65 &  6.15 &  8.67 &  1.0 &  2.41 & MW &  2.0 &     1.68 & 12.9 &  1.71 \cr
   ngc4254 &   Sc & 20.00 &     9.86 &    10.42 &  8.55 & 10.91 &  4.5 &  1.68 & MW &  1.5 &     0.96 &  7.9 &  0.72 \cr
   ngc4321 &  Sbc & 20.00 &     9.67 &    10.32 & \oldtext{8.53}\newtext{8.57} 
                                                & \oldtext{10.80}\newtext{10.82}
                                                & \oldtext{3.5}\newtext{4.2} 
                                                & \oldtext{1.34}\newtext{1.33} & MW &  1.2 
                                                & \oldtext{0.95}\newtext{0.85} 
                                                & \oldtext{7.8}\newtext{7.1} 
                                                & \oldtext{1.59}\newtext{0.39} \cr
   ngc4450 &  Sab & 20.00 &     8.61 &     9.45 &  7.78 &  9.78 &  2.4 &  0.73 & MW &  0.7 &     0.36 &  3.1 &  0.30 \cr
   ngc4559 &  Scd & 11.60 &    10.05 &      --- &  7.57 &  9.92 &  3.4 &  1.64 & MW &  1.5 &     0.73 &  6.1 &  0.69 \cr
   ngc4579 &   Sb & 20.00 &     8.91 &     9.75 &  8.18 & 10.25 &  3.5 &  0.86 & MW &  0.8 &     0.58 &  5.0 &  1.37 \cr
   ngc4594 &   Sa &  9.25\tablenotemark{g}
                          &     8.43 &     8.65\tablenotemark{h}
                                                &  7.56 &  9.55 &  3.7\tablenotemark{k}
                                                                       &  0.72 & MW &  0.7 &     0.24 &  2.1 &  1.00 \cr
   ngc4625 &   Sm &  9.50 &     9.02 &      --- &  6.35 &  8.81 &  4.4 &  2.13 & MW &  2.0 &     0.53 &  4.6 &  0.22 \cr
   ngc4725 &  Sab & 17.10 &     9.87 &     9.95 &  8.20 & 10.18 &  4.5 &  0.71 & MW &  0.7 &     0.16 &  1.4 &  0.57 \cr
   ngc4736 &  Sab &  5.30 &     8.71\tablenotemark{d} 
                                     &     9.05 &  7.11 &  9.89 &  4.1 &  4.37 & MW &  4.0 &     0.81 &  6.8 &  0.33 \cr
   ngc5033 &   Sc & 13.30 &     9.97 &     9.66 &  7.95 & 10.28 &  4.1 &  1.61 & MW &  1.5 &     0.58 &  5.0 &  0.35 \cr
   ngc5055 &  Sbc &  8.20 &     9.78\tablenotemark{d}
                                     &     9.77 &  8.19 & 10.35 &  4.1 &  1.06 & MW &  1.0 &     0.48 &  4.1 &  0.31 \cr
   ngc5194 &  Sbc &  8.20 &     9.43\tablenotemark{d} 
                                     &    10.03 &  8.26 & 10.61 &  4.5 &  1.66 & MW &  1.5 &     0.84 &  7.0 &  0.40 \cr
    tolo89 &  Sdm & 15.00 &     9.11 &      --- &  6.46 &  9.15 &  0.4 &  3.55 & MW &  2.0 &     6.41 & 33.2 &  2.50 \cr
   ngc5408 &   Im &  4.81\tablenotemark{m}
                          &     8.51 &      --- &  4.67 &  8.28 &  0.4 & 29.45 & MW & 20.0 &     4.81 & 30.1 &  2.12 \cr
   ngc5474 &  Scd &  6.90 &     9.10 &      --- &  6.39 &  8.74 &  1.9 &  1.63 & MW &  1.5 &     0.71 &  6.0 &  0.46 \cr
    ic4710 &   Sm &  8.50 &     8.51 &      --- &  5.71 &  8.60 &  2.0 &  5.73 & MW &  5.0 &     1.30 & 10.4 &  0.49 \cr
   ngc6822 &   Im &  0.49\tablenotemark{n}
                          &     8.13 &      --- &  5.04 &  7.62 &  0.7 &  2.77 & MW &  2.5 &     0.89 &  7.4 &  1.13 \cr
   ngc6946 &  Scd &  5.50 &     9.37\tablenotemark{i}
                                     &     9.77 & \oldtext{7.75}\newtext{7.74} 
                                                & \oldtext{10.36}\newtext{10.35} 
                                                & \oldtext{4.5}\newtext{4.3} 
                                                & \oldtext{3.04}\newtext{3.03} & MW &  2.5 
                                                & \oldtext{1.80}\newtext{1.77} 
                                                & \oldtext{13.7}\newtext{13.5} 
                                                & \oldtext{0.28}\newtext{0.51} \cr
   ngc7793 &   Sd &  3.20 &     8.77\tablenotemark{d}
                                     &      --- &  6.92 &  9.14 &  3.6 &  1.25 & MW &  1.2 &     0.31 &  2.7 &  0.85 \cr
\hline

\tnt{a}{All models assume MW dust, with $\Umax=10^6$.}

\tnt{b}{$D$, 21~cm flux, and CO~1-0 flux from \citet{Kennicutt+Armus+Bendo_etal_2003} 
                    unless otherwise noted.}

\tnt{c}{$M({\rm H}_2)$ calculated using $\XCO=4\times10^{20}\cm^{-2}(\K\kms)^{-1}$}

\tnt{d}{21~cm flux from \citet{Walter_2005}}

\tnt{e}{CO~1-0 flux from \citet{Sheth+Vogel+Regan_etal_2005}}

\tnt{f}{\citet{Tosi+Sabbi+Bellazzini+etal_2001}}

%
\tnt{g}{$D$ is weighted mean of \citet{Ford+Hui+Ciardullo_etal_1996} and
                  \citet{Ajhar+Lauer+Tonry_etal_1997}}

\tnt{h}{CO~1-0 flux from \citet{Bajaja+Krause+Wiebelinski+Dettmar_1991}}

\tablebreak

\tnt{i}{21~cm flux for radius $<6\arcmin$ from
                  \citet{Boulanger+Viallefond_1992}}

\tnt{j}{CO~1-0 flux from \citet{Leroy+Bolatto+Simon+Blitz_2005}}

\tnt{k}{$\qpah$ very uncertain -- see \S\ref{sec:PAH abundances - scuba plus nonscuba}}

\tnt{l}{\citet{Karachentsev+Dolphin+Geisler_etal_2002}}

\tnt{m}{\citet{Karachentsev+Sharina+Dolphin_etal_2002}}

\tnt{n}{\citet{Cannon+Walter+Bendo_etal_2005}}

\tnt{o}{Estimated from dust model.}

\tnt{p}{Fraction of dust mass contributed by PAHs with $N_C<10^3$ C atoms.}


\tnt{r}{Lower cutoff for starlight intensity scale factor $U$.}

\tnt{s}{Dust-weighted mean starlight intensity scale factor.}

\tnt{t}{Fraction of dust luminosity from regions with $U>10^2$}

\tnt{u}{$\chi^2/(N_b-5)$, where $N_b=$ number of bands used for fitting.}
\enddata
\end{deluxetable}


Models have been proposed to attempt to predict the
dust-to-gas ratio in galaxies as they evolve in metallicity
\citep{Dwek_1998,
       Lisenfeld+Ferrara_1998,
       Edmunds_2001,
       Hirashita+Tajiri+Kamaya_2002}.
\citet{James+Dunne+Eales+Edmunds_2002} compared models to dust
masses estimated from IR and submm observations, and argued
that the fraction of the metals condensed into dust grains
appeared to be the same in dwarf and giant galaxies -- i.e.,
dust-to-gas mass ratios consistent with
eq.\ (\ref{eq:Md/MH envelope}) -- for
$8.1\ltsim\OH\ltsim9.0$.

\begin{figure}[h]
\begin{center}
\includegraphics[angle=0,width=0.9\figwidthd]%
                 {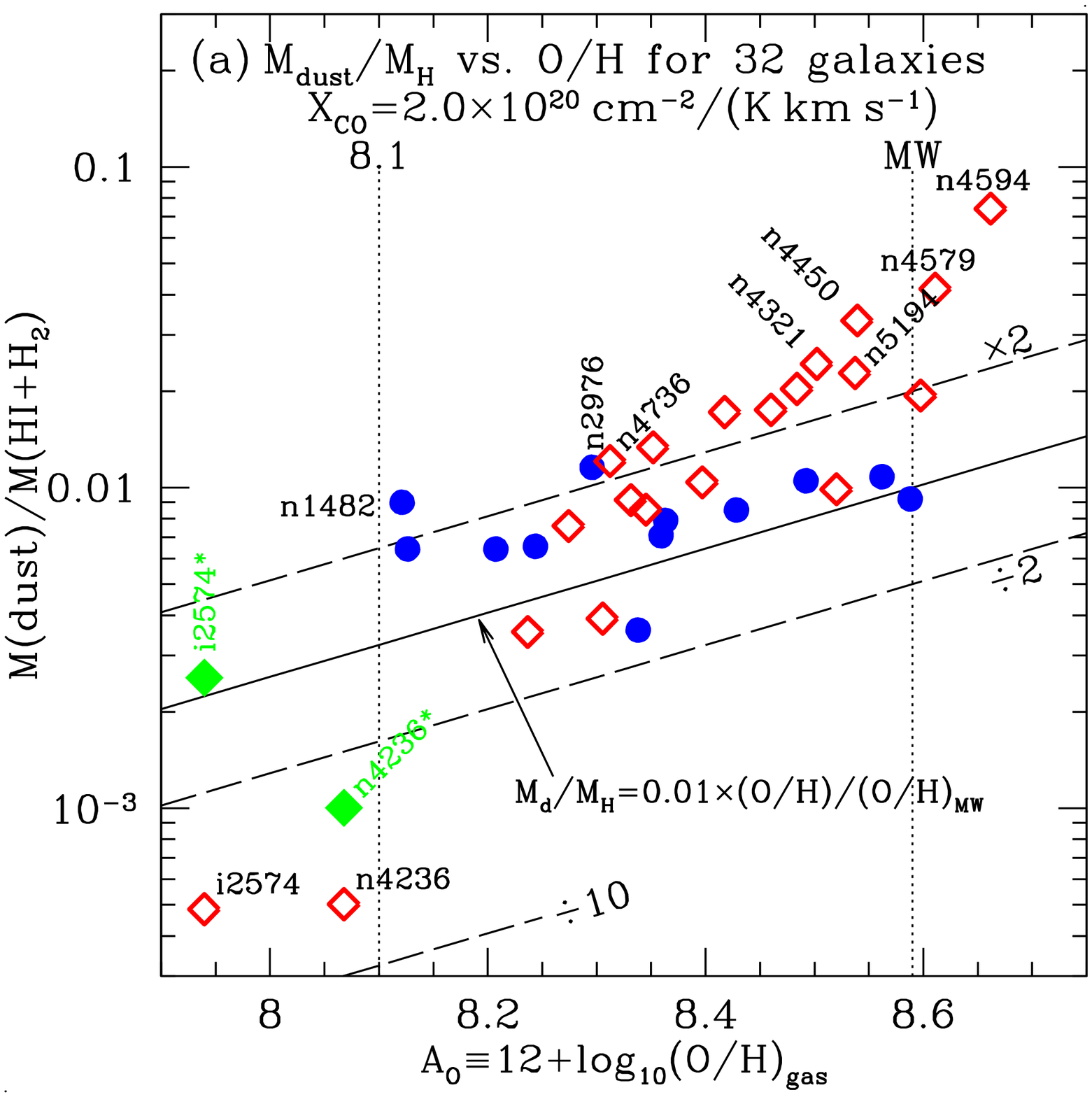}
\includegraphics[angle=0,width=0.9\figwidthd]
                {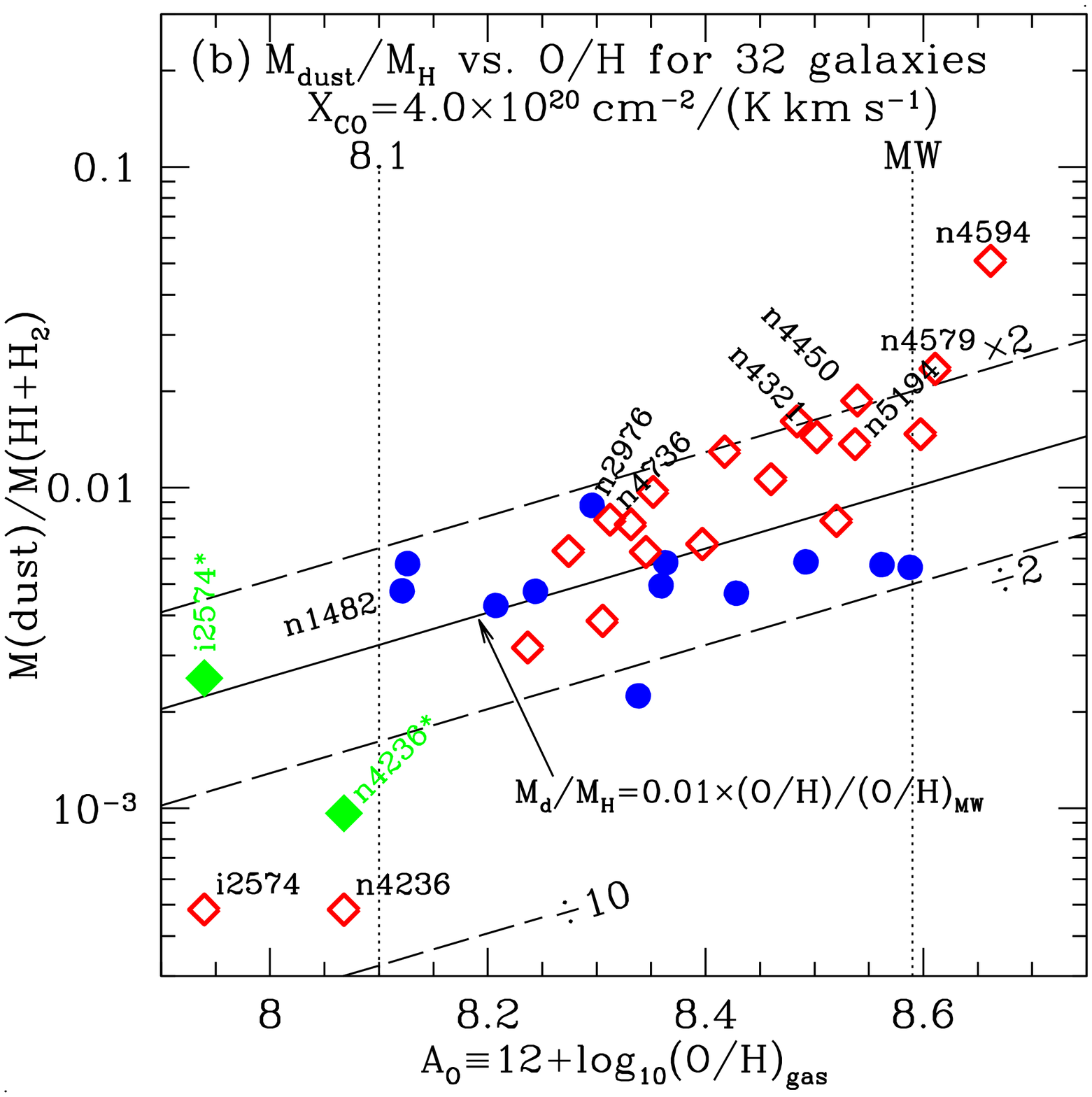}
\includegraphics[angle=0,width=0.9\figwidthd]
                {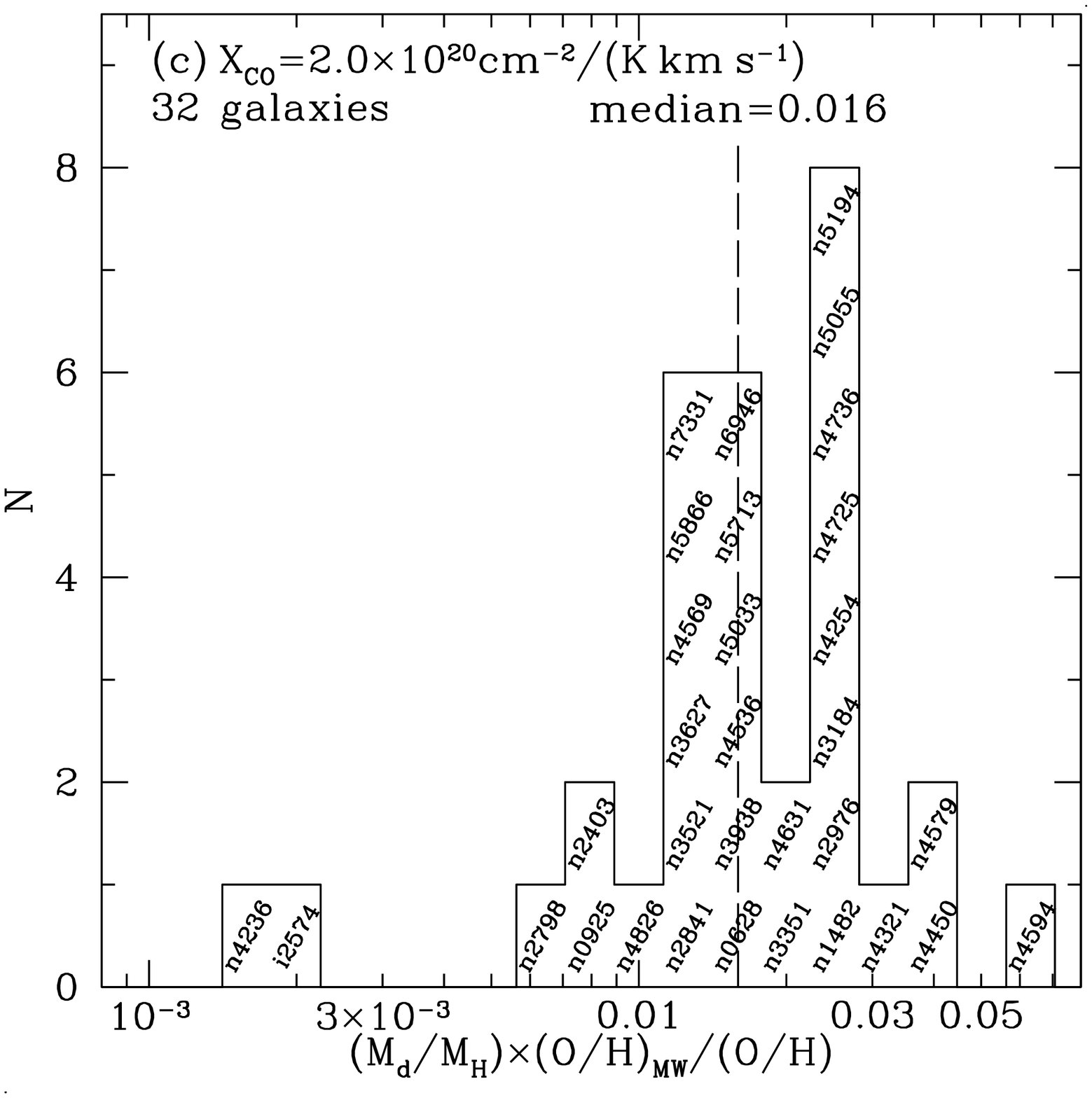}
\includegraphics[angle=0,width=0.9\figwidthd]
                {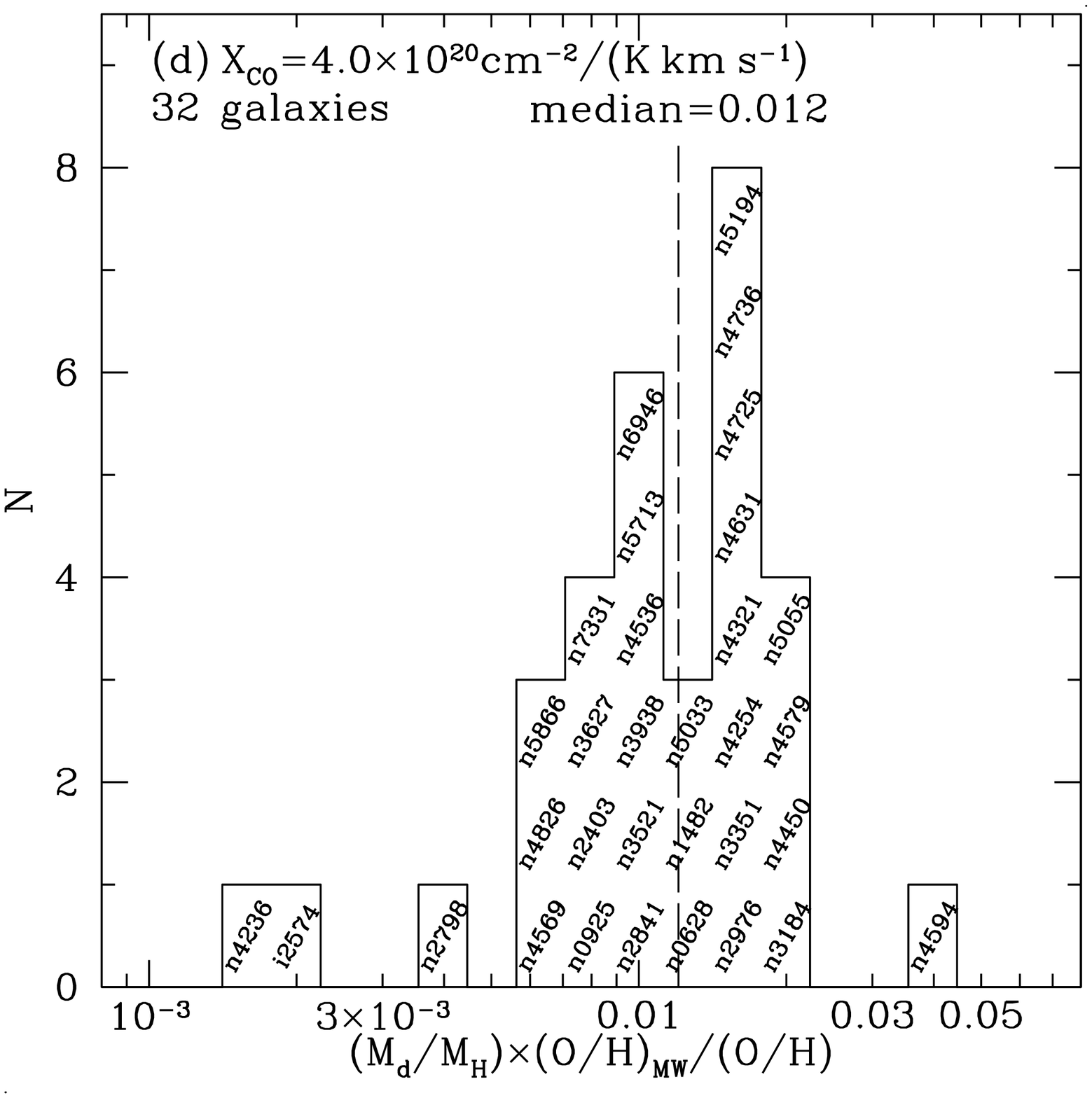}
\caption{\label{fig:Mdust/Mgas vs metallicity}\footnotesize
         Ratio of $\Mdust$ to $\MH$
	 vs. oxygen abundance (see text)
	 for galaxies with and without SCUBA fluxes (circles and diamonds,
	 respectively), showing 32 galaxies where gas mass is known,
	 for two values of $\XCO/[\cm^{-2}(\K\kms)^{-1}]$: 
	 (a) $2\times10^{20}$,
	 and (b) $4\times10^{20}$.
	 Filled diamonds: $\Mdust/\MH$ for regions where
	 IR emission is detected (see text).
	 Solid line shows eq.\ (\ref{eq:Md/MH envelope}); broken
	 lines show factor of 2 variations around the solid line.
	 Histogram of normalized dust-to-gas mass ratio
	 for $\XCO/[\cm^{-2}(\K\kms)^{-1}]=$
	 $2\times10^{20}$ (c), and
	 $4\times10^{20}$ (d).
	 }
\end{center}
\end{figure}
\begin{figure}[h]
\begin{center}
\includegraphics[angle=270,width=\figwidthw]%
                {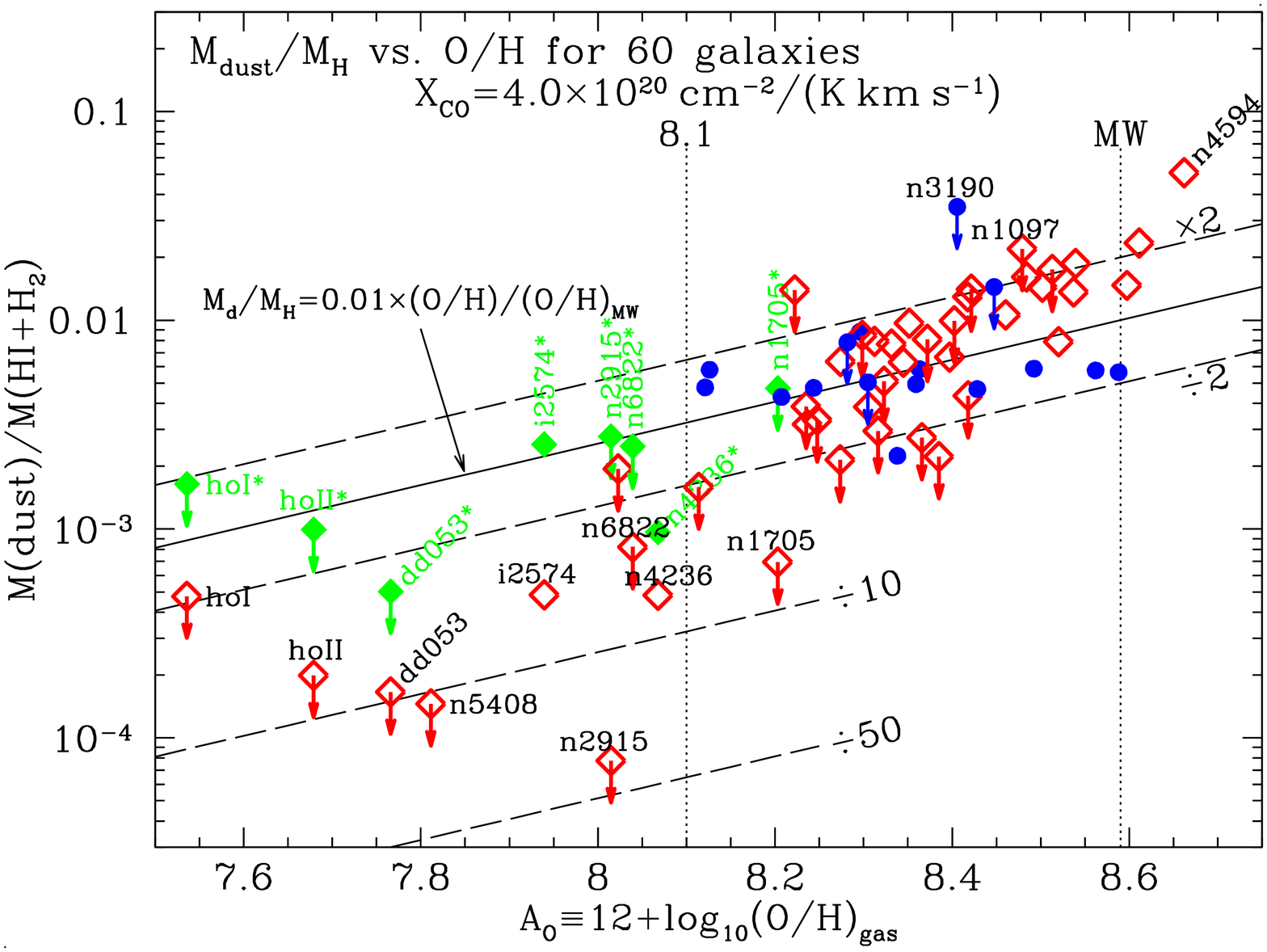}
\caption{\label{fig:Mdust/Mgas with ul vs metallicity}\footnotesize
	 Galaxies with and without SCUBA fluxes (circles and diamonds,
	 respectively), including 28 galaxies with only lower limits
	 on the gas mass.
	 Solid line shows eq.\ (\ref{eq:Md/MH envelope}); broken
	 lines show factor of two variations around the solid line.
	 Note that the upper bounds are all galaxies where $\Mdust$
	 has been determined, but either \ion{H}{1} or CO has not been
	 observed.
	 Filled diamonds show $\Mdust/\MH$ {\it for regions
	 where IR emission is detected} 
	 (see \S\ref{sec:dust/gas in low Z galaxies}).
	 }
\end{center}
\end{figure}

\oldtext{\citet{Galliano+Madden+Jones_etal_2005} estimated the dust 
         masses for 3}%
\newtext{\citet{Galliano+Madden+Jones_etal_2003,
                Galliano+Madden+Jones_etal_2005} estimated the dust
         masses for 4}
dwarf galaxies, and found that 
\oldtext{one of them, He~2-10, had 
         $\Mdust/\MH$ a factor of
         2--10 below eq.\ (\ref{eq:Md/MH envelope}).}%
\newtext{NGC~1569 and He~2-10 had $\Mdust/\MH$ a factor of
         4--7 and 2--10, respectively, below eq.\ (\ref{eq:Md/MH envelope}).}
\citet{Hunt+Bianchi+Maiolino_2005} used IR and submm observations to
estimate dust masses for a sample
of seven low-metallicity blue compact dwarfs.
For SBS~0335-052, with metallicity $\OH=7.29\pm0.01$
\citep{Izotov+Thuan_1998},
\citet{Hunt+Bianchi+Maiolino_2005} found an extremely low
dust-to-gas mass ratio $\Mdust/\MH\approx10^{-7}$, far
below the value $\sim2\times10^{-4}$ given by eq.\ (\ref{eq:Md/MH envelope}).
The results of \citet{Galliano+Madden+Jones_etal_2005}
and \citet{Hunt+Bianchi+Maiolino_2005} 
are clearly inconsistent with the conclusion of 
\citet{James+Dunne+Eales+Edmunds_2002} that dwarf galaxies have the
same dust-to-metals ratio as the Milky Way.

Figure \ref{fig:Mdust/Mgas vs metallicity}a,b shows
how $\Mdust/\MH$ correlates with oxygen abundance for
the \SINGS\ galaxy sample, where we use the
``characteristic'' 
oxygen abundances\footnote{%
    The characteristic abundance is intended to be representative
    of the luminosity-weighted ISM material.  For normal spirals,
    it is taken to be the metallicity at $0.4R_{25}$
    \citep{Moustakas+etal_2007}.
    The quoted uncertainties in oxygen abundances do not include systematic
    uncertainties in the calibrations.
    }
from \citet{Moustakas+etal_2007}, based on only strong nebular lines.
\citet{Moustakas+etal_2007} show that
different ``strong-line'' methods 
[e.g., \citet{Kobulnicky+Kewley_2004} vs.
\citet{Pilyugin+Thuan_2005}] give differing estimates for (O/H),
with the \citet{Pilyugin+Thuan_2005} values tending to be smaller
than the \citet{Kobulnicky+Kewley_2004} values by a factor of four, on
average.
Here we adopt abundances based on the \citet{Pilyugin+Thuan_2005}
method, and refer the reader to 
\citet{Moustakas+etal_2007} for discussion of the various
systematic uncertainties.
For the local MW we take the gas-phase $\OH=8.59$ as determined by
absorption line studies of low-density sightlines
\citep{Cartledge+Lauroesch+Meyer+Sofia_2004}.
Figures \ref{fig:Mdust/Mgas vs metallicity}a and b differ only in the
assumed value of $\XCO$, used to estimate the mass of H$_2$.
For both trial values of $\XCO$,
galaxies with $8.1\ltsim\OH\ltsim8.5$ appear to conform to
eq.\ (\ref{eq:Md/MH envelope}) to within a factor $\sim$2.

The two lowest metallicity galaxies in Figure 
\ref{fig:Mdust/Mgas vs metallicity}, IC~2574 and NGC~4236, both fall well
below eq.\ (\ref{eq:Md/MH envelope}), seemingly showing these systems
to have only a small fraction of their condensible elements in dust.
However, we will see below that this is not necessarily the case.
The points labelled i2574* and n4236* show estimates for
the dust/gas ratio 
{\it for the regions in these galaxies where IR emission is observed}
(see \S\ref{sec:dust/gas in low Z galaxies} for further details).
IC~2574 now falls right on eq.\ (\ref{eq:Md/MH envelope}).
NGC~4236 is low by a factor $\sim2.5$, but the estimate for the mass of gas
in the IR-emitting region is very uncertain 
(see \S\ref{sec:dust/gas in low Z galaxies}).

The highest metallicity system in our sample
is NGC~4594 (the Sombrero galaxy, M104).
NGC~4594 also has the highest dust-to-metals ratio;
even for $\XCO=4\times10^{20}\cm^{-2}(\K\kms)^{-1}$,
$(\Mdust/\MH)$/(O/H) is a factor of $\sim$4 above
eq.\ (\ref{eq:Md/MH envelope}).
Although this galaxy harbors a massive AGN, the dusty disk and ring
are strongly detected at 24, 71, and 160$\micron$
\citep{Bendo+Buckalew+Dale_etal_2006}.
The dust-to-gas ratio estimated for the disk of this galaxy is quite high,
$\Mdust/\MH=0.05$; the reason
for this high value is not clear.
Perhaps the dust properties in this system are significantly different from
the Milky Way dust properties that have been assumed in the modeling.
Alternatively, 
perhaps the molecular mass in the dusty ring has been underestimated.
The dusty ring has only been observed at a few positions in CO; additional
CO observations would be valuable.
It is also possible that the ratio of CO luminosity to molecular mass
in this galaxy may differ from that in spiral galaxies where the so-called
``X factor'' has been calibrated.
Unfortunately, the dust mass estimate for NGC~4594 
is not yet constrained by submm data --
while NGC~4594 is strongly detected at 850\um, the flux is dominated by
the AGN, and the 850\um\ flux from the dusty disk has not yet been
determined
\citep{Bendo+Buckalew+Dale_etal_2006}.
We have seen above (Fig.\ \ref{fig:scuba_v_descuba}) that dust mass
estimates made with and without SCUBA data can differ by
factors of $\sim$2.
Measurement of the submm continuum 
emission from the dust ring in NGC~4594 would be of great value.


\subsection{\label{sec:dust/gas in low Z galaxies}
            Dust-to-Gas Mass Ratios in Low-Metallicity Galaxies}

There are many galaxies for which we have only lower limits on the gas
mass, because either HI or CO has not been observed.
The resulting upper limits on $\Mdust/\MH$ are included in Figure
\ref{fig:Mdust/Mgas with ul vs metallicity}, which now includes systems
with $\OH$ as low as $7.54\pm\oldtext{0.34}\newtext{0.35}$ (HoI).
For $\OH>8.1$, upper limits on
the global dust-to-gas ratio 
are generally consistent with eq.\ (\ref{eq:Md/MH envelope}), to
within a factor $\sim$2.
However, {\it every} 
galaxy in Figure \ref{fig:Mdust/Mgas with ul vs metallicity}
with $\OH<8.1$ has a global dust-to-gas ratio that falls below
eq.\ (\ref{eq:Md/MH envelope}), sometimes by a very large factor.
IC~2574 is low by a factor of 5,
and 4 galaxies 
(Ho~II, DDO~053, NGC~5408, NGC~2915) are factors of 8 or more
below eq.\ (\ref{eq:Md/MH envelope}).
The most extreme case is NGC~2915, below 
eq.\ (\ref{eq:Md/MH envelope}) by a factor of 40.
At face value, this might suggest that
the balance between dust grain formation
and destruction in low-metallicity galaxies such as
NGC~2915 is 
much less favorable to grain formation and survival
than in the Milky Way.  However, we will argue here that 
(1) the dust-to-metals
ratio in the central regions of these galaxies is in fact normal
(i.e., consistent with eq.\ \ref{eq:Md/MH envelope}),
and
(2) the dust-to-metals ratio outside the central regions is not yet known.

\citet{Walter+Cannon+Roussel+etal_2007} compare
the stellar, IR, and \ion{H}{1} morphologies of the dwarf irregular (Im)
galaxies
Ho~II, Ho~I, IC~2474, and DDO~053:
in each of these galaxies,
most of the \ion{H}{1} is located outside the region where IR
emission is detected.
For example, only a fraction $f({\rm H~I})\approx0.19$ of the \ion{H}{1} in IC~2574 
lies within the regions where 70 and 160\um\ emission
is seen.  
Similarly, the fraction of the \ion{H}{1} within the contours where
dust emission is detected is $f({\rm H~I})\approx 0.2$, 0.29, 
and 0.33 for Ho~II,
Ho~I, and DDO~053. 

We have estimated the radii $\theta_{\rm IR}$ containing the
bulk of the infrared emission for four additional low-metallicity dwarfs: 
NGC~1705 \citep{Cannon+Smith+Walter+etal_2006},
NGC~2915,
NGC~4236, and
NGC~6822 \citep{Cannon+Walter+Armus+etal_2006}.
Our estimates for $\theta_{\rm IR}$, and for
the fraction 
$f({\rm H~I})$ of the \ion{H}{1} located at $\theta<\theta_{\rm IR}$, are
given in Table \ref{tab:dwarfs}, with $f$ ranging from 0.028 (for NGC~2915) to
0.5 (for NGC~4236).
The dust-to-gas mass ratios {\it for the regions where IR emission is detected}
are shown in Figure \ref{fig:Mdust/Mgas with ul vs metallicity}.
Six of the eight galaxies in Table \ref{tab:dwarfs} now have upper limits 
consistent with (\ref{eq:Md/MH envelope}) to within a factor two, with 
only two (DDO~053 and NGC~4236) falling low by a factor $\sim$2.5.
Given the uncertainties in estimation of both $\Mdust$ and $f({\rm H~I})$,
it is entirely possible that $>50\%$ of the interstellar 
Mg, Si, and Fe may be in solid grains, at least in
systems with $\OH\gtsim 7.5$.

What can we say about the dust/metals ratio in the extended \ion{H}{1} envelopes?
Both the metal abundances and the dust masses in these regions are
uncertain.
The metallicities are estimated
using \ion{H}{2} regions, and therefore apply only to regions with
recent star formation.
The outer \ion{H}{1} envelopes lack bright \ion{H}{2} regions, 
and therefore the metallicity there
is unknown.  
It would not be surprising if it were subtantially lower than
the ``characteristic'' metallicities used in 
Figure \ref{fig:Mdust/Mgas with ul vs metallicity}.
For the outer regions of 
some spiral galaxies, oxygen has been found to be underabundant,
relative to the ``characteristic'' metallicity of the central regions, 
by factors of four or more
\citep[e.g.,][]{Ferguson+Gallagher+Wyse_1998,Tullman+Rosa+Elwert+etal_2003}.

In the case of NGC~1705, UV absorption line studies
show that the \ion{H}{1} envelope has $\OH=7.43\pm0.22$
\citep{Heckman+Sembach+Meurer+etal_2001}: the oxygen abundance in
the \ion{H}{1} envelope
is a factor $6$ below the oxygen abundance determined from the \ion{H}{2}
regions \citep{Moustakas+etal_2007}.
For this oxygen abundance, $\Mdust/\MH$ for NGC~1705 is
consistent with eq.\ (\ref{eq:Md/MH envelope}).

\begin{deluxetable}{l c c c c c c c c}
\tabletypesize{\footnotesize}
\tablewidth{0pt}
\tablecolumns{9}
\tablecaption{\label{tab:dwarfs}
              Parameters for Selected Dwarf Galaxies
             }
\tablehead{
  \colhead{galaxy} &
  \colhead{morph} &
  \colhead{$D$} &
  \colhead{$M_B$} &
  \colhead{$(\nu L_\nu)_B$} &
  \colhead{$\theta_{\rm IR}$} &
  \colhead{$f({\rm H~I})$} &
  \colhead{$\theta_{0.5}({\rm H~I})$} &
  \colhead{$U_{0.5}({\rm H~I})$}
  \\
  \colhead{} &
  \colhead{type} &
  \colhead{Mpc} &
  \colhead{} &
  \colhead{$\Lsol$} &
  \colhead{arcmin} &
  \colhead{$\theta<\theta_{\rm IR}$} &
  \colhead{arcmin} &
  \colhead{} 
  }
\startdata
%
DDO~053 & Im & 3.56 & $-13.44$ & $2.1\times10^7$ & --  & 0.33\tablenotemark{a}&  1  & 0.11  \cr  

NGC~6822& Im &0.49  & $-14.03$ & $3.7\times10^7$ & 17  & 0.33\tablenotemark{e}& 33\tablenotemark{e} 
                                                & 0.034 \cr
Ho~I    & Im & 3.84 & $-15.12$ & $1.0\times10^8$ & --  & 0.29\tablenotemark{c} & 2 & 0.11 \cr  

NGC~1705& \ot{S0}\nt{Im} & 5.10 & $-15.76$ & $1.8\times10^8$ & 0.33\tablenotemark{g}
                                                             & 0.15& 0.69\tablenotemark{f}
                                                & 4.0   \cr

NGC~2915& Im & 2.70 & $-16.41$ & $3.3\times10^8$ & 0.5 & 0.028\tablenotemark{d}& 4.75\tablenotemark{a} 
                                                & 0.13  \cr

Ho~II   & Im & 3.39 & $-16.69$ & $4.2\times10^8$ & --  & 0.20\tablenotemark{c}& 4 & 0.15  \cr  

IC~2574 & Sm & 4.02 & $-17.38$ & $8.0\times10^8$ & --  & 0.19\tablenotemark{c}& 6 & 0.089 \cr  

NGC~4236& Sdm & 4.45  & $-17.94$ & $1.3\times10^9$ & 4.25 & $\sim$0.5& 4.2\tablenotemark{b}  &0.18   \cr

\enddata
\tablenotetext{a}{\citet{Becker+Mebold+Reif+vanWoerden_1988}}
\tablenotetext{b}{Estimated from \citet{Rots_1980} map, assuming FWHM=$10\arcmin$ gaussian beam.}
\tablenotetext{c}{\citet{Walter+Cannon+Roussel+etal_2007}}
\tablenotetext{d}{Estimated from \ion{H}{1} data in \citet{Meurer_1997}}
\tablenotetext{e}{Estimated from \ion{H}{1} profile in 
                        \citet{Cannon+Walter+Armus+etal_2006}}
\tablenotetext{f}{\citet{Meurer+Stavely-Smith+Killeen_1998}}
\tablenotetext{g}{\citet{Cannon+Smith+Walter+etal_2006}}
\end{deluxetable}

The dust content of the envelope is often also uncertain.  
The stellar densities in the outer
regions of these galaxies are low, and the starlight 
will be dominated by light coming from the central regions.
Table \ref{tab:dwarfs} lists $(\nu L_\nu)_B$, 
the total stellar luminosity in the B band.
For each of these galaxies we give the estimated 
half-mass radius $\theta_{0.5}$
for the HI,
and we have estimated the starlight intensity scale factor $U$
at the \ion{H}{1} half-mass radius,
\beq \label{eq:U_0.5}
U_{0.5} = \frac{1}{(1.982\times10^{-13}\erg\cm^{-3})}
          \frac{(\nu L_\nu)_B}{4\pi c(D\theta_{0.5})^2} ~~~,
\eeq
where $1.982\times10^{-13}\erg\cm^{-3}$ is $(\nu u_\nu)_B$ for the local
interstellar radiation field \citep{Mathis+Mezger+Panagia_1983}.
Equation (\ref{eq:U_0.5}) neglects internal extinction within the
dwarf galaxy 
[illumination of the outer envelope will suffer the same attenuation
as radiation escaping the galaxy, 
from which $(\nu L_\nu)_{\rm B}$ was obtained].
Only one galaxy in Table \ref{tab:dwarfs} -- 
the dwarf starburst NGC~1705 -- has $U_{0.5}>0.2$.

The $a \gtsim 0.03\micron$ grains that normally dominate the 160\um\
emission are expected to have temperatures 
$T\approx 17 U^{1/6}\K = 12(U/0.1)^{1/6}\K$,
with peak 
$\nu p_\nu$ at $\lambda\approx hc/6kT \approx 180 (0.2/U)^{1/6}\micron$.
Therefore, for the seven galaxies in Table \ref{tab:dwarfs} with 
$U_{0.5}<0.2$,
even the MIPS 160\um\ band (with half-response points at
140 and 174\um)
is relatively insensitive to whatever dust may be present at 
and beyond 
$\theta_{0.5}$ 
except for a small fraction that may be close to local stellar sources.
MIPS photometry for these galaxies is dominated
by the dust in the luminous regions, and the SED fitting
is therefore fitting the parameters for this dust.
If there is dust present beyond the half-mass radius, 
submm observations will be required to detect it
in all of the galaxies in Table \ref{tab:dwarfs} except NGC~1705.

In the case of SBS~0335-052, where \citet{Hunt+Bianchi+Maiolino_2005}
report an extraordinarily low dust-to-gas mass ratio
$\Mdust/\MH=1\times10^{-7}$, a factor $5000$ below
eq.\ (\ref{eq:Md/MH envelope}), we note that
this galaxy has not been detected at wavelengths between 100\um\
and 1~cm.  The dust models are poorly constrained,
and it is possible that they may have significantly underestimated the
mass of cool dust.  In addition, much of the HI mass in SBS0335-052
may be in an extended envelope, where the metallicity is unknown and
where the dust temperatures would be very low.  
Deep submm observations would be of value to better determine
the dust content of SBS~0335-052.

\begin{figure}[h]
\begin{center}
\includegraphics[angle=0,width=\figwidth]%
                {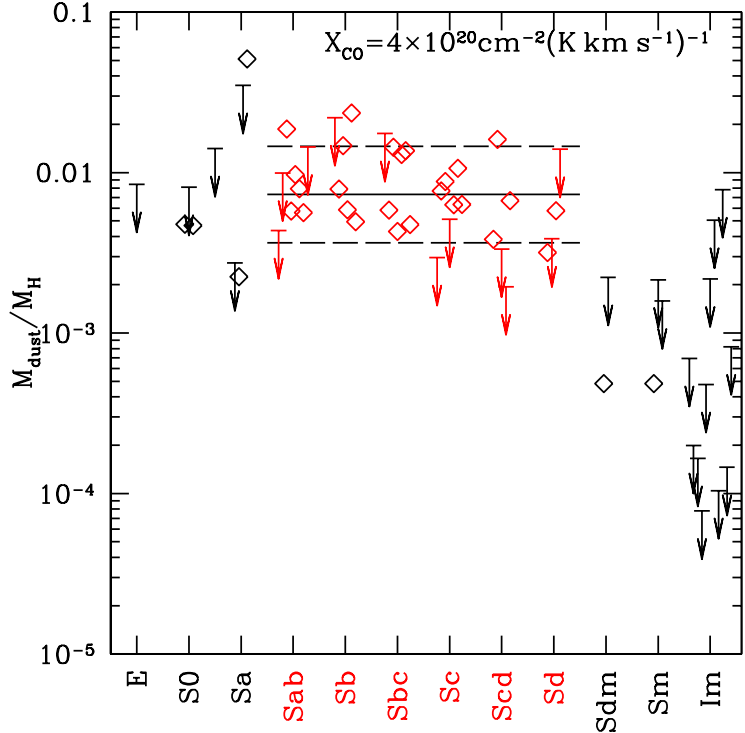}
\caption{\label{fig:mdmg vs morphology}\footnotesize
         Dust-to-gas mass ratio as a function of Hubble type,
	 assuming $\XCO=4\times10^{20}\cm^{-2}/(\K\kms)$.
	 The horizontal line corresponds to
	 $\Mdust/\MH=\oldtext{0.007}\newtext{0.0073}$, and the broken lines to factor-of-two
	 variations around this.
	 Note that the upper bounds are all cases where the dust mass
	 has been estimated, but only lower bounds are available
	 for the gas mass
	 because CO has not yet been observed.
	 }
\end{center}
\end{figure}

\subsection{Dust-to-Gas Mass Ratio and Galaxy Type}

The dependence of the dust-to-gas mass ratio on galaxy type is
explored in Figure \ref{fig:mdmg vs morphology}.
Spiral galaxies (Sab, Sb, Sbc, Sc, Scd, Sd) routinely have
$\Mdust/\MH\approx 0.007$ to within a factor of two.

We have dust mass estimates for 3 elliptical galaxies
(NGC~0855, NGC~3265, and NGC~4125), but
lack information on the gas content of NGC~0855 and NGC~4125.
Detection of 21cm emission 
gives a lower bound on the gas mass in NGC~3265.  If the molecular and ionized
gas mass in NGC~3265 is small compared to the HI mass, then
this elliptical galaxy has a dust/gas ratio consistent with
$\sim$solar metallicity.

\oldtext{Three of the 4}\newtext{The three} 
S0 galaxies in our samples have dust/gas ratios that
are consistent with the range seen for normal spirals:
NGC~3773 has $\Mdust/\MH < 0.01$ and 
NGC~1482 
and NGC~5866 have $\Mdust/\MH\approx 0.005$.
\oldtext{NGC~1705 has a very low estimate for the global
$\Mdust/\MH\ltsim 0.0007$ -- however, as discussed in 
\S\ref{sec:dust/gas in low Z galaxies},
much of the gas mass in NGC~1705 is in an envelope with very low O/H,
and the global dust-to-gas
ratio is consistent with eq.\ (\ref{eq:Md/MH envelope}) when
the appropriate oxygen abundance is used.}

\begin{figure}[h]
\begin{center}
\includegraphics[angle=0,width=\figwidth]%
                {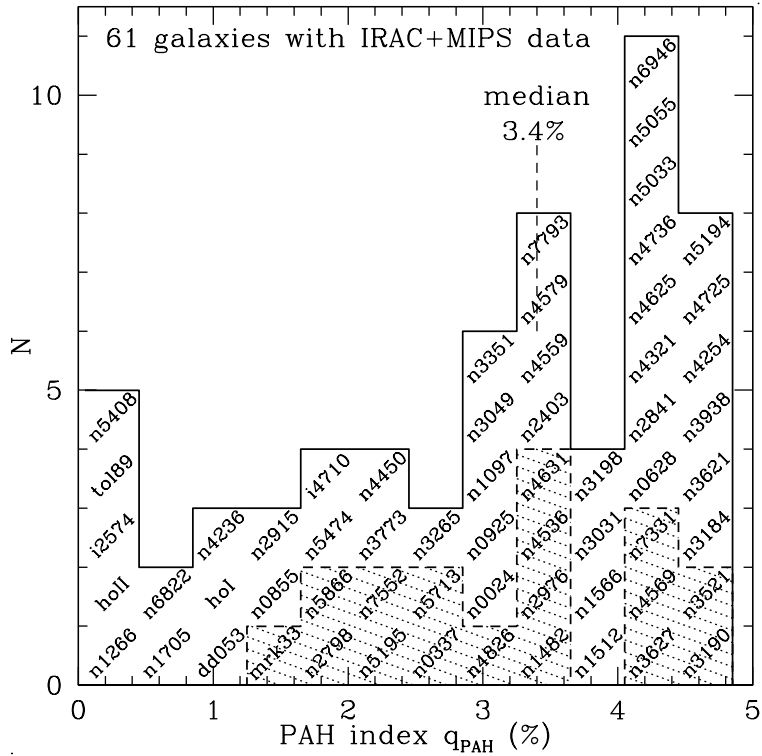}
\caption{\label{fig:overall pah histogram}\footnotesize
         The dotted histogram (shaded) 
	 shows the \Nscuba\ galaxies with SCUBA data for which we can
	 estimate the PAH index $\qpah$.
	 The solid histogram includes 44 additional galaxies without
	 SCUBA fluxes for which $\qpah$ can be estimated.
	 Four galaxies for which $\qpah$ cannot be determined have
	 been omitted (see text).
	 }
\end{center}
\end{figure}

\subsection{\label{sec:PAH abundances - scuba plus nonscuba}
            PAH Abundance}
The set of \Ntot\ \SINGS\ galaxies includes 
galaxies with strong PAH emission detected
in IRAC bands 3 and 4, as well as cases where there is little evidence
of nonstellar emission in these bands.
For some galaxies 
the dust emission (relative to starlight) is so weak that 
$\qpah$
\newtext{%
(the fraction of the total dust mass contributed by PAHs with
$N_{\rm C}<10^3$)
        }
cannot be reliably determined from the broadband photometry.
The DL07 models that are used here have 
\beq
\langle\nu F_\nu^{\rm ns}\rangle_{7.9\mum} \approx 0.070 \left(\frac{\qpah}{0.01}\right)
\left(\frac{\Ldust}{4\pi D^2}\right)
~~~.
\eeq
With the existing uncertainties in IRAC calibration and 
foreground/background subtraction, we do not attempt to estimate
$\qpah$ unless 
\beq
\left[\nu B_\nu(T_\star)\right]_{7.9\mum}\Omega_\star < 
0.15 \left(\frac{\Ldust}{4\pi D^2}\right) ~~~,
\eeq
so that the PAH power in the IRAC 7.9\um\ band 
would be $>$50\% of the stellar contribution
if $\qpah=0.01$.
This implies no estimate of $\qpah$ 
for four galaxies -- NGC~1291, NGC~1316, NGC~4125, and NGC~4594 --
all galaxies with relatively large bulge/disk ratios.
\begin{figure}[htb]
\begin{center}
\includegraphics[angle=0,width=\figwidth]%
                {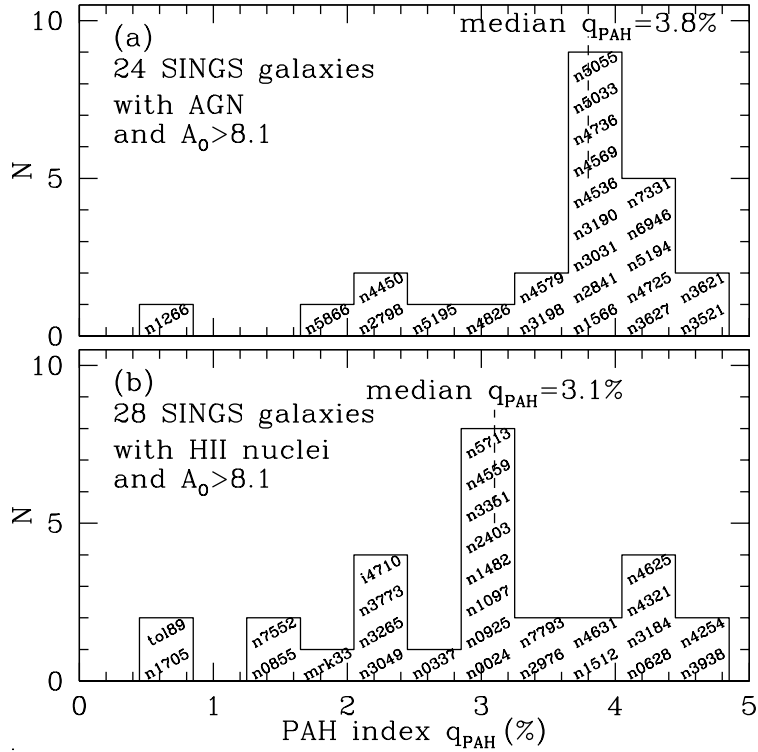}
\caption{\label{fig:pah_v_agn}\footnotesize
         Histogram of global $\qpah$ for SINGS galaxies 
	 with $A_{\rm O}>8.1$, (a) with, and (b) 
         without an AGN component.
	 }
\end{center}
\end{figure}
Figure \ref{fig:overall pah histogram} shows the distribution of
$\qpah$ 
for the remaining \Npah\ galaxies for which we have both IRAC
and MIPS data.  The distribution is broad -- 7/\Npah\ of the galaxies
have $\qpah<0.75\%$, while 30/\Npah\
of the galaxies have $\qpah>3.25\%$.

\newtext{\citet{Roche+Aitken+Smith+Ward_1991} noted that the 8--13\um\ spectra
of AGN were often deficient in the PAH emission features commonly
seen in \ion{H}{2} region galaxies. 
The spectrum of PAH emission around AGNs has since been found to often exhibit
unusual band ratios, with some AGN systems showing extremely low values of
$L(7.7\micron)/L(11.3\micron)$
\citep{Smith+Draine+Dale+etal_2007}.
The suppression of 7.7\um\ emission in AGNs may be attributable to
destruction of small PAHs by, e.g., X-rays from the AGN
\citep{Voit_1992}.
Figure \ref{fig:pah_v_agn} shows the distribution of global values of
$\qpah$ for SINGS galaxies with (a) low-luminosity AGN (Seyfert or LINER), 
or (b) \ion{H}{2} nuclei
[the AGN/\ion{H}{2} nuclear classification is from \citet{Moustakas+etal_2007}].  
Only
galaxies with characteristic metallicities $A_{\rm O}>8.1$ are
included.  The galaxies with AGN tend to have {\it higher} values
of $\qpah$ than the \ion{H}{2} galaxies: the global SEDs show no evidence for
PAH suppression.  This is presumably because the SINGS sample
excludes powerful AGN, and low-luminosity AGN
have little effect on the global SED.
Note that 
the 5 galaxies in Figure \ref{fig:pah_v_agn}b
with $\qpah<2\%$ include two irregulars (NGC~1705, Mrk~33),
one elliptical (NGC~0855), and a dwarf starburst (Tol~89).}

\begin{figure}[h]
\begin{center}
\includegraphics[angle=0,width=\figwidth]%
                {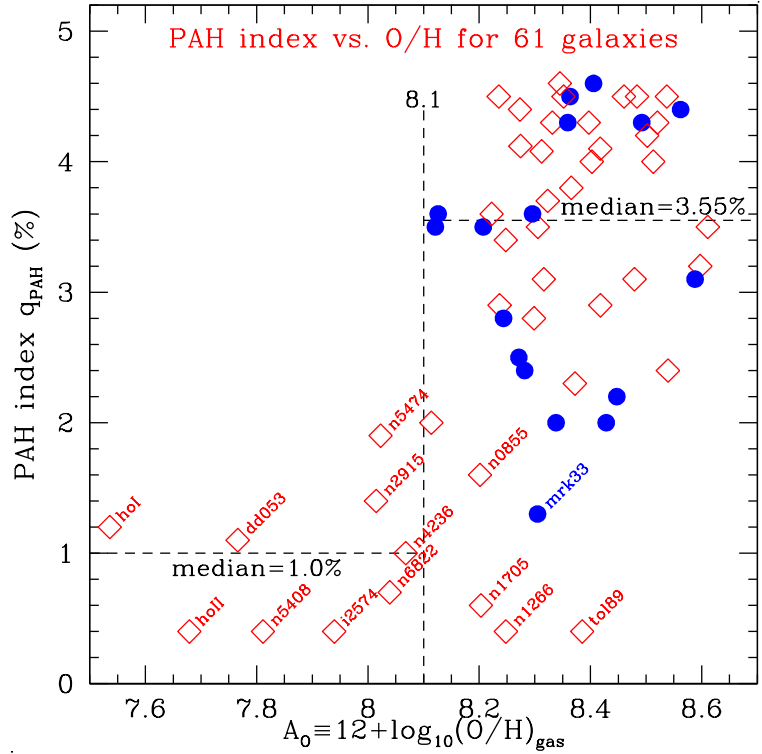}
\caption{\label{fig:pah index vs metallicity}\footnotesize
         PAH index (percentage of dust mass contributed by
	 PAHs with $N_{\rm C}<10^3$ C atoms) vs. galaxy metallicity (see text).
	 Low metallicity galaxies always have low PAH index $\qpah$.
	 Filled circles are SINGS-SCUBA galaxies; diamonds are
	 SINGS galaxies lacking submm data.
	 }
\end{center}
\end{figure}

Figure \ref{fig:pah index vs metallicity} shows $\qpah$
versus metallicity for the \Npah\ galaxies in the present study
for which $\qpah$ has
been determined.
We are able to estimate $\qpah$ for galaxies spanning a range of at least
0.9 dex
in metallicity, from $\OH=7.68\pm0.03$ (Ho~II) to 
$8.60\pm\oldtext{0.05}\newtext{0.06}$ (NGC~3351).
[$\OH=7.54\pm0.34$ of Ho~I 
is nominally lower than for Ho~II, but is uncertain.]
A clear separation is seen: all nine of the galaxies
with $\OH<8.1$ have 
$\qpah\leq\oldtext{2.0}\newtext{1.9}\%$, with median 
$\qpah=\oldtext{1.1}\newtext{1.0}\%$.
When $\OH\gtsim8.1$ , higher PAH abundances
are seen, with
median 
$\qpah=\oldtext{3.5}\newtext{3.55}\%$ 
for the 52 galaxies with $\OH>8.1$,
although note that only three of the 52 galaxies with
$\OH>8.1$ have 
$\qpah<\oldtext{1.1}\newtext{1.0}\%$ (the median for $\OH<8.1$). 
PAH emission strength -- as a fraction of total IR
emission -- therefore appears to always be low for $\OH<8.1$.

The weakness of PAH emission from low-metallicity galaxies was 
\oldtext{noted by \citet{Boselli+Lequeux+Sauvage_etal_1998} and
         \citet{Sturm+Lutz+Tran+etal_2000}.}%
\newtext{first noted by 
         \citet{Roche+Aitken+Smith+Ward_1991}
	 using ground-based spectroscopy,
         and further investigated with ISO data by
         \citet{Boselli+Lequeux+Sauvage_etal_1998},
         \citet{Sturm+Lutz+Tran+etal_2000}, and
         \citet{Madden_2000}.}
The SMC, with $\OH=8.0$
\citep{Kurt+Dufour_1998, Garnett_1999} falls just below the
apparent threshold $\OH=8.1$ 
seen in Figure \ref{fig:pah index vs metallicity}.
\citet{Li+Draine_2002c} concluded
that the SED of the SMC Bar
was consistent with the ``SMC'' dust model of
\citet{Weingartner+Draine_2001a}, with
a PAH index $\qpah$ of only 0.1\%
\footnote{
  This result is controversial: for the SMC Bar,
  \citet{Bot+Boulanger+Lagache_etal_2004} have argued that the PAH
  abundance as a fraction of the dust mass is similar to the MW value -- which
  we estimate to correspond to $\qpah\approx5\%$.
  Observations of the SMC with \Spitzer\ should 
  soon clarify this issue.
  }
(see Table \ref{tab:dust models}).
ISO observed PAH emission from molecular cloud
SMC B1 No. 1 
\citep{Reach+Boulanger+Contursi+Lequeux_2000}
outside the SMC bar;
Li \& Draine showed that the observed emission
could be reproduced with a model with
only 3\% of the SMC carbon in PAHs -- corresponding to 
$\qpah\approx0.8\%$.  Thus the SMC appears to have $\qpah<1\%$.

%

\citet{Hunt+Bianchi+Maiolino_2005} observed PAH emission to be
weak in a number of low metallicity blue compact dwarf galaxies.
In a photometric study of 34 galaxies spanning two decades in metallicity,
\citet{Engelbracht+Gordon+Rieke_etal_2005} concluded that there was a
sharp difference
in the ratio of 8\um\ emission from PAHs to 24\um\ emission
from warm dust for metallicities above and 
below a threshold value $\OH\approx8.2$.\footnote{%
   Note that the oxygen abundances
   used by Engelbracht et al. (2005) were based on a heterogeneous
   compilation of measurements from the literature, whereas the
   abundances in our study have been derived self-consistently, and
   placed on a common abundance scale 
   \citep[see][in prep., for details]{Moustakas+etal_2007}}
\citet{Hogg+Tremonti+Blanton+etal_2005}
compared PAH emission with starlight for a sample of 313 SDSS galaxies,
and noted that low-luminosity (presumably low metallicity) 
star-forming galaxies tended to have
low ratios of PAH emission to starlight.
\newtext{Using ISOCAM data, \citet{Madden+Galliano+Jones+Sauvage_2006} 
         concluded that PAH abundance,
         relative to larger grains, was positively correlated 
         with metallicity, and suppressed in systems with hard 
         radiation fields (i.e., high [\ion{Ne}{3}]/[\ion{Ne}{2}] ratio).}
In a spectroscopic study of blue compact dwarf galaxies, 
\citet{Wu+Charmandaris+Hao+etal_2006} found that the equivalent width of the
PAH 6.2 and 11.2\um\ features appeared to be suppressed for 
$Z/Z_\odot\ltsim0.2$, or
$\OH\ltsim7.9$.

The threshold $\OH=8.1$ found here is close to the value 8.2 found
by 
\citet{Engelbracht+Gordon+Rieke_etal_2005}, 
and the value 8.0 found by 
\citet{Wu+Charmandaris+Hao+etal_2006}.
The reason for the small value of $\qpah$ when $\OH<8.1$ is not
clear at this time.  A number of possibilities exist:
\begin{itemize}
\item Low $\OH$ might imply more rapid destruction of PAHs by UV radiation 
in an interstellar medium
with reduced shielding by dust.
\item Low $\OH$ might imply more effective destruction
of PAHs by thermal sputtering in 
shock-heated gas that cools slowly because of the reduced metallicity.
\item Galaxies with low O/H have lower values of C/O
      \citep{Henry+Worthey_1999}; the reduced PAH abundance in these
      galaxies might be due to a
      deficiency of PAH-producing carbon stars and C-rich planetary nebulae.
\item PAH formation and growth in the interstellar medium would
      be suppressed by low gas-phase C abundances.
\item We also call attention to the relatively low value
of $\qpah\approx1.3\%$ for Mrk~33, which with $\OH=8.30\pm\oldtext{0.01}\newtext{0.10}$
\citep{Moustakas+etal_2007} is well above the threshold $\OH\approx8.1$.
The relatively low value of $\qpah$ in Mrk~33 may be a consequence of
the vigorous star formation in Mrk~33 which -- via the combined
effects of hard UV and supernova blast waves -- may be enhancing the
rate of destruction of PAHs, thus acting to lower 
the steady-state PAH abundance.  
\item The strong 7.6\um\ emission feature requires free-flying
PAHs with $N_{\rm C}<10^3$ C atoms (DL07), and emission can
therefore be suppressed if small PAHs coagulate with larger grains.
Such coagulation can take place in molecular clouds.
(How this might correlate with metallicity is uncertain, as the
metallicity-dependence of molecular cloud properties
is not understood.)
\end{itemize}
While we can speculate about various effects that may be important,
our limited understanding of the dynamics of the interstellar medium in
other galaxies,
together with our limited knowledge regarding many of the processes acting to
form and destroy PAHs in the interstellar medium,
precludes definite explanations of the observed
galaxy-to-galaxy variations in $\qpah$. 

\section{\label{sec:starlight parameters - scuba plus nonscuba}
          Starlight Properties for \Ntot\ SINGS Galaxies}

\begin{figure}[h]
\begin{center}
\includegraphics[angle=0,width=\figwidthd]%
                {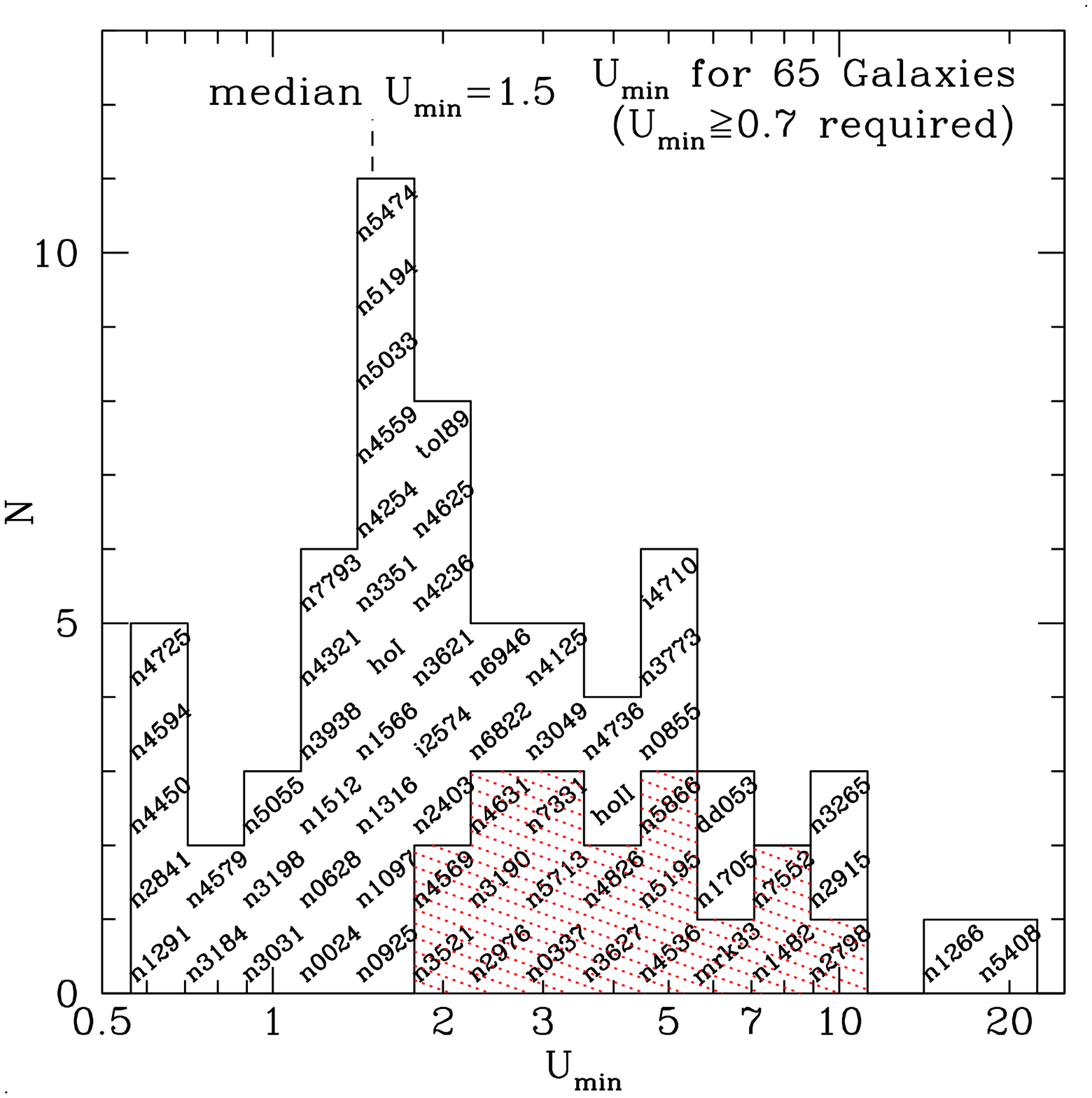}
\includegraphics[angle=0,width=\figwidthd]%
                {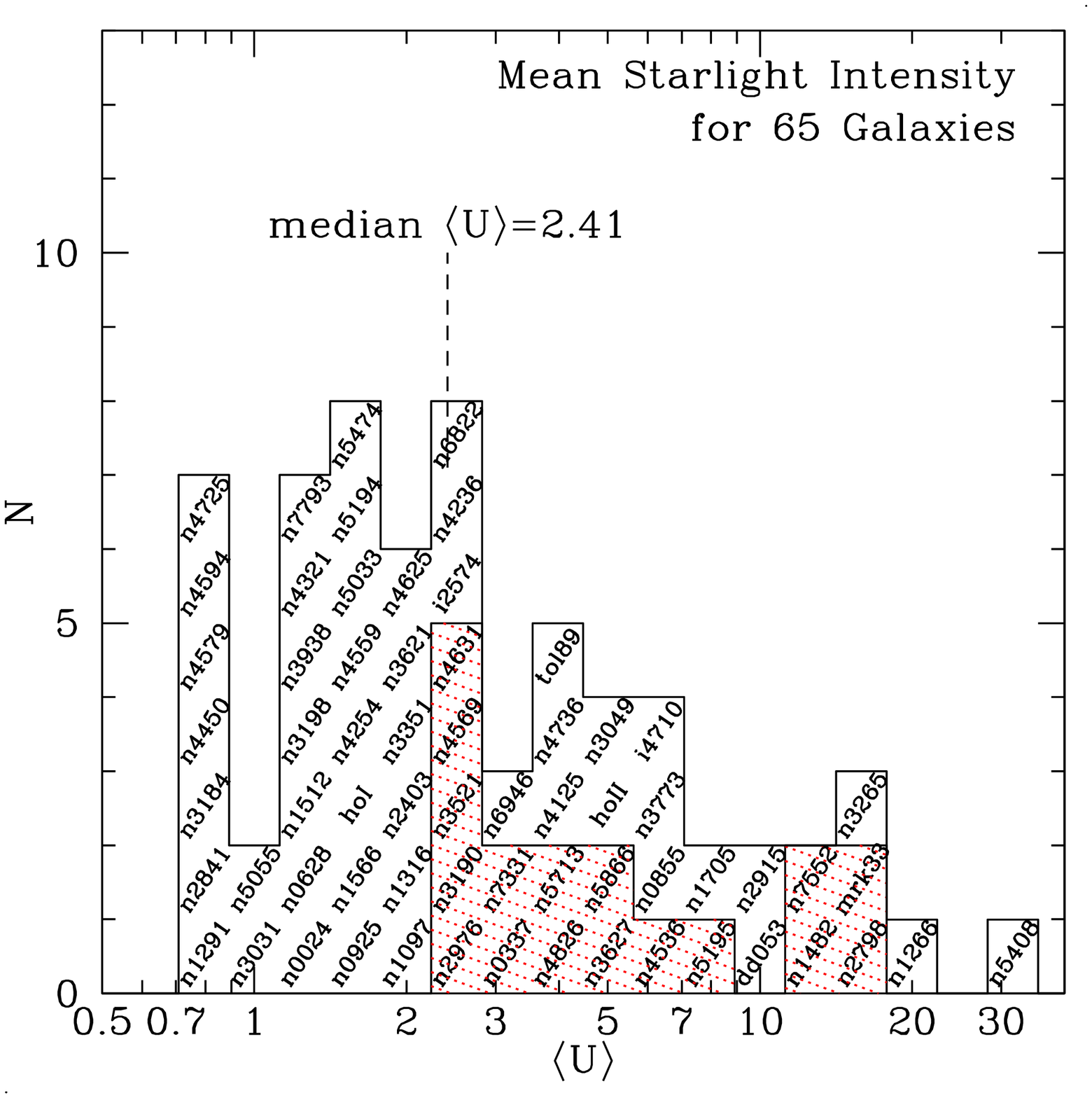}
\caption{\label{fig:umin and ubar for 70 galaxies}\footnotesize
         $\Umin$ and $\langle U\rangle$ for \Ntot\ galaxies (see text).
         The shaded portion of the histogram consists of the 
	 \Nscuba\ galaxies
	 for which SCUBA global fluxes constrain the dust models.
	 }
\end{center}
\end{figure}
The distribution of $\Umin$ values is shown in 
Figure \ref{fig:umin and ubar for 70 galaxies}.
The shaded portion of the histogram is contributed by the 
\Nscuba\ galaxies for
which SCUBA data were available and used in the model-fitting
with fixed $\Umax=10^6$; 
for these galaxies, there was no
lower limit imposed on $\Umin$.
Two of the \Nscuba\ galaxies have
$\Umin=2.0$, and the other 15 have $\Umin\geq2.5$. 
The remaining \Nnoscuba\ galaxies lack submillimeter data, and for these the
restricted fitting procedure allows only $\Umin\geq 0.7$.  This lower limit
on $\Umin$ results in a small ``pile-up'' of five galaxies at $\Umin=0.7$ in
Figure \ref{fig:umin and ubar for 70 galaxies} -- possibly some of these
should have smaller values of $\Umin$.
What is needed, of course, are submillimeter observations so that 
the restricted fitting procedure becomes unnecessary.

The form of the starlight distribution function (eq.\ \ref{eq:Udist})
is based on the expectation that the bulk of the IR luminosity
would be radiated by grains in a diffuse interstellar medium with
a more-or-less uniform starlight background, with $U\approx\Umin$.
This assumption appears to be validated by the success of the model in
reproducing the observed dust emission extending from 5.7\um\ to 850\um.
It is interesting that the median $\Umin=1.5$ is only slightly more
intense that our local interstellar radiation field ($U=1$), suggesting that
interstellar processes (star formation, absorption of starlight by dust)
regulate the diffuse starlight intensity to have $U\approx 1.5$ to within
a factor of two.

The distribution of mean starlight intensities is shown in Figure
\ref{fig:umin and ubar for 70 galaxies}b.
As before, the galaxies with SCUBA data are shaded.
The galaxies with SCUBA data and those without SCUBA data both show a
tendency to have $\langle U\rangle\approx 2.5$, although about 
\oldtext{15}\newtext{12}\%
of the sample has $\langle U\rangle>10$.

\begin{figure}[h]
\begin{center}
\includegraphics[angle=270,width=\figwidthww]%
             {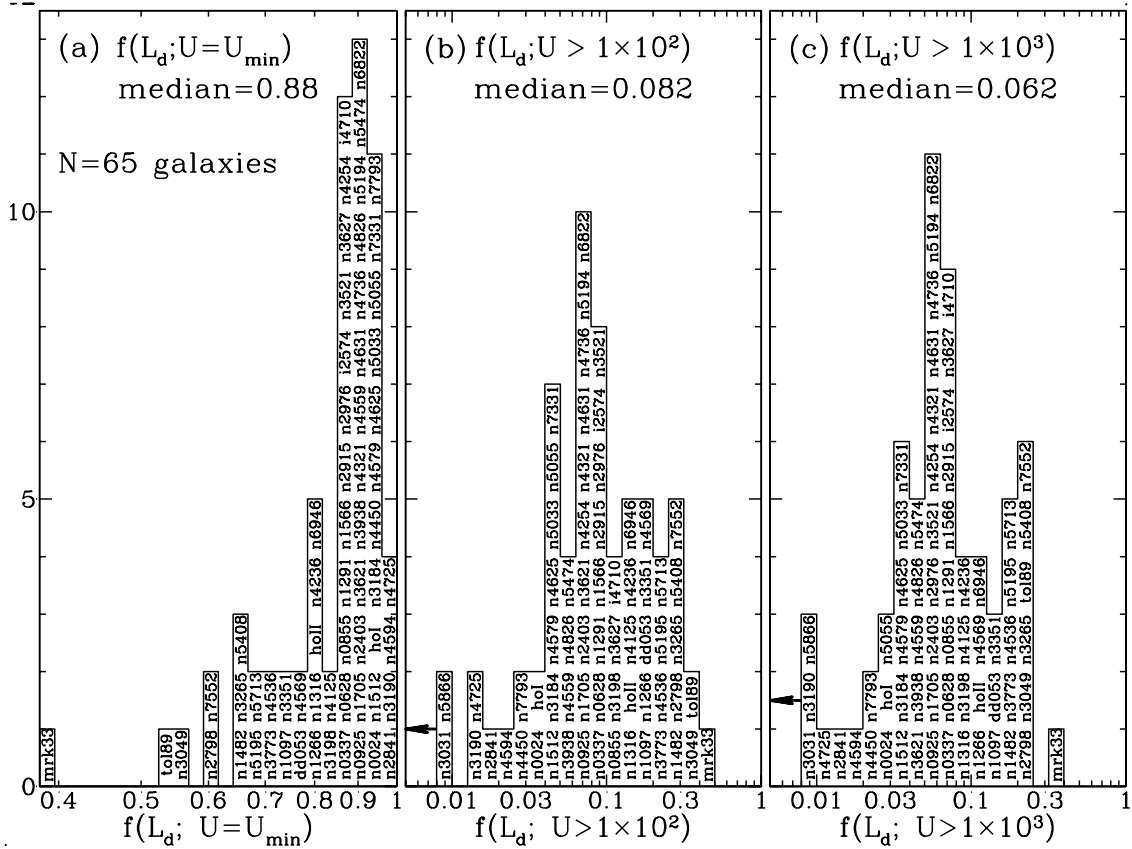}
\caption{\label{fig:pdr power distribution}\footnotesize
         Fraction of the dust luminosity originating in 
	 regions with (a) $U=\Umin$ , (b) $U>10^2$, and 
	 (c) $U>10^3$ (see text).
	 Galaxies with $f\leq0.01$ have been grouped in
	 bin at 0.01.
	 }
\end{center}
\end{figure}
\begin{figure}[h]
\begin{center}
\includegraphics[angle=0,width=\figwidth]
                {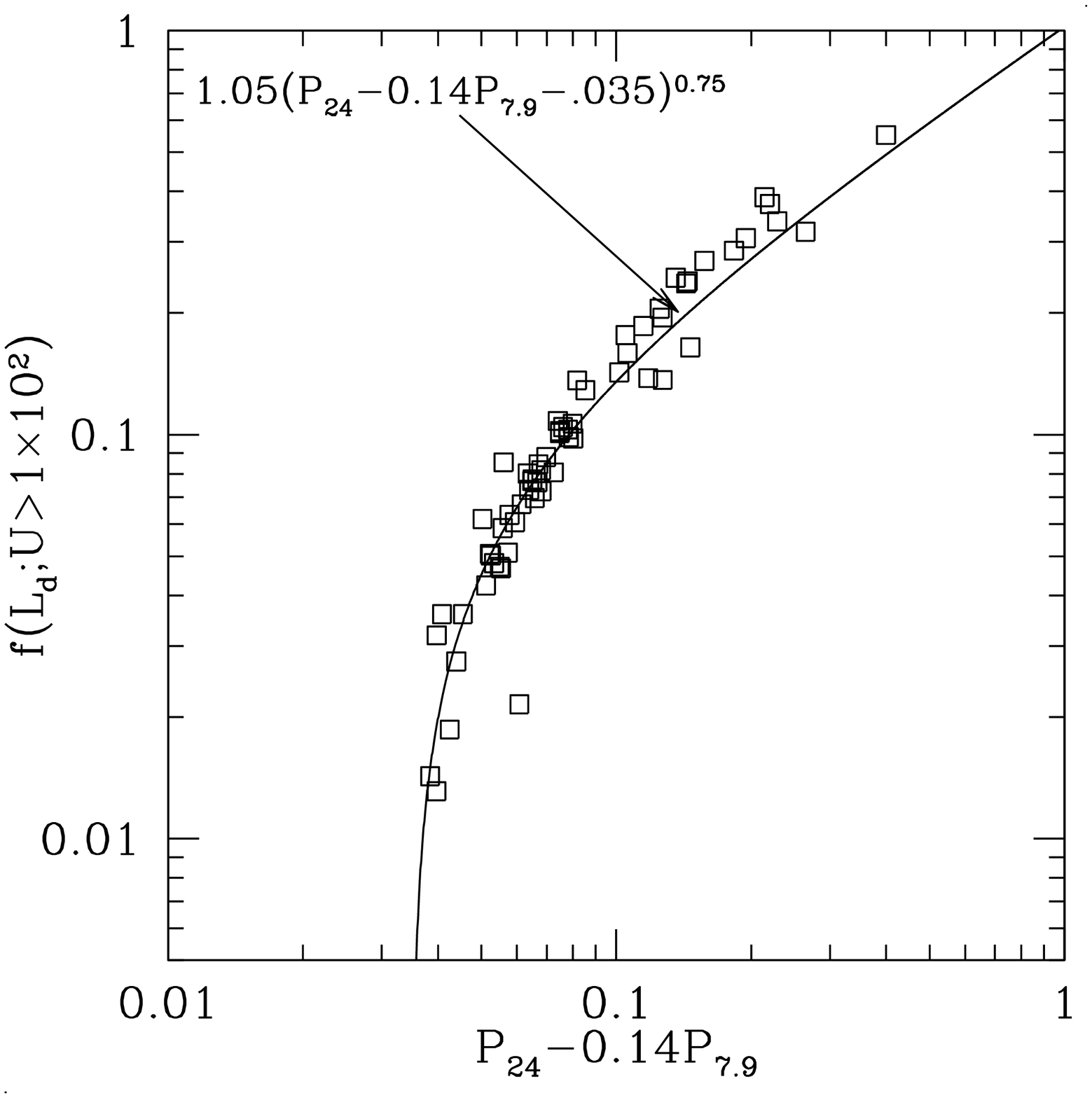}
\caption{\label{fig:gamma_vs_fluxratios}\footnotesize
        Estimated fraction of the dust luminosity 
	contributed by dust in regions
	with $U>10^2$ vs.\ observed value of
	$P_{24}-0.14P_{7.9}$ (see text).
	Also shown is the fitting function proposed by
	DL07.
	}
\end{center}
\end{figure}

The contribution of high intensity regions to the dust heating is shown
in Figure \ref{fig:pdr power distribution}.
In many of the sample galaxies, 
the observed 24\um\ emission appears to require that a significant fraction of
the dust heating take place in regions with high starlight intensities.
Figure \ref{fig:pdr power distribution} shows that the median galaxy
has $\sim$8\% of the total IR power originating from regions with
$U>10^2$, and $\sim$6\% of the IR power originating from regions with
$U>10^3$.  In most cases these high-intensity 
regions are probably PDRs bounding dense
\ion{H}{2} regions.  For example, the dust just outside the ionization front in the
Orion Bar PDR is heated by starlight with
$U\approx3\times10^4$ \citep{Allers+Jaffe+Lacy_etal_2005}, and similar
regions should be present in other star-forming galaxies.
For steady star formation with the IMF observed in the Milky Way,
perhaps $\sim$10\% of the bolometric luminosity 
\btdnote{does anyone know a ref where this number is calculated?}
is contributed by the massive
($M\gtsim20\Msol$)
stars that tend to spend most of their luminous existences close to their
natal clouds, photodissociating and photoionizing the gas, 
and heating the dust.
Therefore the median galaxies in Figure \ref{fig:pdr power distribution}
do not seem surprising.  

The galaxies with the highest values of $f(U>10^3)$
are Mrk~33 (Im), Tolo~89 (Im), NGC~5408 (Im),
NGC~2798 (Sa), 
NGC~3049 (Sab), 
and NGC7552 (Sab).
All of these galaxies appear to be forming stars at a high rate per unit
area (see, e.g., discussion of Mrk~33 in \S\ref{sec:Mrk33}), and
all have high ratios of
$(\nu F_\nu)_{24\mu{\rm m}}/(\nu F_\nu)_{160\mu{\rm m}}$,
ranging from 0.52 for Tolo~89 to 1.47 for Mrk~33,
indicative
of a substantial fraction of the dust power originating in regions
where the grains are hot enough to radiate at 24\um.

DL07 showed that, for their dust models, the fraction of the starlight
heating contributed by high intensity regions could be related to the
observed quantity
\beq \label{eq:P24_corr}
P_{24}-0.14P_{7.9} \equiv 
\frac{\langle\nu F_\nu^{\rm ns}\rangle_{24}-
0.14 \langle\nu F_\nu^{\rm ns}\rangle_{7.9}}
{\langle\nu F_\nu\rangle_{71}+\langle\nu F_\nu\rangle_{160}} ~~~,
\eeq
where $F_\nu^{\rm ns}$ is the nonstellar contribution to the flux density.
The term $0.14 \langle\nu F_\nu^{\rm ns}\rangle_{7.9}$ in the
numerator of eq.\ (\ref{eq:P24_corr}) is intended to approximate
the contribution to the 24\um\ flux arising from single-photon heating
of PAHs, so that the numerator of eq.\ (\ref{eq:P24_corr})
is the 24\um\ flux contributed by larger grains, which will only radiate
at 24\um\ when exposed to high intensity starlight.
Figure \ref{fig:gamma_vs_fluxratios} shows the estimated fraction
of the dust power originating in regions with $U>10^2$ vs.\ 
the observed value of the quantity $(P_{24}-0.14P_{7.9})$.
The galaxies fall in an approximately 1-dimensional sequence, and the
fraction of the starlight absorption taking place in high-intensity
PDRs  can be estimated from the observed 
$\langle F_\nu^{\rm ns}\rangle_{7.9}$,
$\langle F_\nu^{\rm ns}\rangle_{24}$,
$\langle F_\nu\rangle_{71}$, and
$\langle F_\nu\rangle_{160}$ using the DL07 fitting function
\beq
f(L_d; U>10^2) \approx 1.05(P_{24} - 0.14P_{7.9} -0.035)^{0.75} ~~~.
\eeq

\section{\label{sec:discussion}
         Discussion}
\subsection{\label{sec:dark clouds}
            What About the Dust in Dark Clouds?}

The adopted ``model'' for the distribution of starlight intensities
makes no allowance for the possibilility of dust in dark clouds, since it
assumes that there is no dust heated by starlight intensities below the
``diffuse ISM'' value $\Umin$.  While such dust may not be important
as a contributor to the total bolometric luminosity, it would not be surprising
if it contained an appreciable fraction of the total dust mass, and
it might have been expected to contribute an appreciable fraction of the
emission at 850\um.

However, we have seen that dust with the physical properties adopted
by DL07 appears to reproduce the observed SEDs of the 17 SINGS-SCUBA galaxies
without a ``cold dust'' component -- indeed, when we fix $\Umax=10^6$,
the best-fit models for the
17 SINGS-SCUBA galaxies all have $\Umin\geq2$.

As discussed by DL07, the dust emissivities employed here reproduce the
FIR and submm emission from our local diffuse ISM at galactic latitudes
$|b|>7^\circ$ as measured by the FIRAS instrument on COBE
\citep{Finkbeiner+Davis+Schlegel_1999}.
To the extent that our models reproduce the observed global FIR and 
submm emission from other galaxies, it suggests that the dust in dark
clouds makes only a minor contribution to the emission.  Suppose that
a fraction $\fcold$ of the dust mass is actually in dark clouds,
with dust temperature $\Tcold$, and the remainder $(1-\fcold)$ is
in diffuse clouds, all at a single dust temperature $\Twarm$.
Then the fraction $p_{\rm cold}$
of the total emission contributed by the dark cloud
dust at wavelength $\lambda$ is just
\beq
p_{\rm cold}(\lambda) = \frac{\fcold/(e^{x_c}-1)} 
                {\fcold/(e^{x_c}-1) + (1-\fcold)/(e^{x_d}-1)} ~~~,
\eeq
where $x_c\equiv hc/\lambda k \Tcold$ and $x_d\equiv hc/\lambda k \Twarm$.
If we take $\Twarm=20\K$ and $\Tcold=12\K$, for example, then
at 850\um\ we have
\beq
p_{\rm cold} = \frac{\fcold}{\fcold + 2.33(1-\fcold)} ~~~.
\eeq
Hence if, hypothetically, 50\% of the dust mass were in the cold clouds
($\fcold=0.5$), they
would contribute only 30\% of the total emission at 850\um;
if this hypothetical cold component were left out,
the model would fall short at 850\um\ by only 30\%, 
comparable to the existing photometric uncertainties.

Because the model does not include a cold dust component, it would not
be surprising if the resulting estimates of the dust mass $\Mdust$ were
low.  However, we have seen that if we adopt 
$\XCO=4\times10^{20}\cm^{-2}(\K\kms)^{-1}$, we obtain dust-to-gas mass
ratios that appear to account for all of the refractory elements in the
ISM, as shown in Figure \ref{fig:Mdust/Mgas with ul vs metallicity}.
If in fact we have substantially underestimated the dust mass in some
of these galaxies, then our adopted gas masses would have to also be
increased, by using an even larger value of $\XCO$.

Note that while
much of the gas in the Milky Way is molecular, much of this molecular
gas is actually in regions with visual extinction $A_V\ltsim 1$, so that
it is not ``dark'', and the dust grains are heated by diffuse starlight.
In dark clouds that are star-forming, 
such as Orion OMC-1 or M17, much of the dust is actually warm, heated by the
embedded stars.
It appears most likely that the fraction of the dust mass that is 
genuinely ``cold''
is probably relatively small (less than, say, 30\%).  
Although the model omits a ``cold dust'' component, it appears
to recover the majority of the global dust mass in the SINGS galaxies.

\subsection{\label{sec:graphical methods}Comparison with Graphical Methods}

\begin{deluxetable}{l c c c c c c}
\tablewidth{0pt}
\tablecolumns{7}
\tablecaption{\label{tab:graphical method}
              Comparison with Results of Graphical Method}
\tablehead{
   \colhead{galaxy} & 
   \multicolumn{2}{c}{$\qpah$} &
   \multicolumn{2}{c}{$\log[\Mdust/\Msol]$} &
   \multicolumn{2}{c}{$\langle U\rangle$}
\\
  &
  \colhead{GM$^a$} &
  \colhead{SEDM$^b$} &
  \colhead{GM$^a$} &
  \colhead{SEDM$^b$} &
  \colhead{GM$^a$} &
  \colhead{SEDM$^b$}
}
\startdata
IC2574   & 0.5\% & 0.4\% & 5.79 & 5.86 & 2.6 & 2.3
\cr
Mrk~33  & 3.2\% & 1.3\% & 6.41 & 6.47 & 18  & 16
\cr
NGC~1266 & 0.5\% & 0.4\% & 7.04 & 7.05 & 16  & 18
\cr
NGC~3521 & 4.6\% & 4.5\% & 8.10 & \ot{7.83}\nt{7.84} & 1.2 & 2.2
\cr
NGC~6822 & 1.1\% & 0.7\% & 5.15 & 5.04 & 2.1 & 2.8
\cr
\hline
\multicolumn{7}{l}{$^a$ Graphical method from DL07}\cr
\multicolumn{7}{l}{$^b$ Results from fitting MW models with 
$\Umax=10^6$ and $\Umin\geq1$.}\cr
\enddata
\end{deluxetable}

The model-fitting carried out here has been done by an automated
fitting procedure, that includes using interpolated models with
values of $\qpah$ intermediate between the seven $\qpah$ values
in the Milky Way dust models examined by DL07.
DL07 presented a straightforward graphical procedure for estimating 
$\qpah$, dust masses
and starlight intensities from IRAC and MIPS photometry, and applied it
to 5 galaxies as examples.
The results of that graphical 
procedure are compared with the present results in 
Table \ref{tab:graphical method}.
For two of the galaxies (Mrk~33 and NGC~3521), the SED model fitting
made use of SCUBA data, which is not used in the DL07 graphical procedure.

Table \ref{tab:graphical method} shows generally good agreement between
the graphical method and the present method for estimation of
$\qpah$, $\Mdust$, and $\langle U\rangle$.
The dust masses agree to within a factor of two in all cases.
The values of $\qpah$ are in generally good agreement except for
Mrk~33, where $\qpah=3.2\%$ estimated 
from the graphical method is considerably
larger than $\qpah=1.3\%$ from the SED fitting procedure used in this
paper.  The reason for this discrepancy is straightforward: the
DL07 graphical fitting procedure advises the user to
estimate $\qpah$ by seeking a model that reproduces the 7.9\um\
emission with the smallest possible value of $\gamma$, the fraction of
the dust mass exposed to starlight with $U>\Umin$.
While this approach is appropriate for most galaxies -- which generally
have $\gamma\ltsim 0.03$ -- Mrk~33 
has an extreme spectrum, with a very large fraction of the
dust exposed to intense radiation fields 
(see the location of Mrk33 in Fig.\ \ref{fig:pdr power distribution}).
The SED modeling in the present paper estimates $\gamma=0.12$ for Mrk~33.
In fact, examination of the graphs shown in DL07 (Figs.\ 20
and 21 of that paper) reveal that a solution with
$\gamma\approx 0.1$ does provide a better fit to the observed flux
ratios.
The DL07 graphical fitting procedure should therefore be employed
with caution for galaxies with $P_{24}\equiv
\langle\nu F_\nu^{\rm ns}\rangle_{24}/
[\langle\nu F_\nu\rangle_{71}+\langle\nu F_\nu\rangle_{160}]\gtsim0.2$,
indicative of substantial amounts of IR power radiated by dust
in intense radiation fields.

\subsection{\label{sec:dust composition}Dust Composition}

The graphite component of the dust model used here produces a broad
emission feature near 33\um\ when the dust is illuminated by starlight
with $U\gtsim10^4$ \citep{Draine+Li_2007}.
This feature is a consequence of the optical properties assumed for graphite.

Visual inspection of the 25--35\um\
spectra presented by \citet{Smith+Draine+Dale+etal_2007} suggest that
the 33\um\ feature is not present, at least not with the strength that would
be expected in some of our global models.
\oldtext{
    Some of the nuclear spectra (e.g., NGC~4321, NGC~4569)
    may have a broad emission feature near 33\um, but it is weak,\footnote{
       The PAHFIT spectrum decomposition program 
       \citep{Smith+Draine+Dale+etal_2007} 
       assigns an equivalent width ~$\sim0.2\micron$ for these galaxies.}
    it is near the long-wavelength limit of the IRS instrument,
    and its reality is uncertain.
    Without spectrophometry covering the 
    25--45\um\ range, it does not appear possible to either confirm or rule out
    the presence of this graphite emission feature.
    }%
\newtext{%
    Some of the central spectra (e.g., NGC~4321)
    have a broad (FWHM\,$\approx1.7\micron$) emission feature near 33\um, 
    but it is weak,\footnote{
       The PAHFIT spectrum decomposition program 
       \citep{Smith+Draine+Dale+etal_2007} 
       assigns an equivalent width ~$\sim0.2\micron$ for these galaxies.}
    and narrower than the very broad (FWHM\,$\approx20\micron$) peak
    found in our models.
    A similar $\sim1\micron$ wide plateau near 33-34\um\ 
    was also seen in the ISO spectra
    of bright starburst galaxies 
    \citep{Sturm+Lutz+Tran+etal_2000} 
    and planetary nebulae \citep{Waters+Beintema+Zijlstra+etal_1998},
    and tentatively attributed to crystalline olivine.}
If future observations find that the 
graphite feature is not present with the
strength predicted by our model calculations, what would be the implications?
First of all, it must be remembered that the present calculations are
based on two highly idealized assumptions: 
(1) the $a > 100\Angstrom$ carbonaceous
grains are spherical and composed of single crystals of graphite, 
and (2) the dielectric tensor in the IR
conforms to a simple free-electron model
\citet{Draine+Lee_1984,Draine+Lee_1987}.
The 33\um\ feature is a consequence of the dielectric function
for ${\bf E}\parallel c$, 
which is directly related to the electrical conductivity
parallel to the $c$-axis (i.e., normal to the graphite basal plane).
This conductivity is modelled by a free electron gas with a plasma
frequency $\omega_p=1.53\times10^{14}\s^{-1}$, and a damping time
$\tau=1.4\times10^{-14}\s$.
The simple free-electron model is not expected to accurately
describe the detailed wavelength dependence of this feature
(located at $\omega=2\pi c/33\micron=5.7\times10^{13}\s^{-1}$).
Furthermore, even if the model were accurate for ideal graphite, it
seems likely that impurities and defects -- which might vary from
grain to grain in a single region, or from
region to region in the galaxy -- would have the effect of
causing $\omega_p$ and $\tau$ to vary over the population of emitting grains
(and perhaps even within a single grain)
resulting in further broadening of this feature.
Therefore, failure to observe this feature would probably not
rule out graphite as a grain constituent -- it would only rule out
monocrystalline graphite grains with a dielectric function resembling the
simple free-electron model used by \citet{Draine+Lee_1984}.

Of course, it would not be surprising if the carbonaceous component of the
interstellar grain population consisted of something quite different from
monocrystalline graphite.  ``Amorphous carbon'', for instance, 
refers to a class of materials with widely varying properties:
for example, the amorphous carbon materials BE, ACAR, and ACH2
studied by
\citet{Colangeli+Mennella+Blanco_etal_1993,
Colangeli+Mennella+Palumbo_etal_1995,
Mennella+Colangeli+Blanco_etal_1995}
are all insulators, with bandgaps of 0.15~eV, 0.52~eV, and 1.32~eV
respectively
\citep{Zubko+Mennella+Colangeli+Bussoletti_1996}.
These carbon solids, each having
different properties in the FIR, would not show
the 33\um\ feature. 
It is quite possible that the carbonaceous material in the larger
($a\gtsim 0.02\micron$) interstellar grains may be better approximated
by some form of amorphous carbon.
However, we have seen that the SINGS photometry appears to \newtext{also}
be compatible with graphite.

\subsection{\label{sec:XCO}
             The H$_2$/CO Conversion Factor $\XCO$}

To estimate the dust-to-gas mass ratio in \S\ref{sec:dust to gas ratio}, we
need the total hydrogen mass $\MH$, including H$_2$.
Because H$_2$ resides 
primarily in its lowest two rotational states ($J=0$ and 1), which
do not radiate, the CO $J=1\rightarrow0$ line is routinely
employed to trace H$_2$.
It is usually assumed that the H$_2$ column density
is proportional to the CO 1--0 surface brightness (see equation 
\ref{eq:XCO}), with an empirical conversion factor $\XCO$ that is based on
some method of estimating the total gas mass.
For the Milky Way, virial mass
estimates based on giant molecular cloud linewidths 
give $\XCO=3.0\times10^{20}\cm^{-2}(\K\kms)^{-1}$
\citep{Young+Scoville_1991}.
Studies using $\gamma$-ray
emission as a mass tracer in the Milky Way found
$\XCO/[\cm^{-2}(\K\kms)^{-1}]=(1.9\pm0.2)\times10^{20}$
\citep{Strong+Mattox_1996} and
$(1.56\pm0.05)\times10^{20}$
\citep{Hunter+Bertsch+Catelli+etal_1997}.
Using IR emission from dust as a mass indicator,
\citet{Dame+Hartmann+Thaddeus_2001} found
$\XCO=(1.8\pm0.3)\times10^{20}\cm^{-2}(\K\kms)^{-1}$ 
for gas at $5^\circ<|b|<30^\circ$.
Because (1) CO 1--0 emission is often optically thick, (2) CO may
be subthermally excited in the outer layers of molecular clouds,
and (3) 
photodissociation of CO results in low CO/H$_2$ ratios in the outer
layers of molecular clouds, there is no reason to expect a single
value of $\XCO$ to apply to all galaxies, or to all regions within a
galaxy.

For low-luminosity spiral galaxies in the Virgo cluster, 
\citet{Boselli+Lequeux+Gavazzi_2002} estimated values of
$\XCO$ ranging from $\sim 10^{20}\cm^{-2} (\K\kms)^{-1}$ 
for giant spirals
to $\sim10^{21}\cm^{-2}(\K\kms)^{-1}$ for dwarf irregulars.
Based on an extensive study of giant molecular clouds in a number of
galaxies in the Local Group,
\citet{Blitz+Fukui+Kawamura_etal_2006} conclude that
$\XCO=4\times10^{20}\cm^{-2}(\K\kms)^{-1}$ is a good value to use
for most galaxies if the individual GMCs are assumed to be virialized.  

The present study has demonstrated that we can
use a physical dust model to reproduce observed IR and FIR
emission from nearby galaxies, allowing us to estimate the dust mass
in these galaxies.
Because we have a-priori expectations for the value of the dust-to-gas
mass ratio, our determination of $\Mdust$ therefore allows us to
estimate the total gas mass,
and thereby estimate $\XCO$ for these galaxies.

Figure \ref{fig:Mdust/Mgas vs metallicity} explores the
relationship between dust-to-gas mass ratio and metallicity.
Observed values of $\Mdust/\MH$ are plotted versus metallicity in
Figures \ref{fig:Mdust/Mgas vs metallicity}a and b, showing only galaxies where
both \ion{H}{1} and CO have been detected, or where the upper limit on either
\ion{H}{1} or CO shows it to be of minor importance compared to the other
gas component.
Equation (\ref{eq:Md/MH envelope})
is plotted in Figure \ref{fig:Mdust/Mgas vs metallicity}a,b, where
we assume
$[\OH]_{\rm MW} = \oldtext{8.60}\newtext{8.59}$.
The figure includes 12 galaxies for which SCUBA data 
constrains the dust models --
these should be the most reliable dust masses.
The figure includes an additional 20 galaxies lacking submm data.
In Figure \ref{fig:Mdust/Mgas vs metallicity}a the H$_2$ mass is estimated
using $\XCO=2\times10^{20}\cm^{-2}(\K\kms)^{-1}$, whereas Figure
\ref{fig:Mdust/Mgas vs metallicity}b
uses $\XCO=4\times10^{20}\cm^{-2}(\K\kms)^{-1}$.

From comparison of Figures \ref{fig:Mdust/Mgas vs metallicity}a and b, we
see that $\XCO=4\times10^{20}\cm^{-2}(\K\kms)^{-1}$ results
in dust-to-gas mass ratios for the \SINGS-SCUBA galaxies 
that are consistent with 
the amount of material actually available to form silicate and
carbonaceous grains, when one 
considers that the dust mass estimates are 
uncertain at the factor-of-two level, and the
metallicity estimates are also uncertain.
There are some cases with mass determinations (not upper limits) that
fall well above eq.\ (\ref{eq:Md/MH envelope}), which was
expected to define an upper envelope for $\Mdust/\MH$.
For all cases where submm data were used in the fitting, the dust
mass estimates were expected to have better than factor of two accuracy;
if we adopt $\XCO=4\times10^{20}\cm^{-2}(\K\kms)^{-1}$, none of the
galaxies with SCUBA data exceed eq.\ (\ref{eq:Md/MH envelope}) by more
than a factor 2.
Conversely, $\XCO=2\times10^{20}\cm^{-2}(\K\kms)^{-1}$ would
imply dust masses for NGC~1482 and NGC~2976 that would be 
uncomfortably large
given their estimated oxygen abundances.
The one galaxy in Figure \ref{fig:Mdust/Mgas vs metallicity}b that
falls well above the expected envelope is NGC~4594, the Sombrero galaxy.
The interstellar medium in this unusual system merits further study.

The histograms in Figures 
\ref{fig:Mdust/Mgas vs metallicity}c,d show the
distribution of the dust-to-gas mass ratio, normalized to 
\oldtext{solar metallicity.}%
\newtext{the characteristic oxygen abundance of the Milky Way.}
Ideally, the distribution would be narrow and peaking at $\sim$0.01, 
possibly with a tail to low values of
$(\Mdust/\MH)/[{\rm (O/H)/(O/H)}_{\rm MW}]$ 
representing galaxies where the
grains do not incorporate a large fraction of the refractory elements.
For $\XCO=2\times10^{20} \cm^{-2}(\K\kms)^{-1}$, the median is at 
\oldtext{0.013}\newtext{0.016}.
For $\XCO=4\times10^{20} \cm^{-2}(\K\kms)^{-1}$, 
the median shifts to 
\oldtext{0.010}\newtext{0.012} -- approximately the value expected 
if grains incorporate most of the refractory elements (as is the case
in the Milky Way).

The above considerations suggest that the value
$\XCO\approx4\times10^{20} \cm^{-2}(\K\kms)^{-1}$
recommended by \citet{Blitz+Fukui+Kawamura_etal_2006} for Local Group 
galaxies
is also appropriate for the \SINGS\ galaxy sample.
This value of $\XCO$ is a factor of $\sim$2 larger
than the value
$(1.8\pm0.3)\times10^{20}\cm^{-2}(\K\kms)^{-1}$ 
found recently by \citet{Dame+Hartmann+Thaddeus_2001} for the local Milky Way.

\section{\label{sec:summary}
         Summary}

The principal results of this paper are as follows:
\begin{enumerate}
\item A physical dust model consisting of PAHs, carbonaceous grains,
      and amorphous silicate grains 
      \citep{
	Draine+Li_2007}
      heated by starlight, with a distribution of
      starlight intensities, 
      successfully reproduces the IR and submm emission for 
      a sample of 17 galaxies with IRAC and MIPS photometry plus
      SCUBA photometry.
      In particular, the SCUBA photometry does {\it not} require
      an additional population of cold dust grains, or grains with
      enhanced submm emissivities.
      Cold dust grains in dark clouds appear to contain only
      \newtext{a}
      minor fraction 
      \newtext{($\ltsim50\%$)}
      of the total dust mass.

\item The dust model allows the mean starlight intensity 
      scale factor $\langle U\rangle$ and the total dust mass $\Mdust$
      to be estimated for each galaxy.
      For the \Nscuba\ galaxies with SCUBA data, $\langle U\rangle$ ranges
      from 2.2 to 16, with a median of 4.3
      (see Fig.\ \ref{fig:ubar_scuba_all}).
      For the 12 galaxies with SCUBA data and CO and \ion{H}{1} observations,
      and assuming $\XCO=4\times10^{20}\cm^{-2}(\K\kms)^{-1}$,
      the median $\Mdust/\MH=0.0052$ 
      (see Fig.\ \ref{fig:histogram MH/Md_scuba_all}) -- this is
      similar to $\Mdust/\MH\approx\oldtext{0.007}\newtext{0.0073}$ estimated for
      the Milky Way based on observed depletions 
      (see Table \ref{tab:MW depleted mass}),
      or the value $\sim$0.010 estimated from dust models that
      reproduce the observed Milky Way extinction (see
      Table \ref{tab:dust models}).
\item When submm observations are unavailable, the mass of cool dust
      in a galaxy is poorly constrained.  However, we use
      the model fits for the \Nscuba\ galaxies where SCUBA data are
      available to test a ``restricted'' fitting procedure
      that can be applied when submm observations are unavailable.
      For the \Nscuba\ galaxies where global SCUBA data are available, the
      restricted fitting procedure recovered
      the dust mass to within a factor $1.5$ in 11/\Nscuba\ cases
      (see Fig.\ \ref{fig:scuba_v_descuba}), and to within a factor 2.2
      in every case.
      This restricted fitting procedure can therefore be used to
      estimate dust masses and starlight intensities in galaxies
      with dust emission detected in the three MIPS bands.
\item For the 20 galaxies
      which lack SCUBA fluxes but
      for which the gas mass is known, the median
      $\Mdust/\MH \approx 0.0088$ 
      (see Fig.\ \ref{fig:noscuba dust to gas histogram}).
      \oldtext{%
      Some galaxies appear to have very low global dust-to-gas mass ratios:
      for example, IC~2574 and NGC~4236 have
      $\Mdust/\MH \approx 0.0005$ 
      (see Fig.\ \ref{fig:noscuba dust to gas histogram}),
      a factor $\sim$15 below the dust-to-gas mass ratio for the
      Milky Way.
      }
\item By comparing dust-to-gas mass ratios computed with different assumed
      values of the CO to molecular mass conversion factor $\XCO$,
      we find that the dust-to-gas mass ratio normalized to galaxy metallicity
      gives dust masses consistent with
      available metal abundances for 
      $\XCO\approx4\times10^{20}\cm^{-2}(\K\kms)^{-1}$
      \newtext{(see Fig.\ \ref{fig:Mdust/Mgas vs metallicity})}.
\item All 49 galaxies 
      in Figure \ref{fig:Mdust/Mgas with ul vs metallicity}
      with $\OH>8.2$
      appear to have dust masses
      consistent (to within a factor $\sim$3)
      with eq.\ (\ref{eq:Md/MH envelope}), corresponding
      to the total mass of dust in the WD01 models for Milky Way dust
      (see Table \ref{tab:dust models})
      scaled by metallicity.  This is close to the mass that
      will be formed if silicates are formed from the bulk of the
      interstellar Mg, Si, and Fe, and 
      $\sim$50\% of the interstellar carbon
      goes into carbonaceous grains
      (see Table \ref{tab:MW depleted mass}).

      All 9 of the galaxies in 
      Fig.\ \ref{fig:Mdust/Mgas with ul vs metallicity}
      with metallicities
      $\OH<8.1$ appear to have {\it global} dust-to-gas ratios 
      that fall below
      eq.\ (\ref{eq:Md/MH envelope}), in some cases by large factors:
      NGC~2915 appears to be at least a factor 40 below 
      eq.\ (\ref{eq:Md/MH envelope}).
      However, the dust-to-gas ratio {\it in the regions where dust
      emission is detected} appears to be close to 
      eq.\ (\ref{eq:Md/MH envelope}).
      The low apparent global dust-to-gas ratios are found in 
      dwarf galaxies with extended \ion{H}{1} envelopes.  
      When allowance is made for the lower metallicity and lower starlight
      intensity in the envelopes, it appears that the
      dust-to-gas mass ratios may be consistent with eq.\ (13).

\item We characterize the PAH abundance by a quantitative ``PAH
      index'' $\qpah$, defined to be the fraction of the total dust mass
      contributed by PAHs containing less than $10^3$ C atoms.
      The \Npah\ galaxies in this sample for which we are able to
      estimate $\qpah$ span the range of PAH index
      from 0.4 to 4.7\%, with a median 
      $\qpah=\oldtext{3.2}\newtext{3.4}\%$
      (see Fig.\ \ref{fig:overall pah histogram}).

\item The PAH index $\qpah$ is strongly correlated with metallicity
      (see Fig.\ \ref{fig:pah index vs metallicity}).
      The nine galaxies with $\OH<8.1$ have $\qpah\leq 2\%$,
      and median 
      $\qpah=\oldtext{1.1}\newtext{1.0}\%$,
      whereas the 52 galaxies with $\OH>8.1$ have
      median 
      $\qpah=\oldtext{3.5}\newtext{3.55}\%$.  
      Metallicity $\OH=8.1$
      appears to mark a transition in the composition of interstellar dust
      in galaxies, from low-PAH to high-PAH.
      Possible reasons for the observed variation in $\qpah$ are discussed
      in \S\ref{sec:PAH abundances - scuba plus nonscuba}.

\newtext{%
\item Except for three starbursting galaxies (Mrk~33, Tolo~89, NGC~3049),
      the infrared emission from dust is dominated by dust grains
      exposed to starlight intensities $U\approx\Umin$, interpreted
      here as dust in diffuse regions heated by the general diffuse
      starlight background (see Fig.\ \ref{fig:pdr power distribution}a).  
      Half of the galaxies have more than 88\%
      of the total dust power provided by this component.
      }

\item The SEDs for the star-forming galaxies in the sample require
      that a significant fraction of the dust luminosity be produced
      by warm dust in regions of intense starlight.  Half of the galaxies
      in the sample have more than 8\% of the total dust luminosity
      originating in regions with $U>10^2$ 
      (see Fig.\ \ref{fig:pdr power distribution}).
      This appears to be consistent with the expected
      heating of dust in PDRs near massive stars.
\end{enumerate}
\acknowledgements
We thank S.~Madden and the anonymous referee for helpful comments.
Support for this work, part of the {\it Spitzer Space Telescope} Legacy
Science Program, was provided by NASA through 
the Jet Propulsion Laboratory under NASA contract 1407.
BTD was supported in part by NSF grant AST-0406883.

BTD is grateful to R.H. Lupton for availability of the SM graphics
program.

\bibliography{btdrefs}

\newpage
\appendix
\section{\label{sec:nonscuba SEDs}
  Spectral Energy Distributions for 48 Galaxies Lacking Submm Data}
\vspace*{-1.0em}
\begin{figure}[h]
\begin{center}
\includegraphics[angle=0,width=15.0cm]{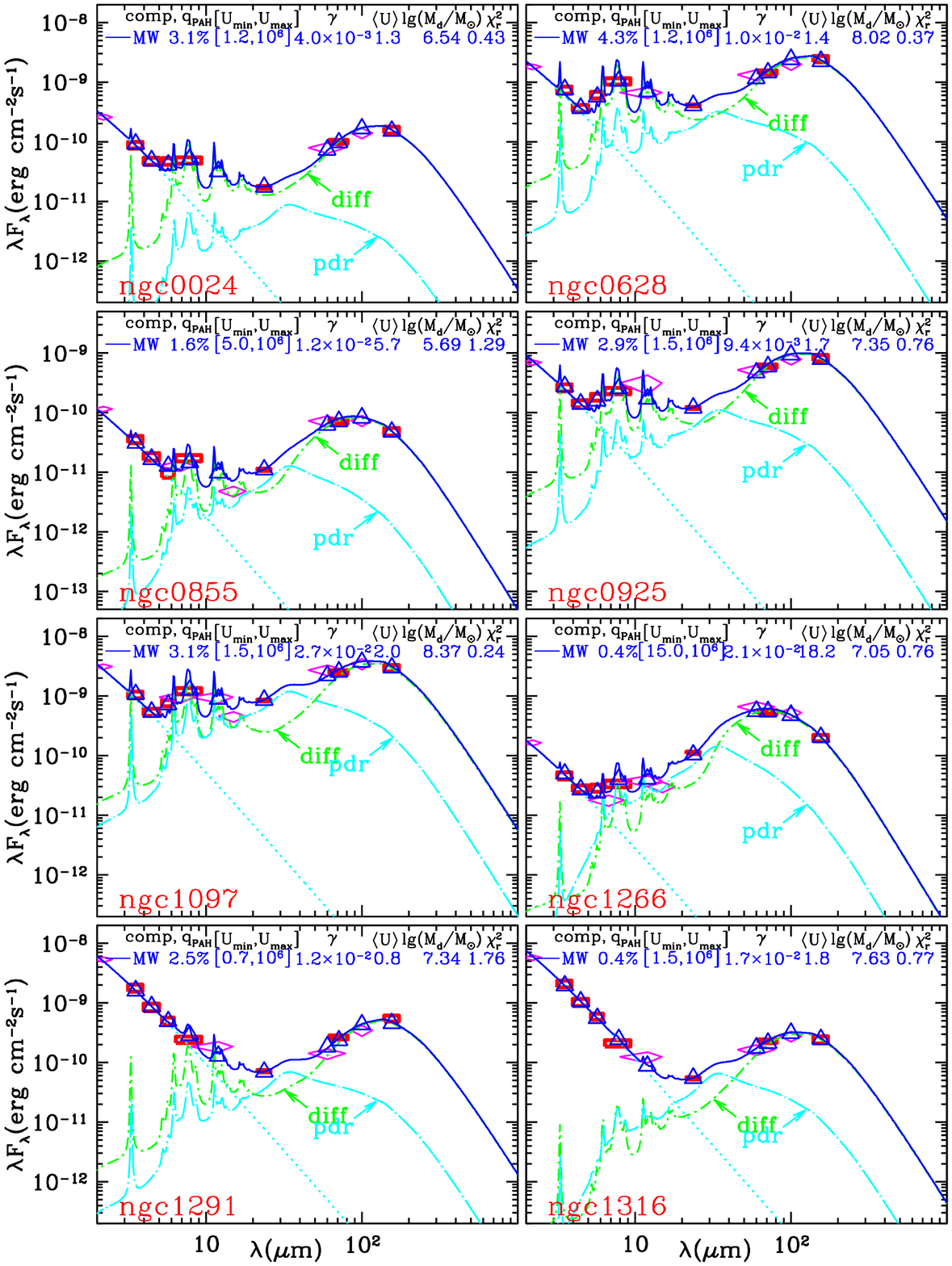}
\vspace*{-1.0em}
\caption{\footnotesize
Same as Fig. \ref{fig:scubagals}, but
for galaxies with only IRAC, MIPS, and IRAS data.
         \label{fig:noscuba}
         \label{fig:n0024}
	 \label{fig:n0628}
	 \label{fig:n0855}
	 \label{fig:n0925}
	 \label{fig:n1097}
	 \label{fig:n1266}
	 \label{fig:n1291}
         \label{fig:n1316}
	 \label{fig:n1512}
         \label{fig:n1566}
	 \label{fig:n1705}
         \label{fig:n2403}
         \label{fig:holmII}
         \label{fig:ddo053}
         \label{fig:n2841}
	 \label{fig:n2915}
	 \label{fig:holmI}
         \label{fig:n3049}
         \label{fig:n3031}
         \label{fig:n3184}
	 \label{fig:n3198}
	 \label{fig:ic2574}
	 \label{fig:n3265}
         \label{fig:n3351}
         \label{fig:n3621}
         \label{fig:n3773}
         \label{fig:n3938}
         \label{fig:n4125}
         \label{fig:n4236}
         \label{fig:n4254}
         \label{fig:n4321}
         \label{fig:n4450}
         \label{fig:n4559}
         \label{fig:n4579}
         \label{fig:n4594}
         \label{fig:n4625}
         \label{fig:n4725}
         \label{fig:n4736}
	 \label{fig:n5033}
         \label{fig:n5055}
         \label{fig:n5194}
         \label{fig:n5398}
         \label{fig:n5408}
         \label{fig:n5474}
         \label{fig:ic4710}
         \label{fig:n6822}
	 \label{fig:n6946}
         \label{fig:n7793}
}
\end{center}
\vspace*{-5.0em}
\end{figure}
\clearpage

\begin{figure}[h]
\begin{center}
\includegraphics[angle=0,width=15cm]{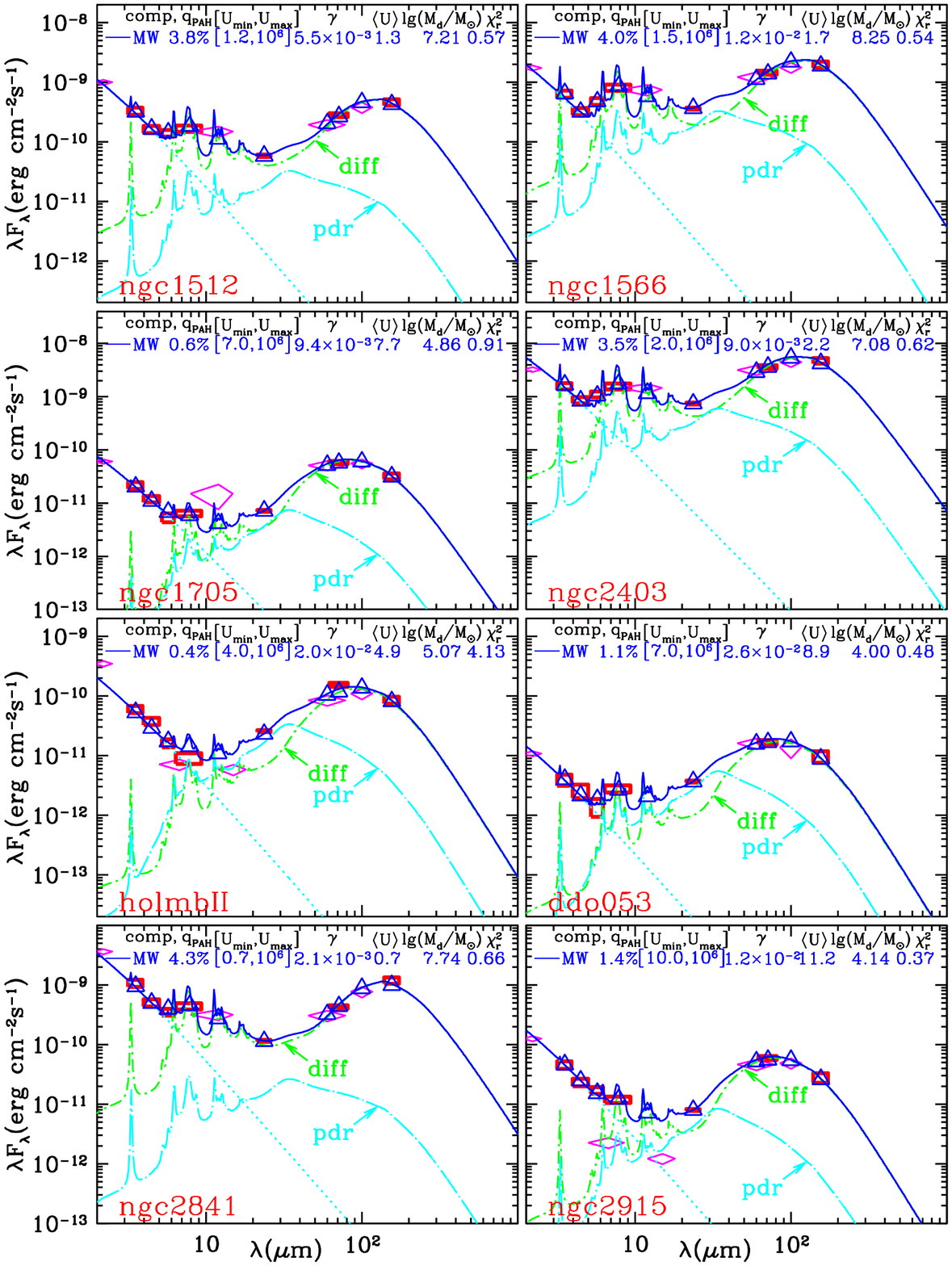}

\vspace*{-0.5em}
{\footnotesize Fig. \ref{fig:noscuba}, continued.}
\end{center}
\end{figure}
\clearpage

\begin{figure}[h]
\begin{center}
\includegraphics[angle=0,width=15.0cm]{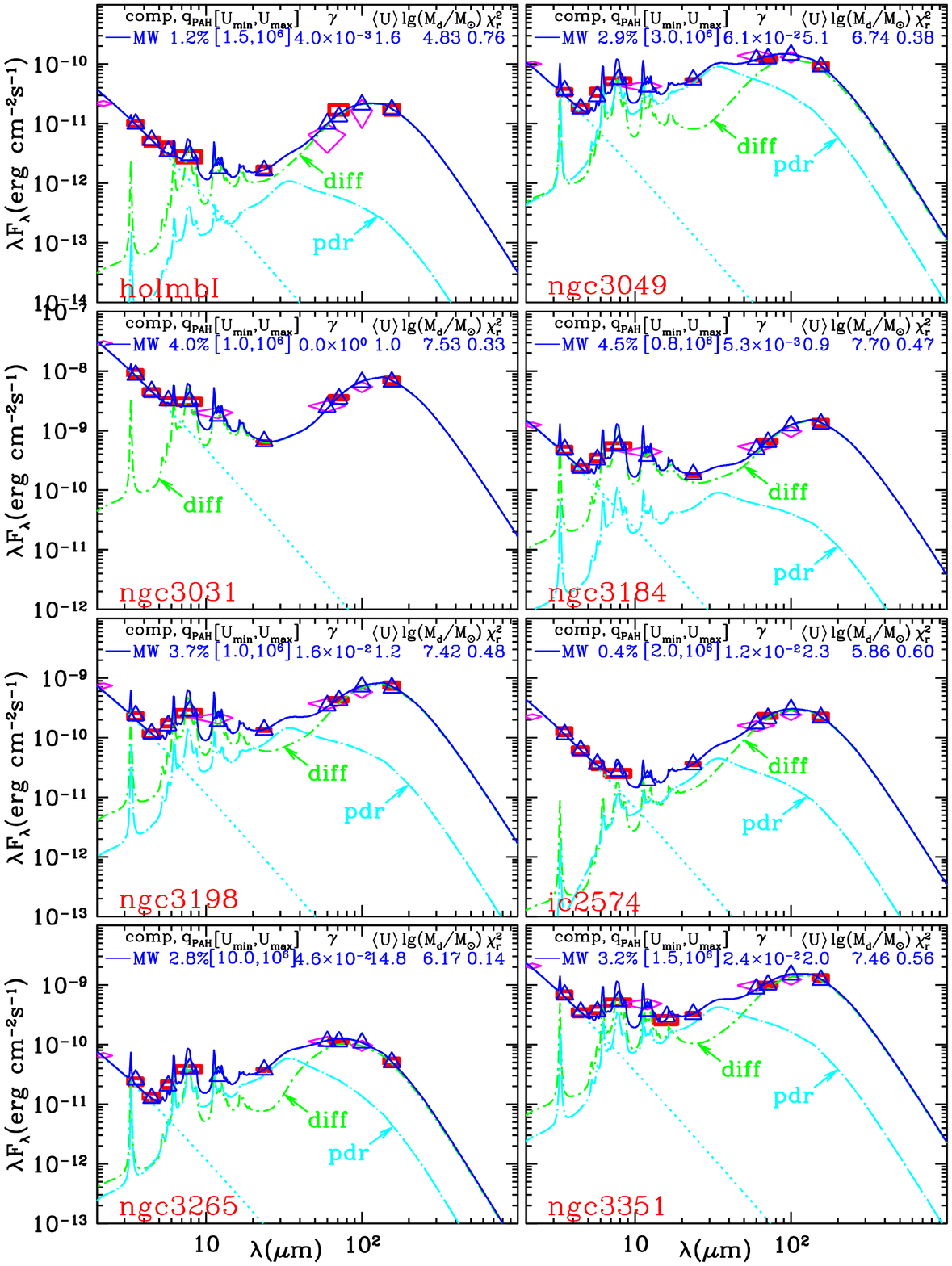}

\vspace*{-0.5em}
{\footnotesize Fig. \ref{fig:noscuba}, continued.}
\end{center}
\end{figure}
\clearpage

\begin{figure}[h]
\begin{center}
\includegraphics[angle=0,width=15.0cm]{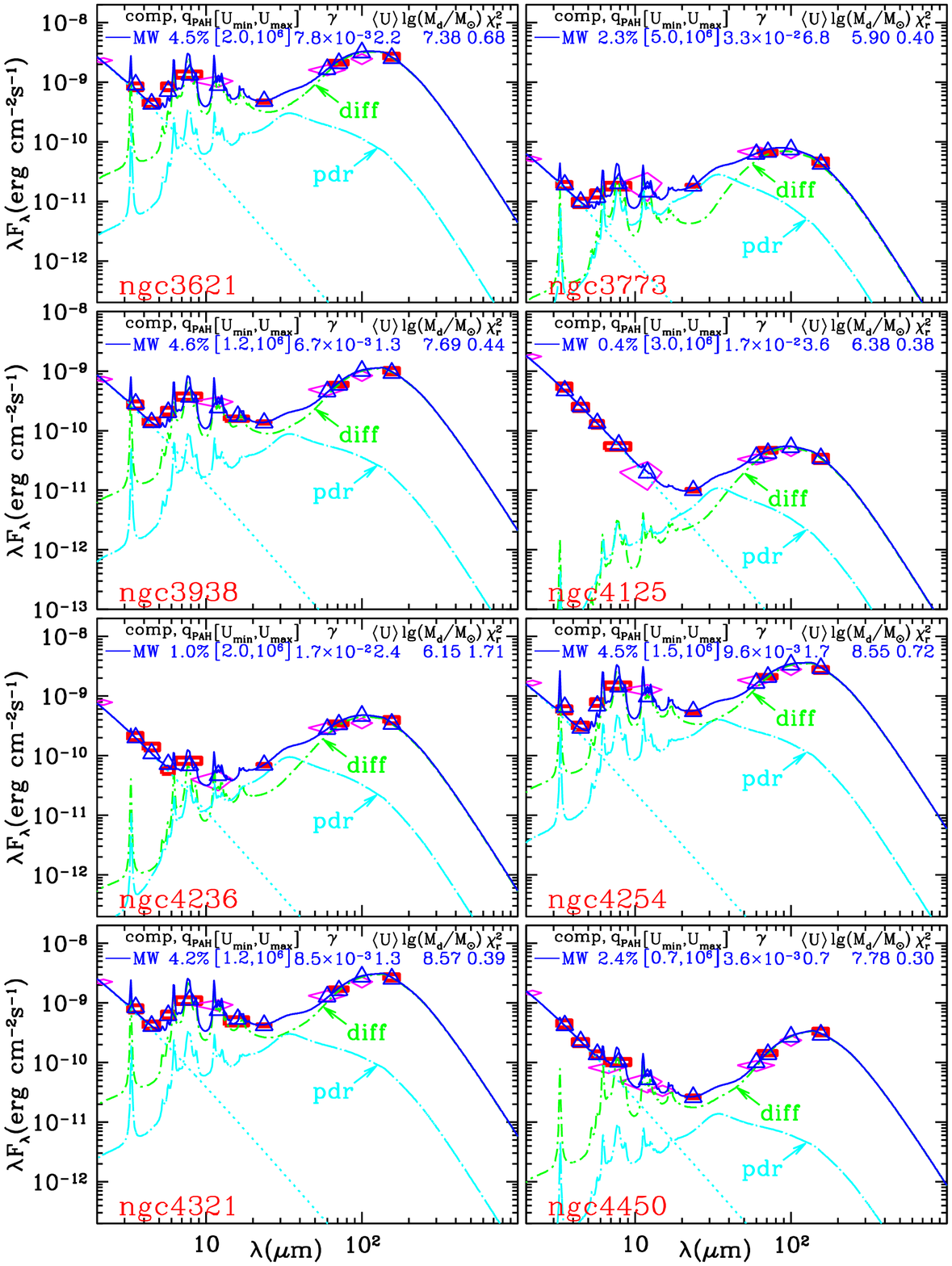}

\vspace*{-0.5em}
{\footnotesize Fig. \ref{fig:noscuba}, continued.}
\end{center}
\end{figure}
\clearpage

\begin{figure}[h]
\begin{center}
\includegraphics[angle=0,width=15.0cm]{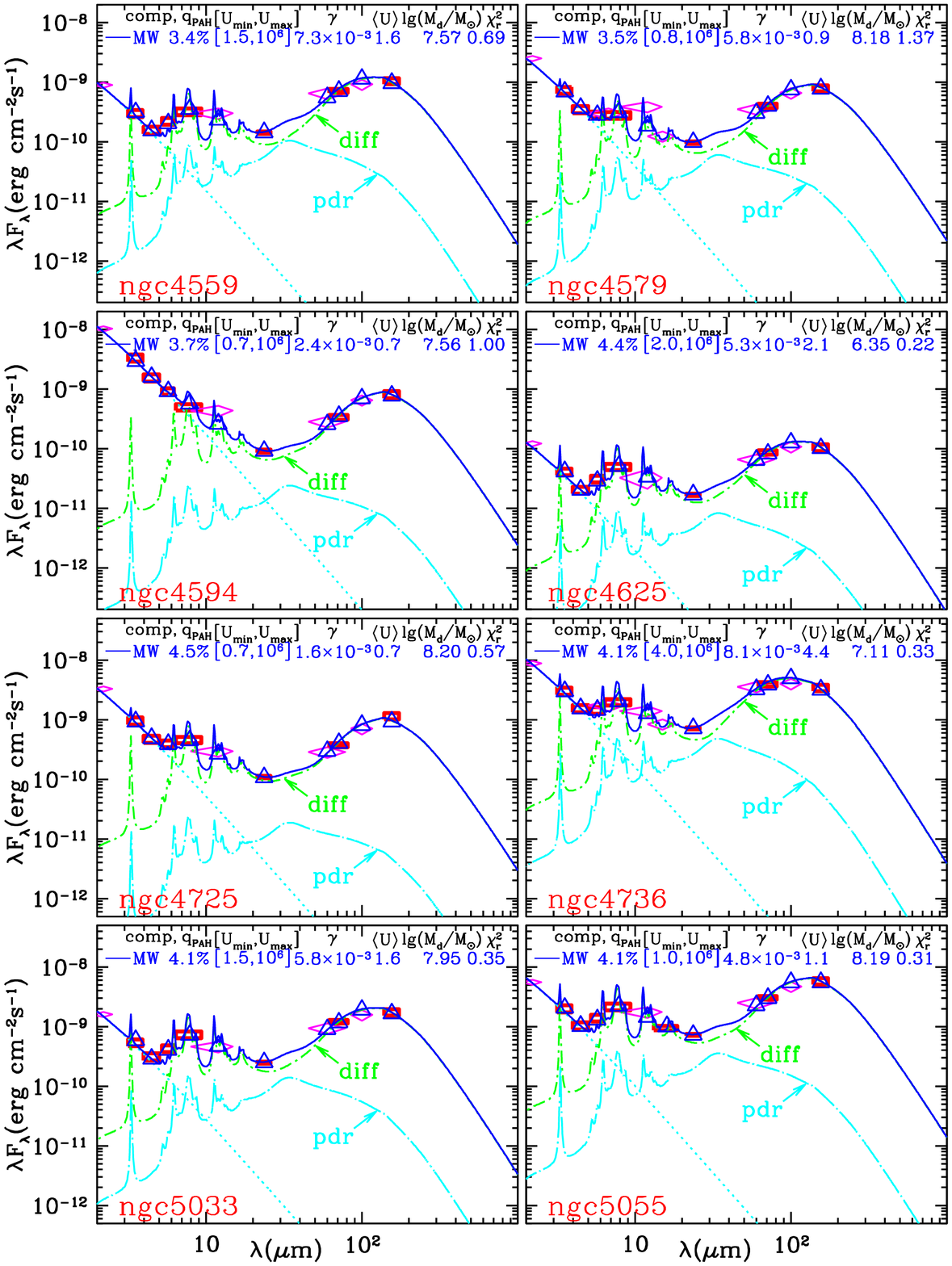}

\vspace*{-0.5em}
{\footnotesize Fig. \ref{fig:noscuba}, continued.}
\end{center}
\end{figure}
\clearpage

\begin{figure}[h]
\begin{center}
\includegraphics[angle=0,width=15.0cm]{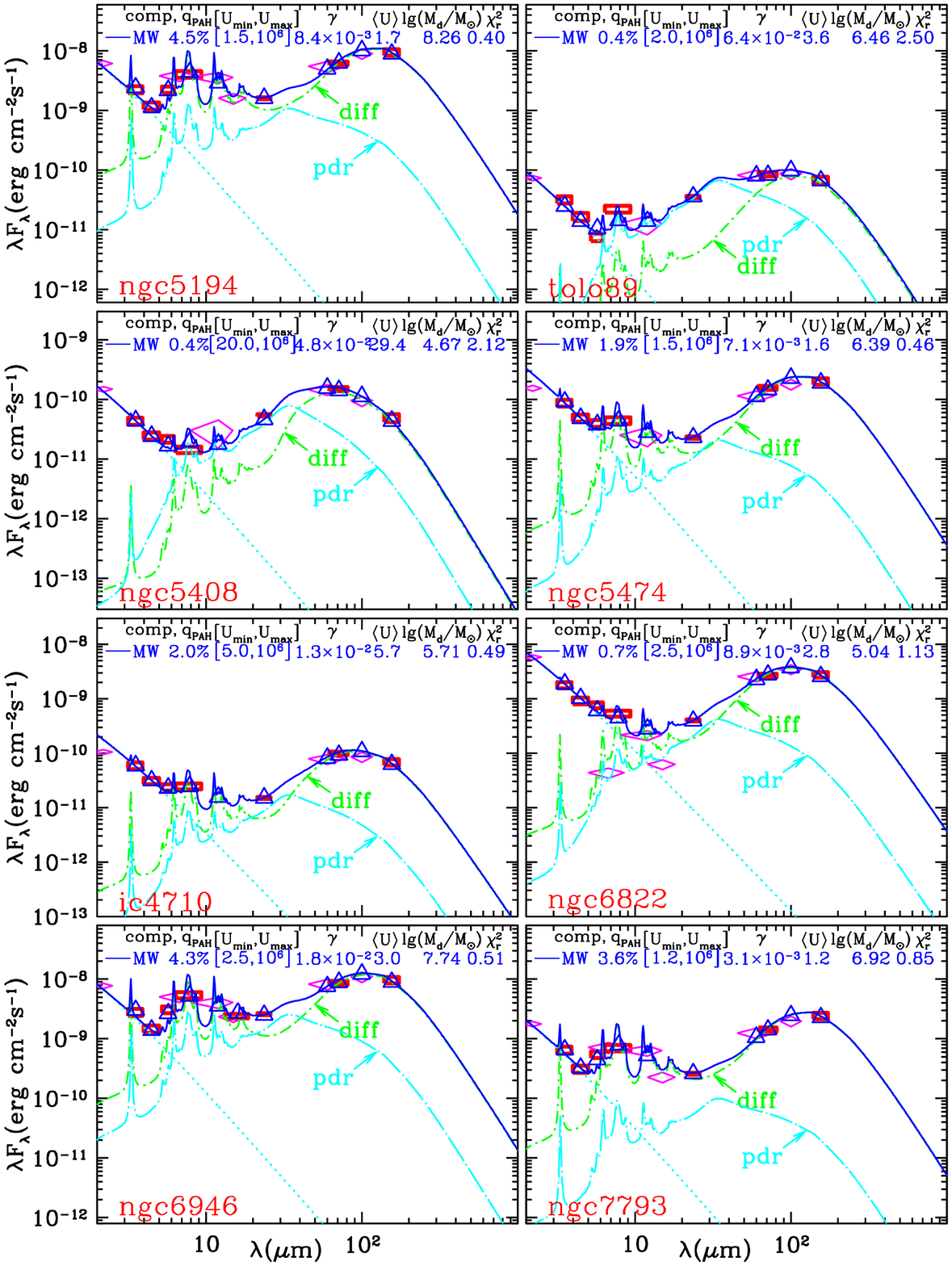}

\vspace*{-0.5em}
{\footnotesize Fig. \ref{fig:noscuba}, continued.}
\end{center}
\end{figure}
\end{document}